\documentclass[11pt,a4paper]{article}
\usepackage{jheppub}
\usepackage{subfigure}
\usepackage{epic}

\newcommand{\beq}{\begin{equation}}
\newcommand{\eeq}{\end{equation}}
\newcommand{\non}{\nonumber\\}

\title{On periodically driven AdS/CFT} 

\author[a]{Roberto Auzzi,}
\author[b]{Shmuel Elitzur,}
\author[b]{Sven Bjarke Gudnason}
\author[a,b]{and Eliezer Rabinovici}

\affiliation[a]{CERN Dep PH-TH, 1211 Geneva 23, Switzerland}
\affiliation[b]{Racah Institute of Physics, The Hebrew University,
 Jerusalem 91904, Israel}

\emailAdd{roberto.auzzi(at)cern.ch}
\emailAdd{elitzur(at)vms.huji.ac.il}
\emailAdd{gudnason(at)phys.huji.ac.il}
\emailAdd{eliezer(at)vms.huji.ac.il}

\abstract{
We use the AdS/CFT correspondence to study a thermally isolated 
conformal field theory in four dimensions which undergoes a repeated
deformation by an external periodic time-dependent source coupled to an operator
of dimension $\Delta$. 
The initial  state of the theory is taken to be at a finite temperature.
We compute the energy dissipated in the system as a
function of the frequency and of the dimension $\Delta$ of the perturbing operator.
This is done in the linear response regime.
In order to study the details of thermalization in the dual field theory,
the leading-order backreaction on the AdS black brane metric is computed.
The evolution of the event and the apparent horizons is monitored;
the increase of area in each cycle coincides with the increase in the equilibrium entropy
corresponding to the amount of energy dissipated.
The time evolution of the entanglement entropy of a spherical region and that of the two-points function 
of a probe operator with a large dimension are also inspected; we find a delay
in the thermalization of these quantities which is proportional to the
size of the region which is being probed.
Thus, the delay is more pronounced in the infrared.
We comment on a possible transition in the time evolution of the energy
fluctuations.
}

\begin{document}

\begin{flushright}
CERN-PH-TH/2013-172\\
\end{flushright}

\maketitle

\section{Introduction}

The framework of gauge/gravity duality is 
 a particularly useful setup to study how classical gravitational singularities
 are resolved in a quantum theory of gravity. 
 Decay of a false vacuum in AdS \cite{Coleman:1980aw}
 generically results in a big-crunch classical geometry.
 Studies have shown that some of such crunches are unresolved
 and lead  to unbounded ill-defined boundary theories \cite{HH1,HH2,Elitzur:2005kz,de Haro:2006nv,Barbon:2010gn}.
 Others seem unresolved with impunity and are nevertheless expressed by well-defined boundary conformal field theories (CFT) \cite{Maldacena:2010un,Barbon:2011ta}.
 The class of periodic time-dependent Hamiltonians which are unbounded
for half of the time and bounded the other half, provide an interesting
situation that is "midway" between the case of bounded and unbounded potentials.
Depending on the frequency of the time-dependent part of the Hamiltonian,
 it can happen that the crunch is shielded by a black hole (BH) horizon \cite{Auzzi:2012ca}.
This process is dual to thermalization in the boundary field theory. 
Finite temperature  can act as a stabilization mechanism  
by providing a positive thermal mass on top of possible tachyonic 
direction(s) in the field space.

Moreover, understanding the dynamics of thermally isolated systems that are
driven or quenched by an external force,  modeled by a
time-dependent Hamiltonian, is an active area of research in condensed
matter physics (see e.g.~\cite{Polkovnikov:2010yn} for a
review). Theoretical research is motivated by many recent advances in
experimental studies of cold atom systems. Near quantum phase
transitions, long-distance physics of many-body quantum systems is
described by conformal field theories (CFT).  
Many universal properties of driven quantum systems can then be
modeled by quantum field theories with time dependent couplings, see
e.g.~\cite{Calabrese:2005in,Calabrese:2006rx,Calabrese:2007rg,Calabrese:2009qy,Sotiriadis:2010si,Hung:2012zr,Hung:2013dka}. 
It remains in general challenging to study such non-equilibrium processes at strong coupling
in higher dimensions; this motivates us to investigate these issues in
the framework of the AdS/CFT correspondence (see 
 e.g.~\cite{Bhattacharyya:2008ji,Bhattacharyya:2009uu,Das:2010yw,Basu:2011ft,Buchel:2012gw,Bhaseen:2012gg,Basu:2012gg,Nozaki:2013wia,newquench,Hartman:2013qma,Basu:2013vva,Li:2013fhw,Buchel:2013gba} for previous work on this topic).

In this paper  we consider a thermally isolated field theory which 
is driven by a periodic external perturbation $\xi(t)$ in the Hamiltonian:
\beq
\mathcal{H}= \mathcal{H}_0 + \xi(t) \delta \mathcal{H} \, , \qquad \xi(t) = \xi(t+2\pi/ \omega) \, .
\label{eq:deformation_term0}
\eeq
One can think of the periodic source $\xi(t)$ also as the vacuum expectation value (VEV)
of some field interacting with the CFT.
We analyze the case in which the undeformed Hamiltonian $\mathcal{H}_0$
belongs to a conformal field theory in $d$ dimensions with an
AdS$_{d+1}$ dual, and $\delta\mathcal{H}$ is a relevant deformation of
the form: 
\beq
\delta \mathcal{H}=\int d^{d-1} x \, \mathcal{O}_{\Delta} \, ,
\label{eq:deformation_term}
\eeq
where $\mathcal{O}_{\Delta}$ is a generic relevant scalar operator
with dimension $\Delta>\frac{d-2}{2}$, i.e.~above the unitarity bound; 
the form of the time dependence is chosen here for simplicity as
$\xi(t)=\xi_0 \cos \omega t$.  
The initial state of the theory is taken to be   at finite temperature
$T$, and we work in the limit in which the total amount of work done
on the system is much smaller than the total initial internal energy. 
The physical response of the CFT to the deformation depends significantly on the
dimension $\Delta$ of the perturbing operator. 
From dimensional analysis, in the limit $\omega \gg T$, the energy dissipated per unit of volume  
in each cycle $\mathcal{W}_c$  is expected to scale 
 as $\omega^\zeta$ where $\zeta=2
\Delta-d$; this is an increasing (decreasing) function of $\omega$ for $\Delta>d/2$ (for $\Delta<d/2$).
For $\omega \ll T$ instead $\mathcal{W}_c$ generically scales as $\omega T^{\zeta-1}$; this
is an increasing (decreasing) function of $T$ for $\Delta>(d+1)/2$ (for $\Delta<(d+1)/2$). 
 Both these expectations are confirmed in strongly-coupled theories with AdS duals.

 We use the AdS/CFT correspondence to study how the
 CFT reacts to the deformation (\ref{eq:deformation_term0}).
The initial thermal state is described by 
an AdS$_{d+1}$  black brane geometry, and the operator 
$\mathcal{O}_\Delta$ is dual to a scalar field $\phi$ with mass squared
$m^2=\Delta(\Delta-d)$. The periodic deformation  (\ref{eq:deformation_term0})
is then implemented as a time-dependent boundary condition for  $\phi$.
In the linear approximation,  $\mathcal{W}_c$
can be determined by computing scalar two-point functions
(see e.g.~\cite{Son:2002sd,Starinets:2002br,Nunez:2003eq,Hartnoll:2005ju}; 
for a review of linear responses in AdS/CFT, see
e.g.~\cite{Son:2007vk}). 
In practice, this amounts to solving the wave equation for $\phi$ in the black brane geometry. 
We specialize to the case of $d=4$ and
compute $\mathcal{W}_c$  numerically as a function of $\omega$ and $\Delta$.
 In both the limits $\omega \gg T$ and $\omega \ll T$, the results are consistent
 with the ones obtained by dimensional analysis and by the Kramers-Kronig relations.
In the limit where the dimension of the driving operator 
approaches the unitarity bound $\Delta\to 1$,  
$\mathcal{W}_c$ becomes a sharply peaked function at a small
frequency $\omega_m$; numerically we find that $\omega_m$
scales as $(\Delta-1)^{1/2}$. 

While the energy $\mathcal{W}_c$ injected into the system  
by the external source during one cycle tends to drive  the system away from its equilibrium thermal state,
the self interactions of the system lead to a restoration
of thermal equilibrium  by entropy production, which
on the gravity side  is dual to an increase of the area of the horizon
\footnote{Some studies on thermalization in AdS/CFT can be found in 
\cite{Chesler:2008hg,Chesler:2009cy,Bhattacharyya:2009uu,Heller:2012je,Hubeny:2007xt,AbajoArrastia:2010yt,Albash:2010mv,Balasubramanian:2010ce,Balasubramanian:2011ur,Aparicio:2011zy,Balasubramanian:2011at,Allais:2011ys,Balasubramanian:2012tu}. }.
In order to study the details of thermalization in the dual field theory,
we compute the leading order gravitational
backreaction in the expansion parameter $\xi_0$  on the bulk AdS$_5$ metric.
The number of cycles is taken to be large but finite, in such a way
that the total work performed on the system is small compared to the initial
temperature; this can be achieved by taking the parameter $\xi_0$ to
be sufficiently small. We indeed find that the thermalization
takes place in a very efficient way.

Equipped with the calculation of the leading correction to the metric,
we monitor the time evolution of a few physical observables.
We follow the time evolution of the event horizon
and we check and confirm that it increases with time. It is also possible to
follow another quantity, the apparent horizon, which was advocated to play
an important role in the CFT \cite{Chesler:2008hg,Figueras:2009iu}.
The area of the apparent horizon is not always a monotonic function of the time $t$
for intermediate values of $t$ \cite{Wald:1991zz}. However in this case we found that also the area of the apparent horizon increases.
 The increase of area in each cycle is the same for both definitions of horizon,
and it coincides  with the increase in the equilibrium entropy
 corresponding to a change in the internal energy given by the work done in a cycle;
 this is true for all the values of $1<\Delta\leq 4$.
 
We next investigate the time evolution of the two-point function of a probe
operator with a large dimension $\Delta_p$, which in the holographic
dual is related to the geodesic length, and of the entanglement
entropy of a spherical region, which is probed by the volume of minimal surfaces.
For both the quantities there is a delay in the thermalization 
which is proportional to the size $L$ of the region which is probed,
in the limit $L \gg 1/T$. 
The thermalization time is longer for infrared observables;
this kind of behavior has been observed before in different systems in  \cite{AbajoArrastia:2010yt,Albash:2010mv,Balasubramanian:2010ce,Balasubramanian:2011ur}. 
The two-point function and the entanglement entropy have also
an oscillatory term in time.
The amplitude of this oscillation is not extensive in the spatial region which is probed,
  and so it is negligible in the limit of large region.

It was pointed  out \cite{nature}  that the energy fluctuations of driven systems are significantly 
affected in a universal way by the protocol by which the periodic driving force is applied.
In our theoretical setting we show that these universal features are determined by $\Delta$ and $\omega$.
The results in \cite{nature} raise the possibility of a transition in the time dependence of energy fluctuations.
We discuss these features in the context of quantum field theory and we find that the
operator spectrum of $\mathcal{N}=4$ Super Yang-Mills (SYM) allows for such a transition.

In section \ref{sec:periodic_cft} 
 we make some general considerations based on dimensional analysis which 
  apply to a generic driven CFT;
we also comment on the relation to the case of quenches.
In section \ref{sec:periodic_adscft} the theoretical setting
is introduced, including the equations used to compute the backreaction of the metric. 
Linear response functions are computed numerically in section
\ref{sec:linear_response}. In section \ref{sec:geometry} we discuss
the apparent and event horizons and we compute the leading correction
to the two-point function and the entanglement entropy. 
In section \ref{sec:fluctuations} we initiate the study of energy fluctuations.
Section \ref{sec:conclusion} concludes with a discussion of the results.
Finally,  appendix \ref{appendix:A} contains some technical details for the special
cases of integer and half-integer $\Delta$s and appendix \ref{appendix:B}
deals with the effective Hamiltonian of a free field theory for large frequencies.

\section{Periodically driven CFT\label{sec:periodic_cft}}

One of the main tools in studying the reaction of a medium to a weak perturbation
which tends to drive it out of equilibrium is the linear response theory.
In this section we review some basic definitions and we make some universal
considerations based on dimensional analysis.
In the linear approximation, the response to the deformation
\eqref{eq:deformation_term} is encoded in the retarded Green function,
which can be defined in terms of a general spacetime-dependent deformation:
\beq
\delta {\mathcal{L}} = -\int d^d x \, \mathcal{O}_{\Delta} J_{\Delta}(x) \, .  
\eeq
The source $J_{\Delta}(x)$ has dimension $d-\Delta$. 
In the linear response regime, where $J_{\Delta}(x)$ is sufficiently small,
the VEV of the operator is given by 
\beq
\langle \mathcal{O}_\Delta(x) \rangle =-\int  d^d y \,  G^R(x-y) J_\Delta(y) \, ,
\eeq
where $G^R$ is the retarded Green function:
\beq
G^R(x-y)=-i \theta_H (x^0-y^0) \langle [ \mathcal{O}_\Delta(x),
  \mathcal{O}_\Delta(y) ] \rangle \, , 
\eeq
and $\theta_H$ is the Heaviside step function.
Here we will focus on a time-dependent source -- which is constant in
space -- of the form $J_\Delta(t)=\xi_0 \,  {\rm Re } ( e^{-i \omega
  t}) $; for small $\xi_0$, we will be in the linear-response regime
where 
\beq
\langle \mathcal{O} \rangle = \xi_0  \, {\rm Re } \left(
G^R(\omega,\vec{p}=0) \, e^{-i \omega t} \right) \, . 
\eeq
The energy per unit volume, which has dissipated in a period
$2\pi/\omega$, denoted by $\mathcal{W}_c$, is proportional to
$\xi_0^2 \, {\rm Im} \, G^R(\omega,0) $. 

\begin{figure}[ht]
\begin{center}
$\begin{array}{c@{\hspace{.2in}}c} \epsfxsize=6.5cm
\epsffile{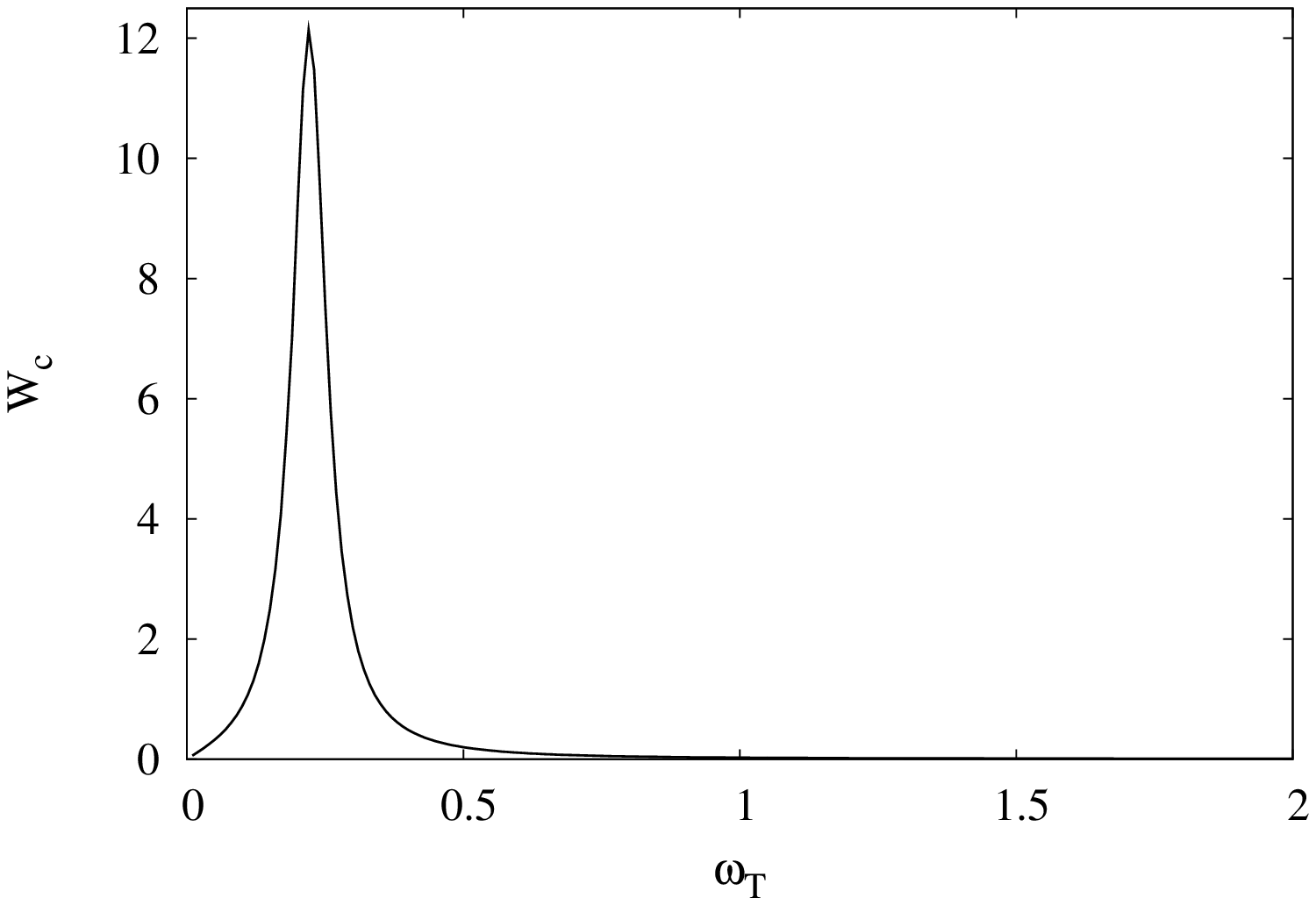} &
     \epsfxsize=6.5cm
    \epsffile{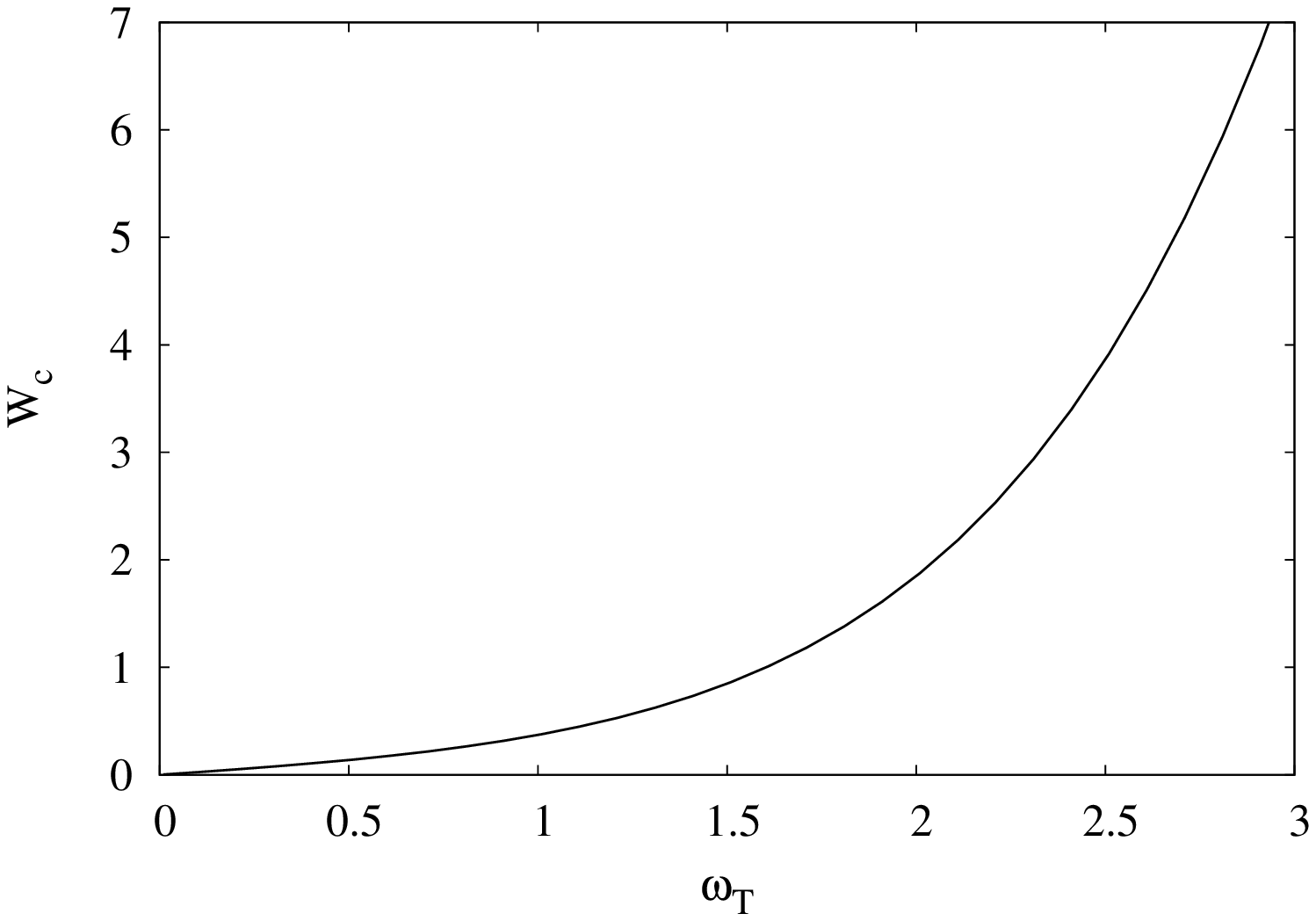}
\end{array}$
\end{center}
\caption{\footnotesize
 $\mathcal{W}_c$ as a function of $\omega_T=\omega/(\pi T)$; in the left panel $\Delta<d/2$ 
and hence $\mathcal{W}_c$ vanishes for large $\omega$; in the right panel $\Delta>d/2$, so $\mathcal{W}_c$ is an increasing function
for large $\omega$. In both cases, $\mathcal{W}_c$ is linear in $\omega$ for $\omega \ll T$.
} \label{esempi}
\end{figure}

The functions ${\rm Re} \,G^R(\omega,0)$ and ${\rm Im} \,G^R(\omega,0)$
are respectively even and odd (this is a consequence of the fact that  $G^R(\omega,0)$
is the Fourier transform of a real function).
 For frequencies $\omega \gg T$,  the behavior $\mathcal{W}_c \propto
\omega^\zeta$ where $\zeta={2 \Delta -d}$ is expected from dimensional
analysis. This expectation has been confirmed in CFTs
 with an  AdS dual,  see e.g. Appendix A of Ref. \cite{Son:2002sd}; 
this  will be reviewed in sections \ref{sec:periodic_adscft} and  \ref{sec:linear_response},
  see eqs.~(\ref{upupa},\ref{grande-omega},\ref{grande-omega-delta-intero}).  
In the same class of theories, in the case of integer  $\Delta$s in even dimensions and of half-integer $\Delta$s in odd dimensions,
 ${\rm Re} \,G^R(\omega,0)$ instead scales as $\omega^\zeta \log(\omega/T)$,
 while it scales as $\omega^\zeta$ for generic dimensions,  see eqs.~(\ref{grande-omega},\ref{grande-omega-delta-intero}).
  For $\frac{d-2}{2}<\Delta<\frac{d}{2}$
the energy dissipated per
unit volume in a cycle, $\mathcal{W}_c$, 
is a decreasing function of $\omega$ for large $\omega\gg T$.
In the opposite limit $\omega\ll T$ one expects that ${\rm Re} \,
G^R(\omega,0)$ approaches a constant (the equilibrium value), while 
${\rm Im} \, G^R(\omega,0)$, being an odd function of $\omega$, is expected to generically start off as
$ \omega \, T^{\zeta-1}$. This behavior will be confirmed in the
numerical analysis of section \ref{sec:linear_response}.
Examples of the behavior of $\mathcal{W}_c$ are shown in figure
\ref{esempi}.

In the case of a free field theory, the retarded propagator is:
\beq G^R(\omega,\vec{k})=-P \frac{1}{\omega^2-\vec{k}^2-m^2} + i \pi \delta(\omega^2-\vec{k} -m^2) \, {\rm sign}(\omega) \, , \eeq
where $P$ denotes the principal value.
In the limit $m \rightarrow 0$, the two $\delta$ functions collide at $(\omega,\vec{k}) \rightarrow 0$.
We will find a similar behavior in AdS/CFT in the limit $\Delta \rightarrow 1$,
with the difference being that the $\delta$s are smeared into a
continuous peak. 

Previous studies on driven systems have focused on the case of quenches,
where the time dependence of the coupling interpolates between
two different asymptotic Hamiltonians, 
i.e.~ $\xi(t)=\xi_i + (\xi_f-\xi_i) \theta_H(t)$.
The effective Hamiltonian approach \cite{Rahav:2003b}, \cite{Auzzi:2012ca}
suggests that for large frequencies, $\omega$, the effect of introducing a periodic deformation
can be described as a quench. A simple case where $\Delta=2$ is discussed in appendix \ref{appendix:B},
where we show that the averaged long-time behavior is that of the quench to leading order in $1/\omega^2$; 
in this approximation the energy dissipation is neglected.

 The authors of \cite{Buchel:2012gw,newquench,Buchel:2013gba} recently
studied quenches in AdS/CFT as a function of the operator dimension
$\Delta$, in $d=4$. In order to approach the limit of $\theta_H (t)$ in a smooth way,
they considered the class of functions: 
\beq
\xi(t)=\xi_0 f(t)=\xi_0 (1+\tanh t/t_0)/2 \, ,
\eeq
 where $t_0$ is a time scale, as a
regularization of $\theta_H$. 
In the limit $t_0 \rightarrow 0$, they found by numerical analysis
that the energy dissipated in the quench is a divergent quantity for
$\Delta \geq 2$, which scales as $1/t_0^{2 \Delta-4}$ (as $\log 1/t_0$
for $\Delta=2$). 

We point out that this divergence is a direct consequence of the fact that
$\mathcal{W}_c$ scales as $\omega^{\zeta}$ in the large $\omega$ limit
of the periodically driven system; the quantity $1/t_0$ acts as a UV
cutoff of the Fourier modes of $\xi(t)$. 
The Fourier transform of the source $f(t)$ is:
\beq
\mathcal{F}(f(t))=\frac{i \sqrt{\pi}}{2 \sqrt{2}} \frac{t_0}{\sinh (\pi t_0 \omega/2)} \, .
\eeq
In the case of $\zeta>0$, the energy dissipated at small $t_0$,
corresponding to a very fast quench, scales as 
\beq
\int^{1/t_0} \frac{{\rm Im} \, G_R(\omega)}{\omega} d \omega \propto
\int^{1/t_0} \frac{\omega^\zeta}{\omega} d \omega \propto \frac{1}{t_0^\zeta} \, 
\eeq
where $1/t_0$ is practically a UV cutoff for
$\mathcal{F}(\theta_H)=\sqrt{\frac{\pi}{2}} \delta(\omega) +
\frac{1}{\sqrt{2 \pi} \omega}$. 
This is in agreement with the numerical result obtained in
\cite{Buchel:2012gw,newquench}.
In the case of $\zeta<0$ there is no divergence, so the energy
dissipated in the process should approach a constant in the limit 
of a sudden quench. 
 For $\zeta=0$, which in $d=4$ corresponds to $\Delta=2$, the divergence
is logarithmic.
We will now turn to the study of periodically driven strongly coupled
systems with the aid of the AdS/CFT correspondence.

\section{Periodically driven AdS/CFT\label{sec:periodic_adscft}}

\subsection{Theoretical setting}

In this section we review work on 
 the gravity dual of a conformal field theory perturbed by a relevant scalar operator
 of dimension $\Delta$ and then adopt it to the purpose of studying the
time dependent periodic driving force dealt with in this work.
We use as an expansion parameter the magnitude of the amplitude
of the time-dependent perturbation. 
The gravitational theory is Einstein gravity coupled to a massive
scalar field: 
\beq
S=\frac{1}{16 \pi G_{N}^{(d+1)}} \int d^{d+1} x \sqrt{-g} \left( R+d(d-1) -\frac{1}{2} (\partial \phi)^2 -\frac{1}{2} m^2 \phi^2 \right) \, .
\eeq
We consider generic values of the mass $m^2 \leq 0$,
which on the boundary corresponds to different operator dimensions
$\Delta(\Delta-d)=m^2$, giving the usual two solutions 
$\Delta_\pm=d/2 \pm \sqrt{m^2+d^2/4}$. 
In the range $m_{BF}=-d^2/4 \leq m^2 \leq -d^2/4+1$ we can either set  
$\Delta=\Delta_+$ or $\Delta=\Delta_-$; for $m^2 > -d^2/4 +1$ we are
forced to set $\Delta=\Delta_+$, in order to respect the unitarity
bound for the dimension of a scalar operator. 
We ignore higher order terms in the potential because they only affect 
dynamics beyond the leading order in our expansion parameter. 

For example, in the case of the AdS$_5$ dual of the $\mathcal{N}=4$
theory, we may consider an operator $\mathcal{O}_2$ which gives mass
to the scalar fields, or an operator $\mathcal{O}_3$ which gives mass
to the fermions.
The operator $\mathcal{O}_2$ corresponds to a field $\phi$ in AdS$_5$
with $m^2=-4$, while $\mathcal{O}_3$ corresponds to a field with
$m^2=-3$. 
The value of the Newton constant $G_N$ is related to the number of
colors $N_c$ of the $\mathcal{N}=4$ theory:
$ G_N=\pi/(2 N_c^2)$.
In the following we will consider the four dimensional case in
detail. 

We choose to study the CFT in flat Minkowski space and so we work in Poincar\'e patch.
In order to model a translational invariant state in the boundary CFT,
we consider the following metric:
\beq
ds^2=-A(v,r) dv^2 + \Sigma (v,r)^2 d \vec{x}^2 +2 dr dv \, .
\label{EF-metric}
\eeq
For convenience, we work in Eddington-Finkelstein (EF) coordinates. 
The AdS boundary is at $r\to\infty$, where $A(v,t)\to r^2$, $\Sigma
\to r$. The variable $v$, for large $r$,
corresponds to the time of the boundary theory. 
The Einstein equations read \cite{Chesler:2008hg,Buchel:2012gw}:
\beq
\Sigma'' +\frac{1}{6} \Sigma (\phi')^2 =0 \, , 
\qquad
\Sigma (\Sigma^\odot)'+2 \Sigma^\odot  \Sigma' -2 \Sigma^2 +\frac{1}{12} m^2 \phi^2 \Sigma^2  = 0 \, .
\label{einst}
\eeq
\[
 A'' - \frac{12}{\Sigma^2} \Sigma^\odot \Sigma' +4 +\phi^\odot \phi' -\frac{1}{6} m^2 \phi^2 = 0 \, , \qquad
 \Sigma^{\odot \odot}-\frac{1}{2}A' \Sigma^\odot +\frac{1}{6} \Sigma (\phi^\odot)^2=0 \, ,
\]
where $h^\odot \equiv \dot{h}+A h' /2$
and $'$ and $\dot{}$  denote derivatives with respect to $r$ and $v$. 
The Klein-Gordon equation reads: 
\beq
 2  \Sigma (\phi^\odot)' +3 \Sigma' \phi^\odot +3 \Sigma^\odot \phi' -m^2 \Sigma \phi  =0 \, .
\eeq
We will solve these equations in a perturbative expansion, taking as a
zeroth order solution a black brane with $\phi=0$. We will use as
expansion parameter the amplitude of the field $\phi$. 

The initial thermal state of the boundary CFT is dual
to a black brane geometry, which corresponds to:
\beq
A=r^2 -\frac{\mu^4}{r^2} \, , \qquad \Sigma=r \, , \qquad \mu=\pi T \, , 
\eeq
where $T$ is the temperature.  The expression in Schwarzschild coordinates is:
\beq
ds^2=-A dt^2 +\frac{dr^2}{A} +r^2 d \vec{x}^2\, , \qquad A=r^2-\frac{\mu^4}{r^2} \, ,
\label{S-metric}
\eeq
which can be obtained from EF by changing variables from $v$ to $t$,
with 
\beq
dt= dv -\frac{dr}{A(r)} \, , \qquad t=v-\frac{\tan^{-1} \frac{r}{\mu} -\tanh^{-1} \frac{\mu}{r} }{2 \mu} +\frac{\pi}{4 \mu} \, ,
\eeq
where the integration constant is chosen in such a way that $v=t$ for
$r\to\infty$. The location of the horizon, $r_0$, is given by the
value of $r$ for which $A$ vanishes; the temperature is then
$T=A'(r_0)/(4 \pi)$.

\subsection{Klein-Gordon equation}

We will expand around the black brane metric, taking the amplitude of the scalar $\phi$ as an expansion parameter.
It is useful to parametrize the scalar field as follows: 
\beq
\phi(v,r)=\lambda \, \tilde{\phi}_1(v,r)=\lambda \, {\rm Re} (\phi_1(r) e^{-i \omega v}),
\eeq
where $\lambda$ is an expansion counting parameter, which is
proportional to the amplitude of the driving force $\xi_0$ on the
field theory side. 
The functions $A,\Sigma$ receive corrections only at order
$\lambda^2$. 
It is convenient to introduce the variables:
 \beq 
 \rho=\mu/r \, , \qquad \tau=\mu v \, , \qquad \vec{X}= \mu x \, , \qquad
   \omega_T=\frac{\omega}{\pi T} \, .
 \label{riscalo}
 \eeq
The equation of motion for the scalar $\phi$ on the black brane background, to leading order in
$\lambda$, is: 
\beq
\partial^2_{\tau \rho} \tilde{\phi}_1 - \frac{1-\rho^4}{2} \partial^2_{\rho \rho} \tilde{\phi}_1 
-\frac{3}{2 \rho} \partial_\tau \tilde{\phi}_1+ \frac{3+\rho^4}{2 \rho} \partial_\rho \tilde{\phi}_1 +\frac{m^2}{2 \rho^2} \tilde{\phi}_1 =0 \, .
\label{wave-eq}
\eeq
Note that this equation contains only a first order time derivative,
while in Schwarzschild coordinates it contains also second order time
derivatives\footnote{In Schwarzschild coordinates, (rescaling
  $\rho=\mu/r$, 
  $\tau=\mu t$), the equation of motion is:
\[
\partial^2_{\tau \tau} \tilde{\phi}_1 + \frac{(1-\rho^4)(3+\rho^4)}{\rho} \partial_{\rho} \tilde{\phi}_1 
-(1-\rho^4)^2 \partial^2_{\rho \rho} \tilde{\phi}_1 + \frac{m^2}{ \rho^2} (1-\rho^4) \tilde{\phi}_1=0 \, .
\]
}.
In practice eq.~(\ref{wave-eq}) will be solved numerically, see section \ref{sec:linear_response}.

In order to set the correct boundary conditions and also to compute one-point functions,
it will be useful to expand $\tilde{\phi}$ as a power series in $\rho$, around the boundary at $\rho=0$.
If $m^2$ is such that $\Delta_{\pm}$ is neither integer nor
half-integer, the scalar $\phi$ can be written as:
\beq
\tilde{\phi}_1=\sum_{j=0}^{\infty}   \rho^{\Delta_{-} +j}   a_{\Delta_{-}+j}+
 \rho^{\Delta_{+} +j}   a_{\Delta_{+} +j} \, .
 \, . \label{scalare-espanso}
\eeq
The case of integer and half-integer values of $\Delta_\pm$ is treated
in appendix \ref{appendix:A}. 
The following iterative relations \cite{newquench}, which are valid
for both $\Delta=\Delta_\pm$, can be derived from the Klein-Gordon equation: 
\beq
a_{\Delta+1}=\dot{a}_{\Delta} \, , \qquad
a_{\Delta+2}=\frac{2 \Delta -1}{4 \Delta -4} \, \dot{a}_{\Delta +1}  \, , \qquad
a_{\Delta+3}= \frac{2 \Delta +1}{6 \Delta -3} \, \dot{a}_{\Delta +2} \, ,
\label{all-the-as}
\eeq
\[
a_{\Delta+s}=\frac{\dot{a}_{\Delta+s-1} (2 \Delta +2 s-5) + a_{\Delta+s-4} (\Delta+s-4)^2}{ 2 s \Delta + s (s-4)} \, ,
\qquad {\rm for } \, \, \, \, s \geq 4 \, ,
\]
where $\dot{}$  denotes a derivative with respect to $\tau$. 
There is no \emph{a priori} algebraic relation between
$a_{\Delta_{-}+j}$ and $a_{\Delta_{+} +j}$; the relative coefficient
is fixed by the requirement that the field $\phi$ has ingoing-wave
boundary conditions at the horizon.

\subsection{Backreaction to leading order}

Next we compute the backreaction of the scalar $\phi$ on the metric 
at leading order in the expansion parameter $\lambda$.
The metric functions $A,\Sigma$ receive corrections at order
$\lambda^2$: 
\beq
A(\tau, \rho)=\mu^2 (\rho^{-2} -\rho^2 +\lambda^2 \tilde{A}_2 (\tau, \rho)) \, , \qquad
\Sigma(\tau, \rho)= \mu (\rho^{-1} + \lambda^2 \tilde{\Sigma}_2(\tau,\rho)) \, ,
\eeq
and the corresponding metric at this order reads:
\beq
ds^2=\frac{-(1-\rho^4+\lambda^2  \rho^2 \tilde{A}_2) d \tau^2 -2  d \rho \, d \tau +(1+2 \lambda^2 \rho \tilde{\Sigma}_2) d \vec{X}^2 }{\rho^2}  \, .
\label{correzione-metrica}
\eeq
From (\ref{einst}), one obtains a set of linearized equations for
$\tilde{A}_2$, $\tilde{\Sigma}_2$ \cite{newquench}: 
\begin{equation}
\tilde{\Sigma}_2'' +\frac{2 \tilde{\Sigma}_2' }{\rho} +\frac{(\tilde{\phi}_1')^2}{6 \rho} = 0 \, , 
\label{prima}
\end{equation}
\begin{equation}
(2 \rho^4-6) \tilde{\Sigma}_2 -4 \rho  \tilde{\Sigma}_2' + (\rho^2-\rho^6)  \tilde{\Sigma}_2''
+4 \rho   \dot{\tilde{\Sigma}}_2
 -2 \rho^2  \dot{\tilde{\Sigma}}_2' 
+2 \rho  \tilde{A}_2 - \rho^2  \tilde{A}_2' + \frac{m^2  (\tilde{\phi}_1)^2}{6 \rho} = 0 \, , 
\label{seconda}
\end{equation}
\begin{equation}
24 \left((\rho^4-1)   \tilde{\Sigma}_2' + \dot{\tilde{\Sigma}}_2 +\frac{\rho^4-1}{\rho}   \tilde{\Sigma}_2 \right)+
2 \left( 6 \tilde{A}_2 -2 \rho  \tilde{A}_2' - \rho^2 \tilde{A}_2'' \right)
\label{terza}
\end{equation}
\[
+(\rho^4-1) (\tilde{\phi}_1')^2 + 2 (\tilde{\phi}_1')(\dot{\tilde{\phi}}_1) +\frac{m^2}{3 \rho^2} (\tilde{\phi}_1)^2 =0 \, ,
\]
\begin{equation}
12 \rho \dot{\tilde{A}}_2 +6 \left( 4 \rho   \ddot{\tilde{\Sigma}}_2   +4 \rho(\rho^4-1)   \dot{\tilde{\Sigma}}_2'
-4 (\rho^4+1)  \dot{\tilde{\Sigma}}_2  \right)
+4 (\rho^4-1) \dot{\tilde{\phi}}_1 \tilde{\phi}_1' +4   \dot{\tilde{\phi}}_1^2 =0 \, .
\label{quarta}
\end{equation}
Here $'$ and $\dot{}$  denote derivatives with respect to $\rho$ and $\tau$. 
These equations have source terms, which are due to insertions of
$\tilde{\phi}_1={\rm Re} (\phi_1 e^{-i \omega_T \tau})$. 
The solutions for $\tilde{\Sigma}_2$ to the homogeneous version of
eq.~(\ref{prima}) are a constant and $1/\rho$; the term proportional
to $1/\rho$ corresponds to a linear rescaling in the $\rho$
coordinate, while the constant in $\tilde{\Sigma}_2$ can be gauged
away using the residual diffeomorphism invariance $\rho\to\rho+
f(\tau)$ and thus we can set both these terms to
zero. Eq.~(\ref{prima}) gives all the information that we need to
solve for $\tilde{\Sigma}_2$.  
We can then insert $\tilde{\Sigma}_2$ into eq.~(\ref{seconda}) and
solve for $\tilde{A}_2$; the solution to the homogeneous version is
$\tilde{A}_2 \propto \rho^2$, so this equation fixes $\tilde{A}_2$
only up to a function of time which is proportional to $\rho^2$ in
space; this is fixed by eq.~(\ref{quarta}). We show that this term is actually
a linear function of time $\tau$. 

 After reviewing the general equations for the leading order backreaction \cite{newquench},
we specialize to the case of a time-dependent periodic source.
It is not consistent within the approximation to consider a
source that is switched on for an infinite number of cycles, because
at some point the total backreaction will grow large, no matter how
small the amplitude $\lambda$ is taken. We can hence imagine switching
on the source at time $\tau_0^*$ and switching it off after a large
number of cycles at time $\tau_f^*$. This while keeping $\lambda$ as
small as 
needed in order to trust the perturbative expansion. 
We neglect also the effect of the transient period, which should be
negligible in the limit of a large number of cycles.
In order to make the time dependence of the problem explicit, we can
use the following Ansatz, valid in the time window
$\tau_0^*<\tau<\tau_f^*$: 
\beq
\tilde{\Sigma}_2= \Sigma_{2,{\rm c}}(\rho)  + 
{\rm Re}\big(\Sigma_{2,{\rm p}}(\rho) e^{-2 i \omega_T (\tau-\tau_0^*)}\big) \, , 
\label{ansaziano}
\eeq
\beq
\tilde{A}_2= A_{2,{\rm c}}(\rho) + A_{2,{\rm l}}(\rho)  (\tau-\tau_0^*) +   {\rm Re}\big(A_{2,{\rm p}}(\rho) e^{-2 i \omega_T (\tau-\tau_0^*)}\big) \, .
\label{ansazianooo}
\eeq
It turns out that $\tilde{A}_2$, as a function of $\tau$, contains a constant term $A_{2,{\rm c}}(\rho)$,
 a linear one $A_{2,{\rm l}}(\rho)$, and a periodic part $A_{2,{\rm p}}(\rho)$; $\tilde{\Sigma}_2$
contains a constant  term $\Sigma_{2,{\rm c}}(\rho)$ and a periodic one $\Sigma_{2,{\rm p}}(\rho)$.

The metric (\ref{correzione-metrica}) resembles the AdS Vaidya metric: 
\beq
ds^2=\frac{-(1-\rho^4 M(\tau)) d \tau^2 -2  d \rho \, d \tau + d
  \vec{X}^2 }{\rho^2} \, ,
\eeq
which is a solution of the Einstein equations in the presence of the 
energy-momentum tensor 
\beq
T_{\tau \tau}=\frac{3}{2} \rho^3 M'(\tau) \, ,
\eeq
with the other components being zero; this describes the absorption of
null dust by a black brane. If we set $\Sigma_{2,{\rm c,p}}=A_{2,{\rm c,p}}=0$ in our
metric, then we would recover a Vaidya metric with 
$M(\tau) \propto \tau$.
Several studies of non-equilibrium physics in the AdS/CFT have focused on
the Vaidya metric, see
e.g.~\cite{Hubeny:2007xt,AbajoArrastia:2010yt,Albash:2010mv,Balasubramanian:2010ce,Balasubramanian:2011ur,Aparicio:2011zy,Balasubramanian:2011at,Allais:2011ys,Balasubramanian:2012tu}.

The function $A_{2,{\rm l}}$ can be obtained by solving
eq.~(\ref{quarta}): 
 \beq
  A_{2,{\rm l}}= \frac{ - \omega_T^2 |\phi_1|^2 +\omega_T (1-\rho^4) 
  ({\rm Re \, \phi_1'} \, {\rm Im \, \phi_1} -{\rm Re \, \phi_1} \, {\rm Im \, \phi_1'} )  }{6 \rho} \, .
  \label{cardellino}
 \eeq
This expression is proportional to $\rho^2$, which can be shown by forming
a linear combination of the real and imaginary part of eq.~(\ref{wave-eq}). 
The functions $\Sigma_{2,{\rm c,p}}$ can be obtained by solving the
following equations: 
\beq
 \Sigma_{2,{\rm c}}''+ \frac{2  \Sigma_{2,{\rm c}}'}{\rho} +\frac{1}{12 \rho} |\phi_1'|^2 =0 \, ,  \qquad
  \Sigma_{2,{\rm p}}''+ \frac{2  \Sigma_{2,{\rm p}}'}{\rho} +\frac{1}{12 \rho} (\phi_1')^2 =0 \, ,
  \label{Sigmarecipe}
\eeq
with the boundary condition that the constant and the $1/\rho$ terms 
in the expansion at small $\rho$ of $\Sigma_{2,{\rm p,c}}$ are zero.
The periodic part $A_{2,{\rm p}}$ can be expressed in terms of
$\phi_1$ and $\Sigma_{2,{\rm p}}$ using eq.~(\ref{quarta}); the result
is 
\beq
A_{2,{\rm p}}=\left( 4 i \omega_T +2\frac{1+\rho^4}{\rho}\right)   \Sigma_{2,{\rm p}} +2(1-\rho^4)   \Sigma_{2,{\rm p}}' +
\frac{(i \omega_T) \phi_1^2 +(1-\rho^4) \phi_1 \phi_1'}{12 \rho} \, .
\label{Aperiodica}
\eeq
The constant part $A_{2,{\rm c}}$ is obtained by solving the following differential equation:
\beq
2 \rho  A_{2,{\rm c}} - \rho^2  A_{2,{\rm c}}' + 
(2 \rho^4-6)  \Sigma_{2,{\rm c}} -4 \rho  \Sigma_{2,{\rm c}}'+ (\rho^2-\rho^6)   \Sigma_{2,{\rm c}}''
+ \frac{m^2  |\phi_1|^2}{12 \rho} = 0 \, , 
\label{Acrecipe}
\eeq

Let us first focus on the case of a generic $m^2$ (except for the cases
of integer and half-integer $\Delta$s, which are treated in appendix
\ref{appendix:A}); 
the following form of the series expansion in $\rho$ can be used
\cite{newquench}: 
\beq
\tilde{A}_2=\sum_{j=0}^{\infty} \alpha_j \rho^{2+j}+\alpha_{\Delta_- +j} \rho^{2 \Delta_- -2 +j} +\alpha_{\Delta_+ +j} \rho^{2 \Delta_+-2 +j} \, ,
\label{sese1}
\eeq
\beq
\tilde{\Sigma}_2=\sum_{j=0}^{\infty} \sigma_j \rho^{3+j}+\sigma_{\Delta_- +j} \rho^{2 \Delta_- -1 +j} +\sigma_{\Delta_+ +j} \rho^{2 \Delta_+ -1 +j} \, ,
\label{sese2}
\eeq
where the $\alpha$s and $\sigma$s are functions of time.
Among all these coefficients, $\alpha_0$ is special; note that
eq.~(\ref{seconda}, \ref{terza}) contain only spatial derivatives of 
$A$, and in a combination which is blind to the coefficient $\alpha_0$. 
The only equation that fixes it is eq.~(\ref{quarta}).
The coefficient $\alpha_0$ is the only one with a secular term which
increases linearly in time; all the other coefficients in
eqs.~(\ref{sese1},\ref{sese2}),  are superpositions of a constant and
a periodic term in time, with period $2 \omega_T$: 
\beq
 \alpha_k=\alpha_{k,c} + {\rm Re} (\alpha_{k,p} e^{-2 i \omega_T \tau}) \, , \qquad
  \sigma_k=\sigma_{k,c} + {\rm Re} (\sigma_{k,p} e^{-2 i \omega_T \tau})  \, .
\eeq
From eq.~(\ref{prima}) one obtains ($j \geq 0$ is an integer): 
\beq
\sigma_j=-\frac{  \sum_{l=0}^j (\Delta_- +l)(\Delta_+ + j -l) (a_{\Delta_- +l}) (a_{\Delta_+ + (j-l)})   }{3 (3+j)(4+j)} \, ,
\label{si1}
\eeq
\beq
\sigma_{\Delta_\pm  +j}= - \frac{  \sum_{l=0}^j (\Delta_\pm  +l)(\Delta_\pm +j-l) (a_{\Delta_\pm +l}) (a_{\Delta_\pm +j-l})    }{6(2 \Delta_\pm  -1+j )(2 \Delta_\pm +j)} \, .
\label{si2}
\eeq
Replacing some of the coefficients of eqs.~(\ref{si1}) in
eq.~(\ref{quarta}), a differential equation which gives the time
evolution of $\alpha_0$ can be found \cite{Buchel:2012gw,newquench}:
\beq
\dot{\alpha}_0 = \frac{\Delta_+ (3-2 \Delta_-)}{9} \dot{a}_{\Delta_-} a_{\Delta_+} + \frac{\Delta_- (3-2 \Delta_+)}{9}   a_{\Delta_-} \dot{a}_{\Delta_+} \, . 
\label{energetica}
\eeq
The coefficient $\alpha_0$ has an important physical significance:
its variation in time is proportional to the total work performed on
the system by the external force. 
Both $\Sigma_{2,{\rm c,p}}$ are divergent at small $\rho$ for
$\Delta_+>7/2$; in this regime it is useful to use the small $\rho$
expansion to set up the right boundary conditions for solving
eq.~(\ref{Sigmarecipe}): 
\beq
\tilde{\Sigma}_2(\rho) \approx a_{\Delta_-}^2 \frac{4-\Delta_+}{12(2 \Delta_+-7)} \rho^{2 \Delta_- -1} +
\frac{5-\Delta_+}{6 (2 \Delta_+-9)} a_{\Delta_-} \dot{a}_{\Delta_-} \rho^{2 \Delta_-} + \ldots \, ,
\eeq
For $\Delta_+>3$, both $A_{2,{\rm c}}$ and $A_{2,{\rm p}}$ diverge at
small $\rho$; it 
is useful to use the small $\rho$ expansion to set up the boundary
conditions for the differential equation (\ref{Acrecipe}): 
\beq
\tilde{A}_2(\rho) \approx a_{\Delta_-}^2 \frac{4-\Delta_+}{6(2 \Delta_+-7)} \rho^{2 \Delta_- -2} +
\frac{\Delta_+-3}{3(7-2 \Delta_+)} a_{\Delta_-} \dot{a}_{\Delta_-} \rho^{2 \Delta_- -1} + \ldots \, 
\eeq
In this subsection we have studied the leading-order
response of the metric to the scalar perturbation;
the general form of the $O(\lambda^2)$ correction to the metric
is parametrized by eqs.~(\ref{ansaziano},\ref{ansazianooo})
in terms of five functions $A_{2,{\rm l,c,p}}(\rho), \Sigma_{2,{\rm c,p}}(\rho)$;
we have shown that the function $A_{2,{\rm l}}(\rho)$ is quadratic in $\rho$.
 In order to fix the boundary conditions
and also to compute the one-point correlators, we have expanded
the metric functions $\tilde{A}_2, \tilde{\Sigma}_2$ as  power series in $\rho$,
see eqs.~(\ref{sese1},\ref{sese2}), and we have written recursion relations
for the coefficients of these expansions.

\subsection{One-point correlators}

Once the background metric and the scalar field solutions have been determined,
one can derive the expectation value of the boundary scalar operator $\mathcal{O}_\Delta$ and 
of the energy-momentum tensor.
The parameter $\xi_0$ which parametrizes the source $J_\Delta = \xi_0
\cos \omega t $ in the boundary field theory is proportional to the
expansion parameter $\lambda$ on the gravity side; by dimensional
analysis $\xi_0 =  \lambda \mu^{4-\Delta}$ (in our conventions we set
the proportionality coefficient equal to one) .
In the case in which the dimension of the source, $\Delta=\Delta_+$,
we use the following parametrization: 
\beq
a_{\Delta_-}={\rm Re} ( e^{- i \omega_T \tau}) \, , \qquad  a_{\Delta_+}={\rm  Re} \big(\chi_\Delta (\omega_T) e^{-i \omega_T \tau}\big) \, ,
\label{chichi1}
\eeq
where the Klein-Gordon equation for the field $\phi$ is solved using
an ingoing boundary condition at the horizon \cite{Son:2002sd}. 
Holographic renormalization \cite{Bianchi:2001kw} can be used to
obtain an expression for the expectation values of the energy-momentum
tensor. 
Explicit expressions for generic $\Delta=\Delta_+$ were found in
\cite{newquench}: 
\beq
8 \pi G_N \langle \mathcal{O}_\Delta \rangle=\lambda \mu^\Delta (\Delta-2) a_{\Delta_+}\,
\label{vevov}
\eeq
\beq
8 \pi G_N \mathcal{E}=\frac{3}{2} \mu^4 -\lambda^2 \mu^4 \left( \frac{3}{2} \alpha_0 +\frac{(2 \Delta -3)(4-\Delta)}{6} a_{\Delta_+} a_{\Delta_-} \right) \, ,
 \label{energydensity}
\eeq
\beq
8 \pi G_N \mathcal{P}=\frac{1}{2} \mu^4 -\lambda^2 \mu^4 \left( \frac{1}{2} \alpha_0 -\frac{(4 \Delta -9)(4-\Delta)}{18} a_{\Delta_+} a_{\Delta_-} \right) \, .
\eeq
In our case, the time dependence of $\mathcal{E}$ and $\mathcal{P}$ has a periodic
part with frequency $2 \omega_T$ and a linear part in time, which
corresponds to the work done on the system. 
These expressions are consistent with the Ward identity
\beq
\partial_t \mathcal{E}= -\langle \mathcal{O}_\Delta \rangle \partial_t J_\Delta \, . 
\label{ward}
\eeq

In the case where the dimension of the source, $\Delta=\Delta_-$, we
use the parametrization: 
\beq
a_{\Delta_-}=-{\rm Re} \big(  \chi_\Delta (\omega_T)   e^{-i \omega_T \tau}\big) \, , \qquad
a_{\Delta_+}={\rm Re} \big(e^{-i \omega_T \tau}\big) \, ,
\label{chichi2}
\eeq
The expressions (\ref{vevov},\ref{energydensity}) for $\langle
\mathcal{O}_\Delta \rangle $ and $\mathcal{E}$ are valid also for
$\Delta=\Delta_-$; this can be checked by inspecting the consistency
of eq.~(\ref{energetica}) with eq.~(\ref{ward}). 
The cases with integer and half-integer dimensions are treated in
detail in appendix \ref{appendix:A}.
 
In our conventions, the retarded Green function and the function
$A_{2,l}$ (both in the case where $\Delta=\Delta_+$ and
$\Delta=\Delta_-$, as well as for $\Delta$ integer and half-integer,  
with the exception of $\Delta=2$ case) are:
 \beq
  G^R(\omega,0) =\frac{|\Delta-2|}{8 \pi G_N}  (\pi T)^{2 \Delta-4} \, \chi_{\Delta}(\omega_T) \, ,
  \qquad
A_{2,{\rm l} }(\rho)= - \left( \frac{\omega_T \, {\rm Im}\,  \chi_\Delta}{3} |\Delta-2| \right)  \rho^2 
\, .
\label{upupa}
\eeq
For $\Delta=2$ these quantities are given in eq.~(\ref{green-delta-2}). 

The work done on the system per unit volume and time is:
\beq
\mathcal{W}_c=- \frac{3}{8 G_N } \frac{\lambda^2    \mu^4}{\omega_T}  A_{2,{\rm l} }(1) \, .
\eeq
In our conventions, this expression is valid for any dimension $\Delta$.
Specializing to different values of the operator dimension:
\begin{equation}
\mathcal{W}_c=\left\{
\begin{array}{rl}  \frac{ \lambda^2    \mu^4  |\Delta-2| }{8 G_N }    {\rm Im}\,  \chi_\Delta(\omega_T) \,   &\text{if} \, \Delta \neq 2 \, ,\\ 
  \frac{ \lambda^2    \mu^4 }{16 G_N }    {\rm Im}\,  \chi_2(\omega_T) \,    &\text{if} \, \Delta = 2 \, .
    \end{array} \right.
\end{equation}

\section{Linear response\label{sec:linear_response}}

We are now prepared to study the physical properties of the system,
by numerically computing the response function $\chi_\Delta(\omega_T)$ for any values
of $\omega_T$ and for operator dimension $1 < \Delta \leq 4$.
We check  the numerical calculations using the Kramers-Kronig relations
 and comparing with the analytical results in both the limits
$\omega_T \gg 1$  and $\omega_T=0$.

\subsection{Numerical study}

\begin{figure}[ht]
\begin{center}
$\begin{array}{c@{\hspace{.2in}}c} \epsfxsize=6.5cm
\epsffile{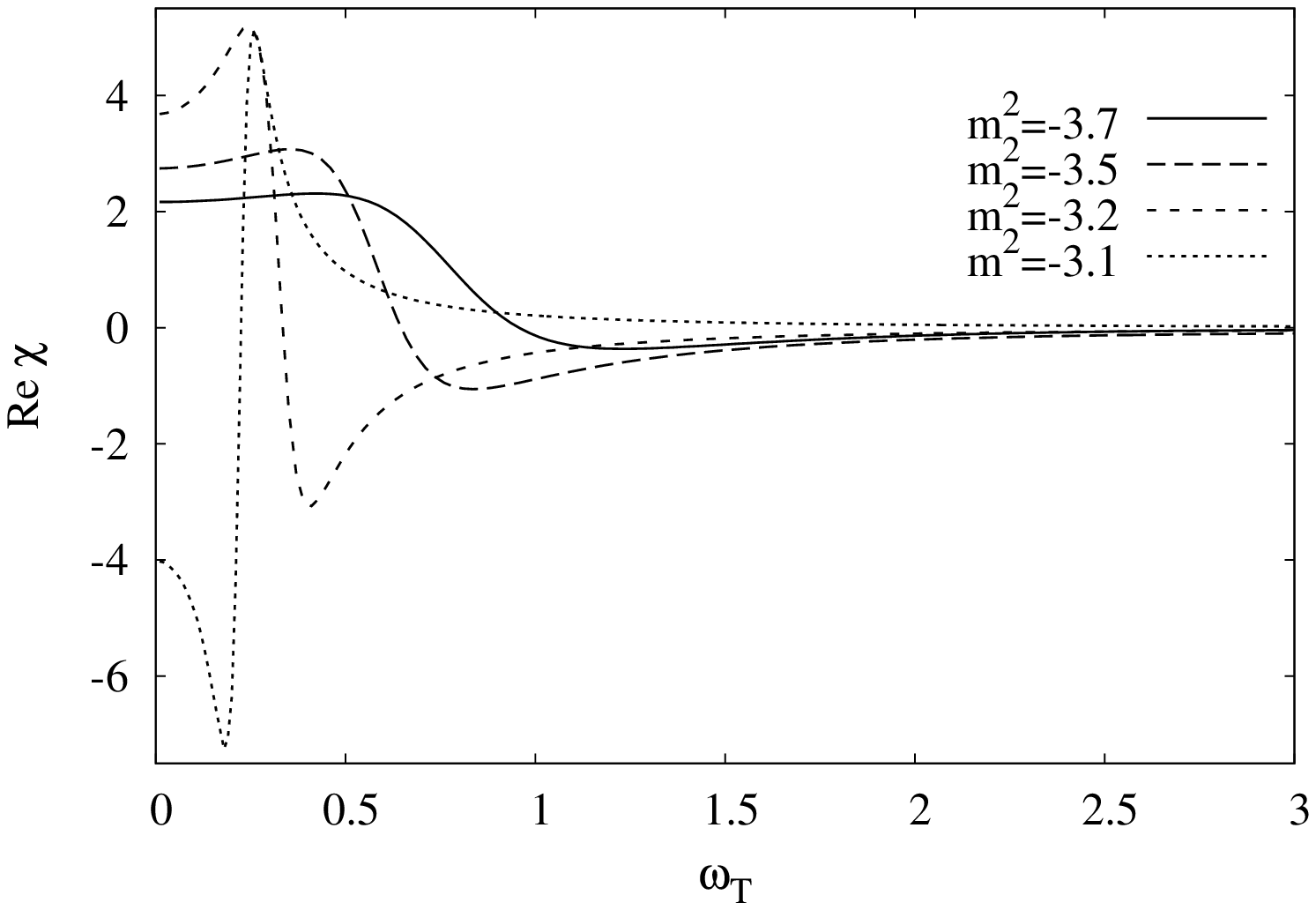} &
     \epsfxsize=6.5cm
    \epsffile{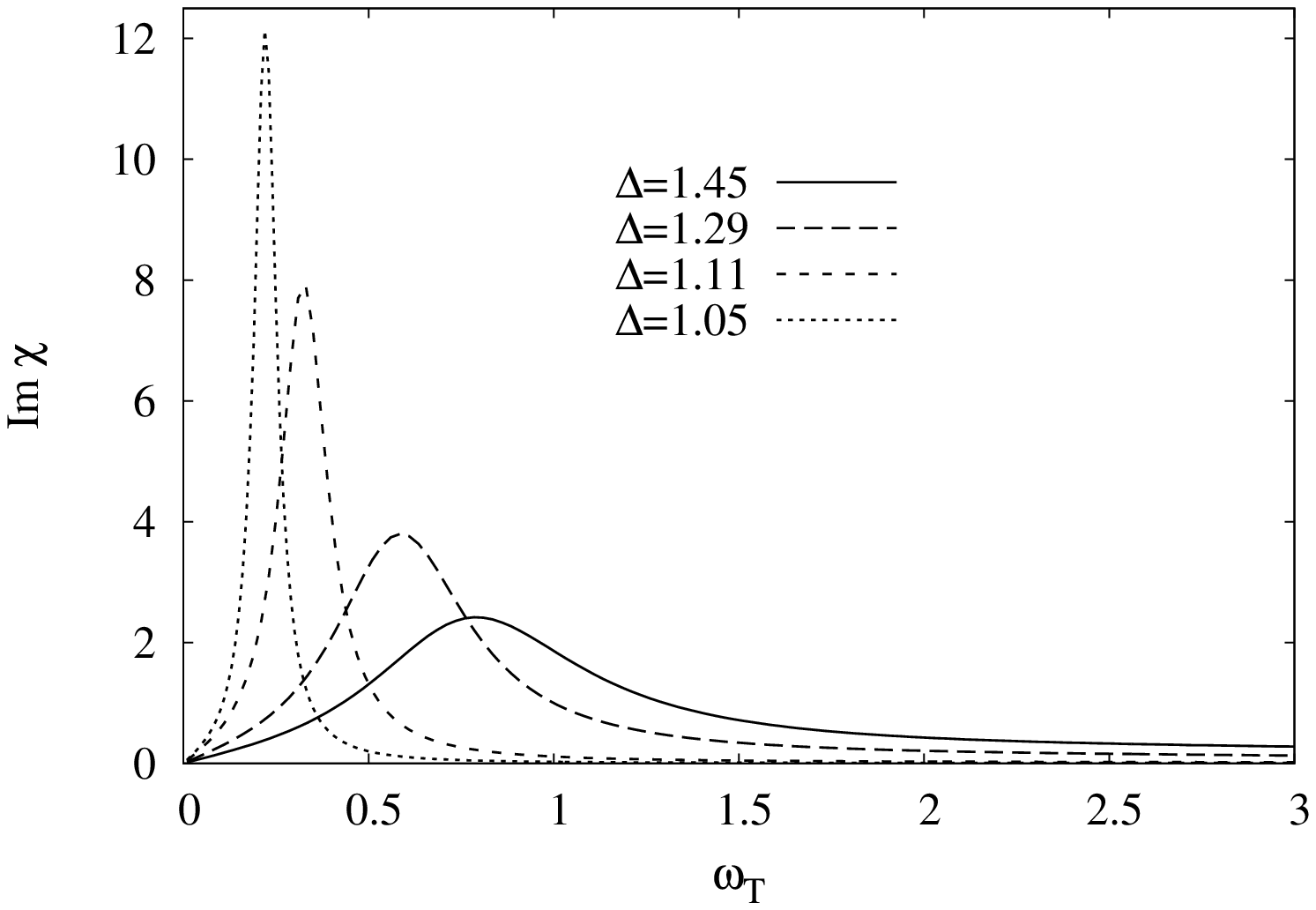}
\end{array}$
\end{center}
\caption{\footnotesize
Real and imaginary parts of the linear response function $\chi$, for
various values of $(m^2(\Delta), \Delta)$, with $\Delta=\Delta_-$.
In the limit $\Delta\to 1$ the function ${\rm Im \,} \chi$ becomes 
sharply peaked at a low frequency.
} \label{chi-mq-varie2-inverted}
\end{figure}

\begin{figure}[ht]
\begin{center}
$\begin{array}{c@{\hspace{.2in}}c} \epsfxsize=6.5cm
\epsffile{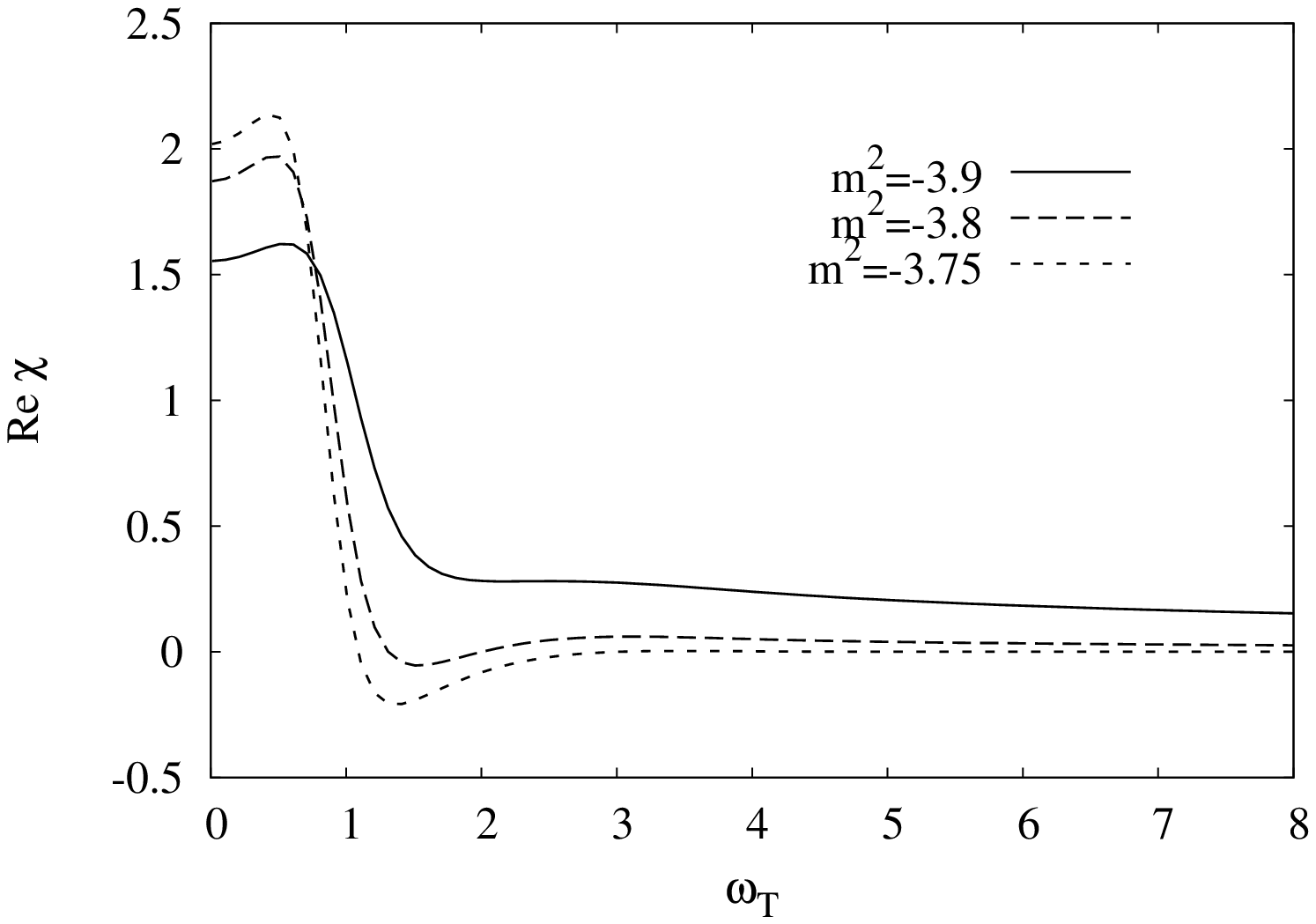} &
     \epsfxsize=6.5cm
    \epsffile{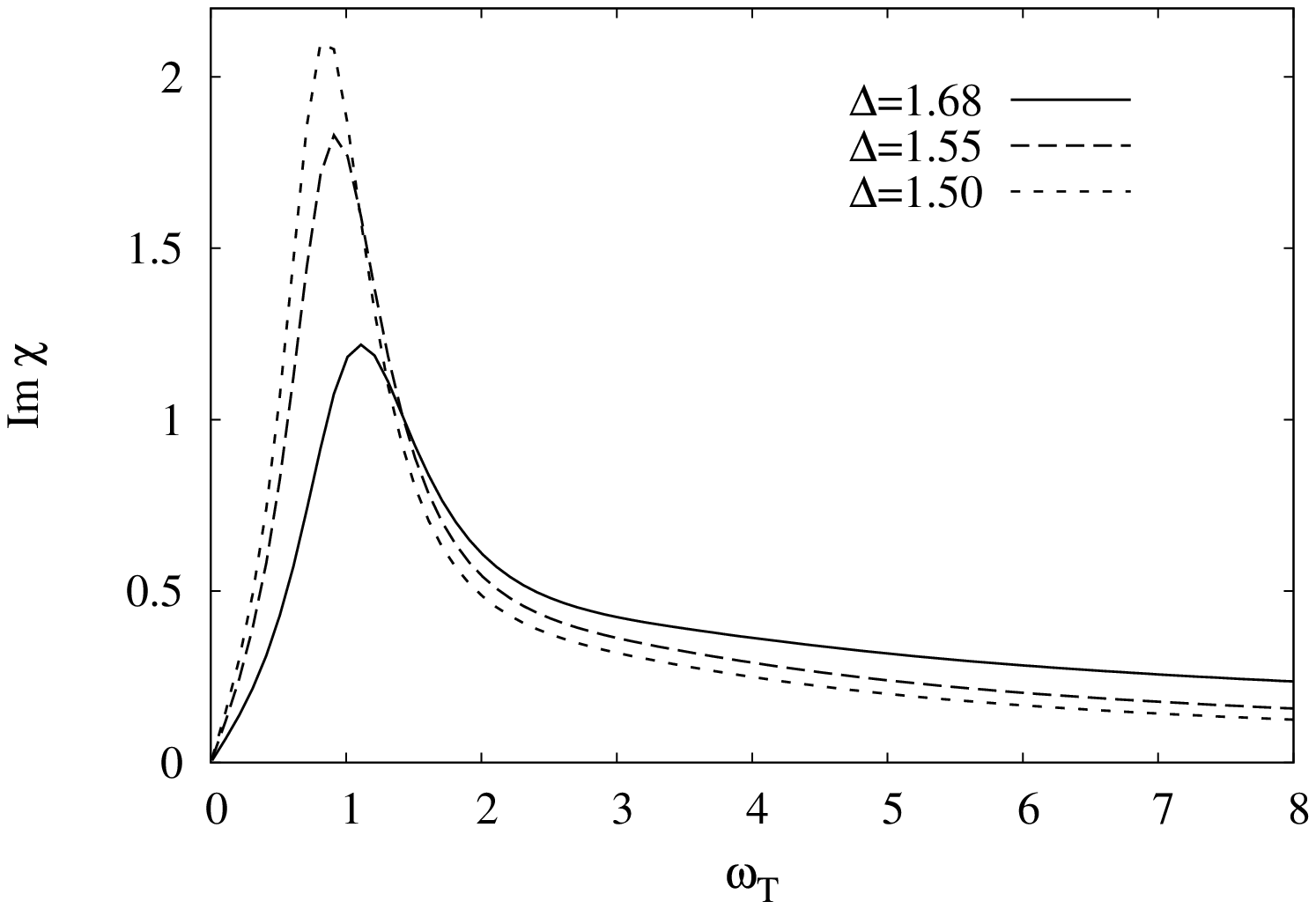}
\end{array}$
\end{center}
\caption{\footnotesize
Real and imaginary parts of the linear response function $\chi$, for
various values of $(m^2(\Delta), \Delta)$, with $\Delta=\Delta_-$. 
} \label{chi-mq-varie1-inverted}
\end{figure}

\begin{figure}[ht]
\begin{center}
$\begin{array}{c@{\hspace{.2in}}c} \epsfxsize=6.5cm
\epsffile{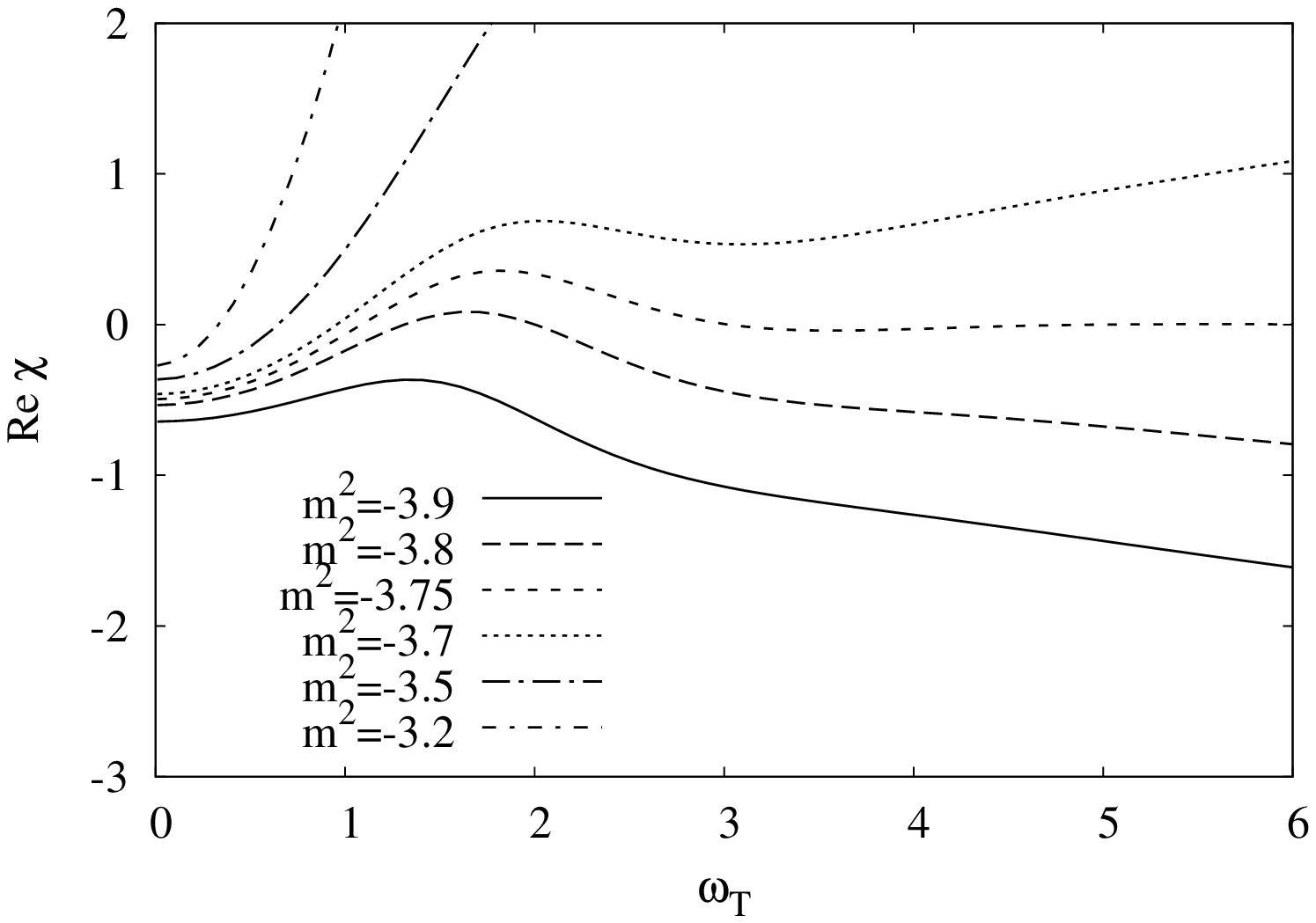} &
     \epsfxsize=6.5cm
    \epsffile{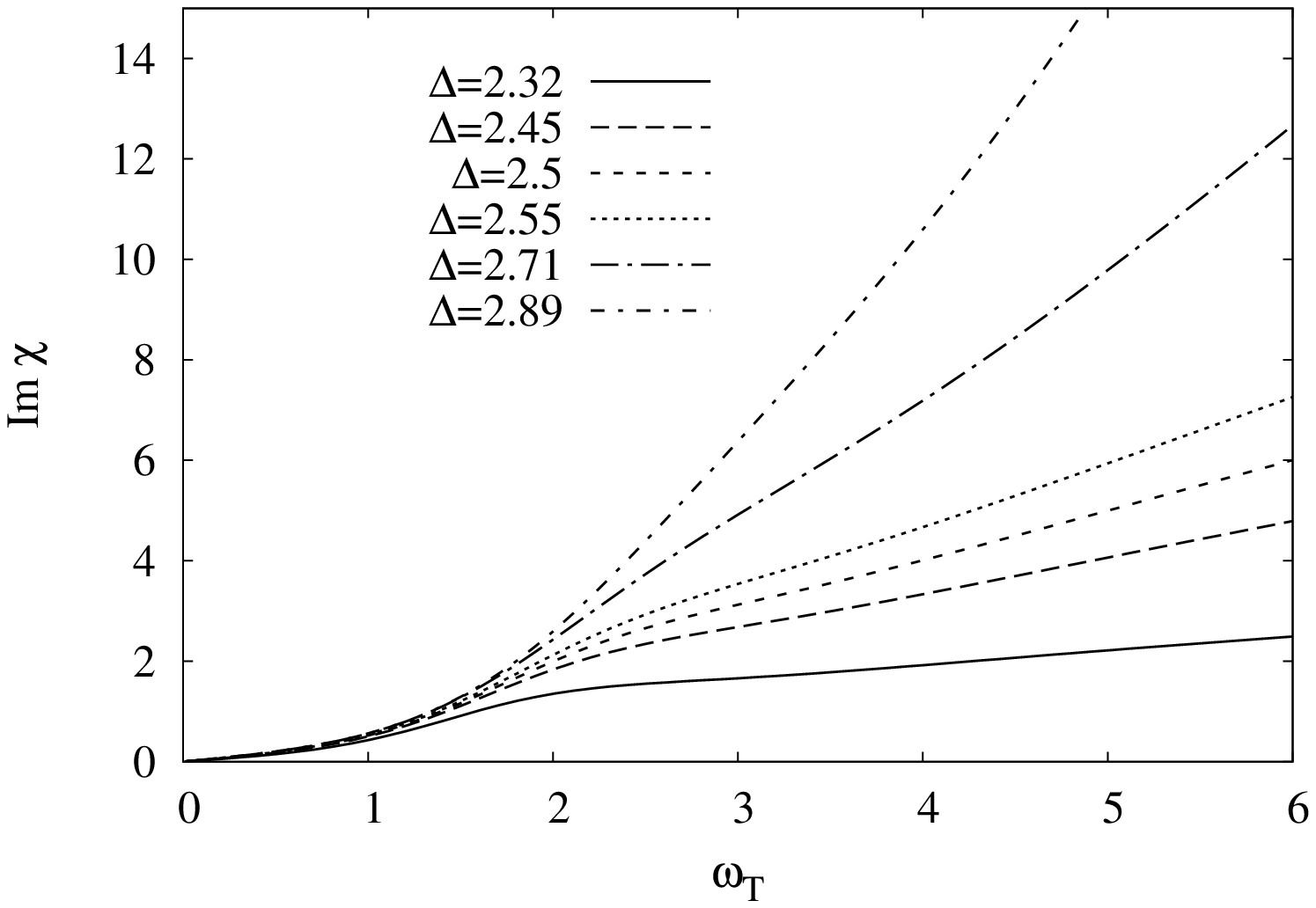}
\end{array}$
\end{center}
\caption{\footnotesize
Real and imaginary parts of the linear response function $\chi$, for
various values of $(m^2(\Delta), \Delta)$, with $\Delta=\Delta_+$. 
} \label{chi-mq-varie1}
\end{figure}

\begin{figure}[ht]
\begin{center}
$\begin{array}{c@{\hspace{.2in}}c} \epsfxsize=6.5cm
\epsffile{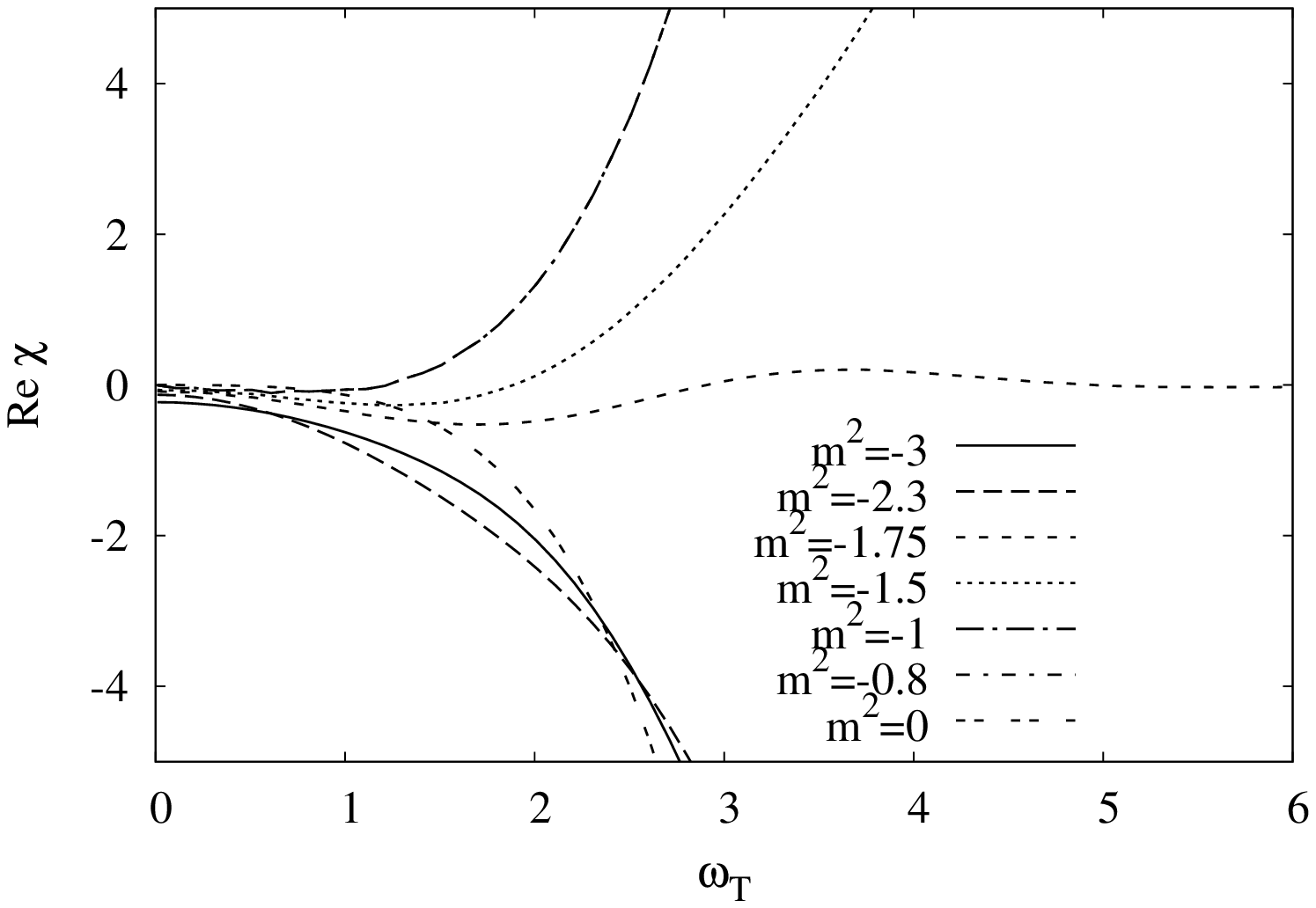} &
     \epsfxsize=6.5cm
    \epsffile{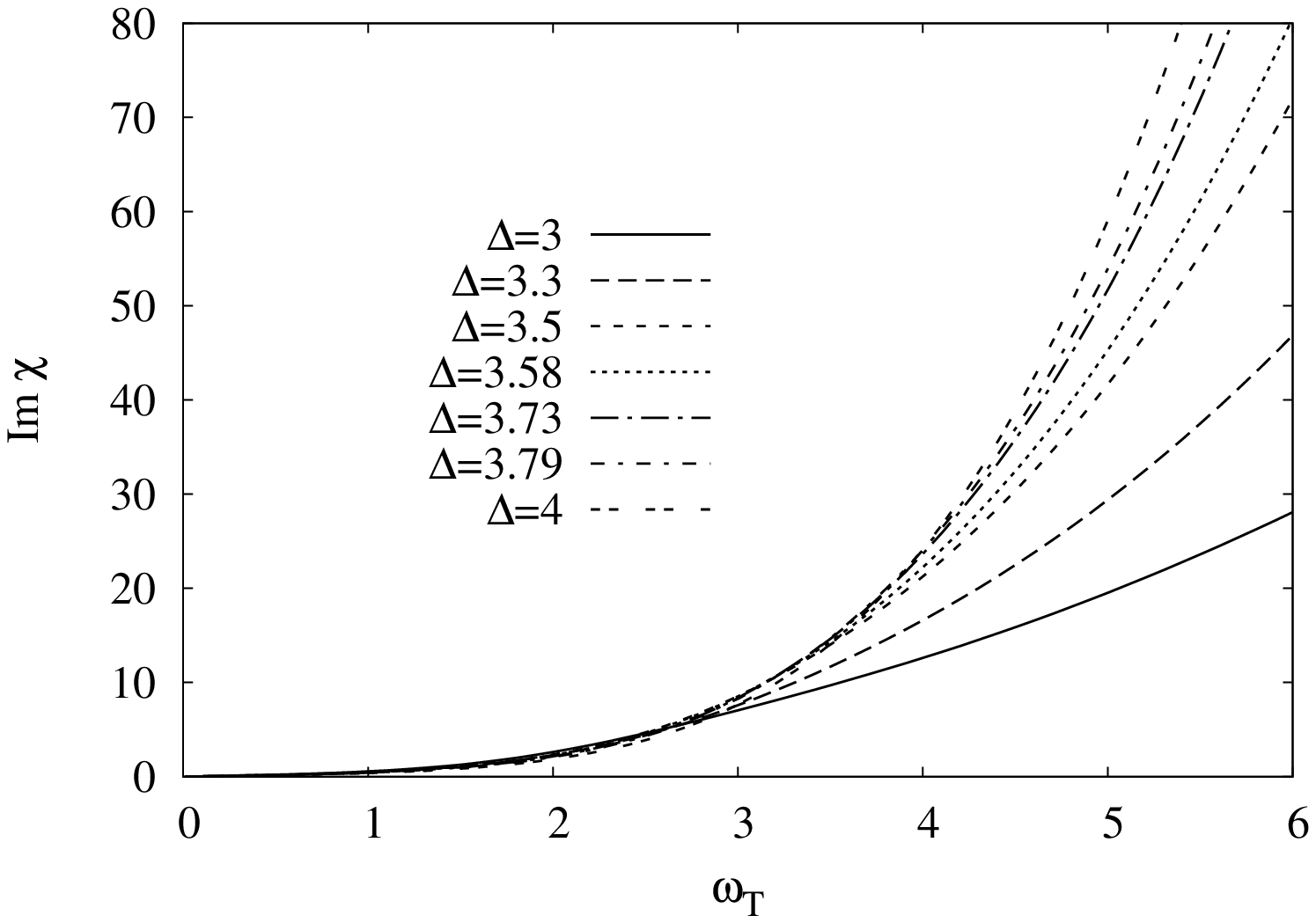}
\end{array}$
\end{center}
\caption{\footnotesize
Real and imaginary parts of the linear response function $\chi$, for
various values of $(m^2(\Delta), \Delta)$, with $\Delta=\Delta_+$.  
} \label{chi-mq-varie5}
\end{figure}

In order to study the time-dependent case numerically, let us write
the field $\phi_1$ in the following form:
 \beq
 \phi_1=\sum_{k=0}^{M} a_{\Delta_- +k} \rho^{\Delta_- + k} + \hat{\phi}_1 \, ,
 \label{ansatz1}
 \eeq
where the integer $M$ is chosen in such a way that the sum includes
(for a given $m^2$) at least all the exponents which are smaller or
equal to $\Delta_+$.
The first coefficients $a_{\Delta_- +k}$ read:
\beq
a_{\Delta_-}=1 \, , \qquad a_{\Delta_- +1}=(- i \omega_T) \, , \qquad
a_{\Delta_- +2}=(-i \omega_T)^2 \frac{2 \Delta_-  -1}{4(\Delta_- - 1)} \, , 
\eeq
\beq
a_{\Delta_- +3}=(-i \omega_T)^3 \frac{2 \Delta_- +1}{12(\Delta_--1) } \, .
\eeq
By substituting the Ansatz of eq.~(\ref{ansatz1}) into eq.~(\ref{wave-eq}),
one finds an equation of the kind: 
\beq
 (-i \omega_T) \hat{\phi}_1' 
- \frac{1-\rho^4}{2}  \hat{\phi}_1'' 
-(-i \omega_T)\frac{3}{2 \rho}  \hat{\phi}_1
+ \frac{3+\rho^4}{2 \rho}  \hat{\phi}_1'
 +\frac{m^2}{2 \rho^2} \hat{\phi}_1 
  +J_0=0 \, ,
  \label{nunu}
\eeq 
where $J_0$ is a source term, whose explicit form depends on the
number of terms $M$ in eq.~(\ref{ansatz1}). 
For example, if $M=3$
\[
J_0=\frac{ (1+2 \Delta_{-}) (3+ 2 \Delta_{-}) }{24 (\Delta_{-}-1)} \rho^{2+ \Delta_{-}}  (-i \omega_T)^4 +
\frac{ (3+ \Delta_{-})^2 (1+ 2 \Delta_{-}) }{24 (\Delta_{-}-1)}  \rho^{5+ \Delta_{-}}   (- i \omega_T)^3
\]
\[
 +\frac{ (2+ \Delta_{-})^2 ( 2 \Delta_{-}-1) }{8 (\Delta_{-}-1)} \rho^{4+ \Delta_{-}}   (-i \omega_T)^2 
 +\frac{ (1+ \Delta_{-})^2  }{2}  \rho^{3+ \Delta_{-}}   (-i \omega_T) 
 +\frac{ \Delta_{-}^2  }{2}  \rho^{2+ \Delta_{-}}     \, .
\]
The equation (\ref{nunu}) can be solved by a shooting method, imposing
that the solution is regular at the horizon and that the expansion at
order $\rho^{\Delta_-}$ of $\hat{\phi}_1$ vanishes at $\rho=0$. 
The value of $\chi(\omega_T)$ can then be extracted from the
coefficient of the term proportional to $\rho^{\Delta_+}$ in
$\hat{\phi}_1$, using eq.~(\ref{chichi1}). 
In the case of $\Delta=\Delta_-$, we can use the same numerical
solution that we used for $\Delta=\Delta_+$, using instead
eq.~(\ref{chichi2}) to compute $\chi(\omega_T)$. 
Numerical results are shown in figures \ref{chi-mq-varie2-inverted},
\ref{chi-mq-varie1-inverted}, \ref{chi-mq-varie1}, \ref{chi-mq-varie5}, \ref{chi-mq-4}.
In the right panel we indicate the dimension $\Delta$ 
of the driving operator, while in the left panel we indicate the corresponding mass $m^2$.

Integer and half-integer values of the dimension $\Delta$ are special; 
details on one-point functions are discussed in appendix \ref{appendix:A}.
The numerical calculation of the Green function proceeds as follows: 
\begin{itemize}
\item $m^2=-4$, where $\Delta_-=\Delta_+=2$.
In order to find $\chi$ numerically, it is convenient to use: 
\beq
\phi_1= - \rho^2 \log \rho - (-i \omega_T) \rho^3 \log \rho + \hat{\phi}_1 \, (\tau,\rho) \, .
\eeq
Numerical results are shown in fig.~\ref{chi-mq-4}.
\begin{figure}[ht]
\begin{center}
$\begin{array}{c@{\hspace{.2in}}c} \epsfxsize=6.5cm
\epsffile{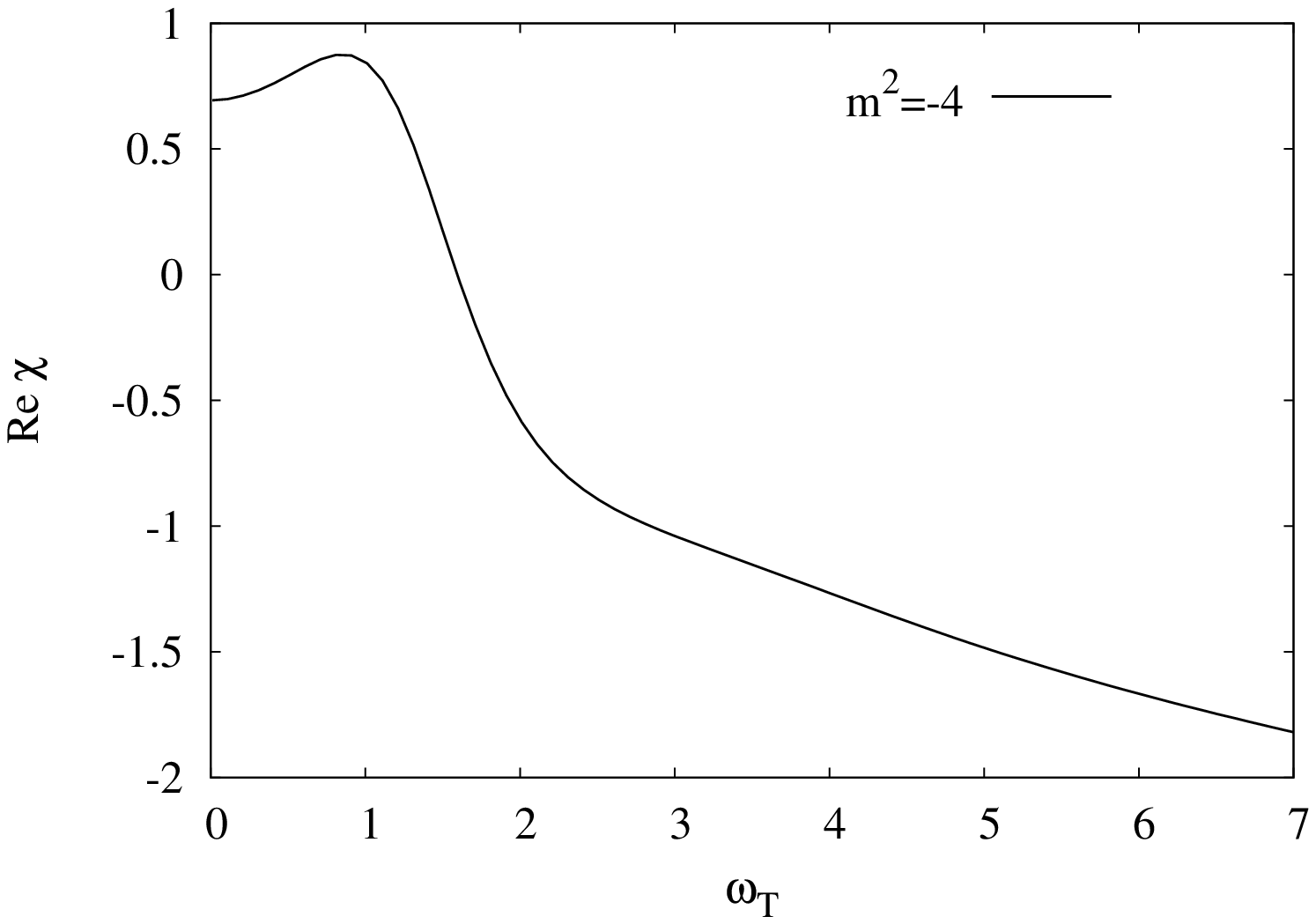} &
     \epsfxsize=6.5cm
    \epsffile{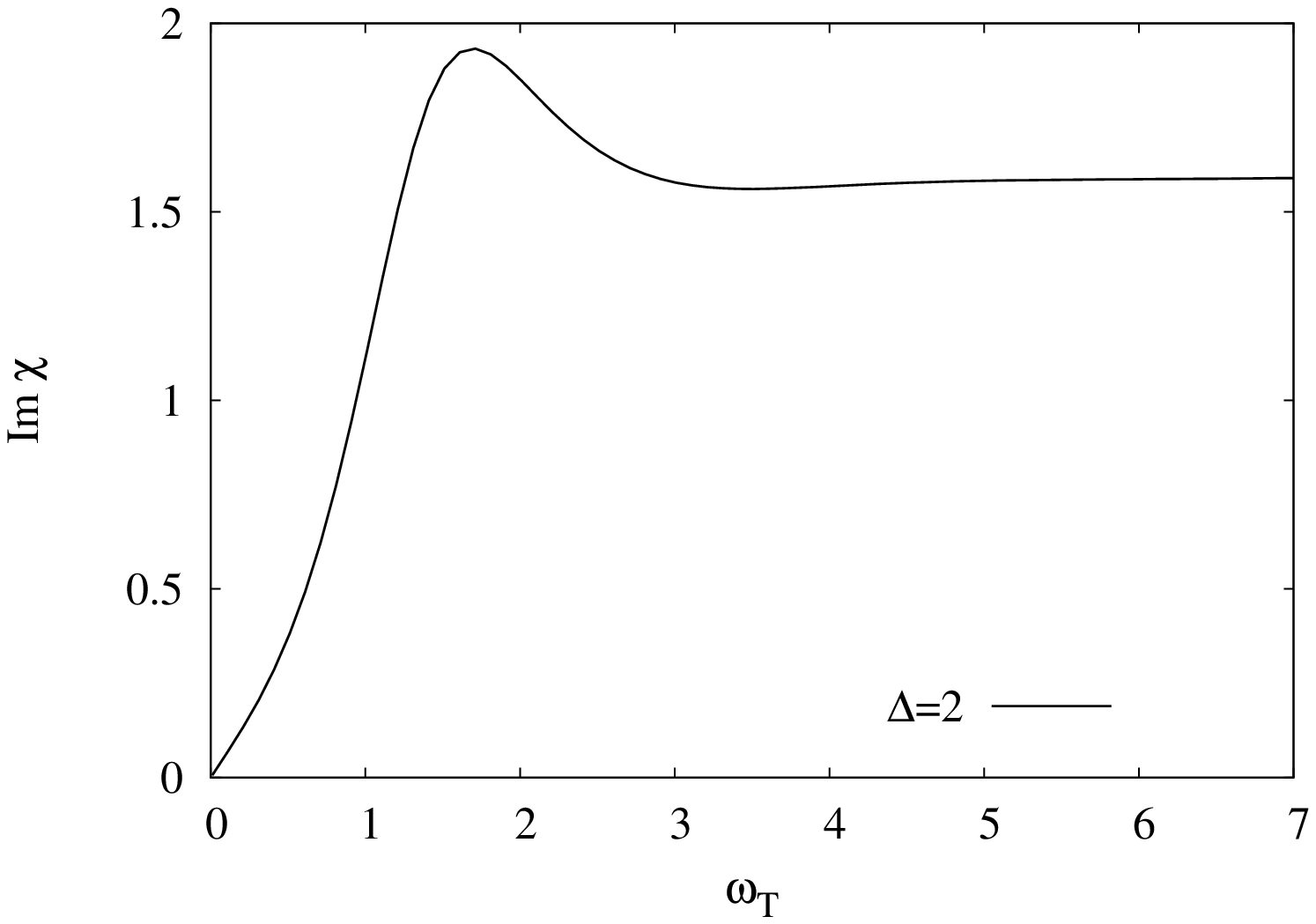}
\end{array}$
\end{center}
\caption{\footnotesize
Real and imaginary parts (solid line) of the linear response function
$\chi$, for $m^2=-4, \Delta=2$; the real part is proportional to $\log
\omega_T$ at large $\omega_T$, the imaginary part goes to a constant.
} \label{chi-mq-4}
\end{figure}

\item $m^2=-15/4$, where  $\Delta_-=3/2$, $\Delta_+=5/2$.
This value (together with $m^2=-7/4$) is at the threshold between
${\rm Re} \chi \rightarrow \pm \infty$ 
at large $\omega_T$; from the numerical calculations, ${\rm Re} \chi
\rightarrow 0$ at large $\omega_T$. 
The following $\hat{\phi}_1 $ is used:
\beq
\phi_1= \rho^{3/2} +   \rho^{5/2} (-i \omega_T) + \hat{\phi}_1 \, ,
\eeq
and then we proceed as in the case of generic $m$.
The function $\chi(\omega_T)$ is shown in figs.~\ref{chi-mq-varie1-inverted} and \ref{chi-mq-varie1}.

\item $m^2=-3$, where $\Delta_-=1$, $\Delta_+=3$
As a tool to enforce the desired boundary conditions, it is then
convenient to introduce $\hat{\phi}_1$: 
\beq
\phi_1= \rho  + (-i \omega_T) \rho^2 + \frac{ (- i \omega_T)^2 }{2} \rho^3 \log \rho 
+  \hat{\phi}_1\, .
\eeq
Numerical results are shown in fig.~\ref{chi-mq-varie5}. There is a
discontinuity in the real part of $\chi$ as a function of $\Delta$,
when it approaches the value $\Delta=3$.

\item  $m^2=-7/4$, where $\Delta_-=1/2$, $\Delta_+=7/2$.
The following numerical Ansatz is used:
\beq
\phi_1= \rho^{1/2} +  (-i \omega_T)  \rho^{3/2}
- \frac{(-i \omega_T)^3}{3} \rho^{7/2}+  \hat{\phi}_1 \, ,
\eeq
and then we proceed as in the case of generic $m$.
The function $\chi(\omega_T)$ is shown in fig.~\ref{chi-mq-varie5}.

\item  $m^2=0$, where $\Delta_-=0$, $\Delta_+=4$, which is the case of
 deformation by a marginal operator.
 For the numerical calculation we set:
 \beq
 \phi_1=\alpha+(-i \omega_T) \rho + \frac{\rho^2}{4} (-i \omega_T)^2 
  -\frac{\rho^3}{12} (-i \omega_T)^3  -\frac{\rho^4 \log \rho}{16} (-i \omega_T)^4 
   +  \hat{\phi}_1 \, ,
 \eeq
and we proceed as in the previous cases.
The numerical results for this case are shown in
fig.~\ref{chi-mq-varie5}. 

\end{itemize}

\subsection{Checks}

A cross check on the numerics comes from the Kramers-Kronig (KK) relations:
\beq
{\rm Re} \, \chi_\Delta (\omega)=
 \frac{1}{\pi} \, {\rm P} \int_{-\infty}^{\infty} \frac{   {\rm Im} \, \chi_\Delta (\gamma) }{\gamma -\omega} d \gamma \, , \qquad
{\rm Im} \, \chi_\Delta (\omega)= -\frac{1}{\pi} \, {\rm P} \int_{-\infty}^{\infty} \frac{ {\rm Re} \, \chi_\Delta (\gamma)}{\gamma -\omega} d \gamma \, , 
\eeq
where P denotes the principal value. 
In the case of $\zeta<0$, we can apply these relations directly, as
$\chi(\omega,0)$ vanishes for large $\omega$. 
In the case of $\zeta \leq 0$, we should first subtract the divergent
part at large $\omega_T$; this is more difficult, because we need to
compute $\chi$ with a rather good accuracy. 
Practically, we managed to check KK relations only up to
$\Delta<3.5$.

For $\omega_T=0$, the Klein-Gordon equation (\ref{wave-eq}) can be
solved in terms of
hypergeometric functions in the static case; the solution with
an ingoing-wave boundary condition at the horizon is
\cite{Buchel:2003ah}:  
\beq
\tilde{\phi}_1=\rho^{\Delta_+} \, _{2}F_1 \left( \frac{\Delta_+}{4} , \frac{\Delta_+}{4} ,1,1-\rho^4\right)=
\rho^{\Delta_-} \, _{2}F_1 \left( \frac{\Delta_-}{4} , \frac{\Delta_-}{4} ,1,1-\rho^4\right) \, .
\label{eq-sol}
\eeq
From this exact solution the following relation can be extracted
\cite{newquench}: 
\beq
a_{\Delta_+}=-\frac{\Gamma \left(\frac{\Delta_-}{2} \right) \Gamma  \left(\frac{\Delta_+}{4} \right)^2 }{\Gamma \left(\frac{\Delta_+}{2} \right) \Gamma  \left(\frac{\Delta_-}{4} \right)^2} a_{\Delta_-} \, .
\label{controllo1}
\eeq
The equilibrium value of the response function $\chi_\Delta(0)$ can be
found using eqs.~(\ref{chichi1},\ref{chichi2}).

The large $\omega_T=\omega/(\pi T)$ regime is obtained in the small temperature limit,
which corresponds to the vacuum AdS metric at $T=0$. 
Calculations are done for example in  \cite{Balasubramanian:1998sn,Son:2002sd}:
\beq
\chi_\Delta(\omega_T)=\frac{- e^{-i \pi |2-\Delta|} \,  \Gamma(3-\Delta) }{\Gamma(\Delta -1) 2^{2 \Delta-4}}  \omega_T^{2 \Delta-4} \, .
\label{grande-omega}
\eeq
The case of integer $\Delta$s is special and contains logarithms: 
\beq
\chi_\Delta(\omega_T) = - A_{\Delta}  \omega_T^{2 \Delta-4} \left( \log \omega_T - \frac{i \pi}{2} \right) \, , \qquad
A_2=1 \, , 
\label{grande-omega-delta-intero}
\eeq
\[
A_\Delta=\frac{1}{(\Delta-2)! \, (\Delta-3) ! \, 2^{2 \Delta -5}} \, ,
\qquad 
{\rm for} \, \Delta>2 \, .
\]    
We find agreement at the per-mille level between the numerical and analytical results
both in the $\omega_T \gg 1$ and in the $\omega_T=0$ regimes.

In the limit $\Delta\to 1$, the theory is just above the unitarity
bound; this is reflected in a singular behavior of ${\rm Im} \,
\chi(\omega_T)$, which becomes a function which is sharply peaked 
at a very low value of the frequency $\omega_{T,0}$. 
The position of the maximum $\omega_{T,0}$ tends to zero for
$\Delta\to 1$; by using a numerical fit we find that $\omega_{T,0} \approx
0.966 (\Delta-1)^{1/2}$. 

The function ${\rm Im} \, \chi_\Delta$ is generically expected to start off with a linear term in $\omega_T$ for $\omega_T \ll 1$,
because it is  an odd function of $\omega_T$. The numerical calculation confirms this behavior.

\section{Probes of the geometry\label{sec:geometry}}

\subsection{Apparent horizon}

We will next use the calculation of the metric backreaction to
study the time evolution of the area of the horizon,
which is related to the entropy in the boundary CFT.
It is still an open question whether a valid notion of local entropy density
exists in out-of-equilibrium situations. 
It has been argued \cite{Bhattacharyya:2008xc} that the area of the
horizon, projected onto the boundary along an in-falling null geodesic 
(which corresponds to a line with  constant $v$ in EF coordinates)
could be identified with the entropy density in the dual field
theory. 
However, it has been suggested in
\cite{Chesler:2008hg,Figueras:2009iu} that apparent horizons provide a 
better definition of entropy density than global horizons. 
We will compute and compare the area of both the apparent and the
event horizon on 
solutions taking the leading order backreaction into account. 

Let us start with the apparent horizon; it is associated with the
presence of a trapped surface.
Considering the following light-like vectors:
\beq
l_a=(-1,0,0,0,0) \, , \qquad n_a=(-A/2,1,0,0,0) \, ,
\eeq
\beq
l^a=(0,-1,0,0,0) \, , \qquad n^a=(1,A/2,0,0,0) \, ,
\eeq
written in the coordinates $(v,r,\vec{x})$, we compute the expansion
of the null geodesics: 
\beq
\theta_w =(g^{ab}+l^a n^b +l^b n^a) \nabla_a w_b \, ,
\eeq
where $w=l,n$. It follows that
\beq
\theta_l = \frac{3 \Sigma'}{\Sigma} \, , \qquad \theta_n = -3 \frac{A \Sigma' +2 \dot{\Sigma}}{2 \Sigma} \, ,
\eeq
where $',\dot{\,}$ are derivatives with respect to $r,v$. 
The position of the apparent horizon is detected by the vanishing of
$\theta_n$. 

Using the variables $\rho,\tau$,  the location
$\rho_a$ of the apparent horizon to  order $\lambda^2$ is determined by:
\beq
(1-\rho_a^4)+\lambda^2( \rho_a^2 \tilde{A}_2 +\tilde{\Sigma}_2' \rho_a^2 (\rho_a^4-1) +2 \dot{\tilde{\Sigma}}_2 \rho_a^2 )=0\, .
\eeq
At this order one can write $\rho_a=1+\lambda^2 \delta \rho_a + \mathcal{O}(\lambda^4)$ and
 the variation of the position of the apparent horizon $\rho_a$ is given by:
\beq
\delta \rho_a =\frac{A_{2,{\rm c}}}{4} +\frac{A_{2,{\rm l}}}{4} \tau +{\rm Re } \left( \left( \frac{A_{2,{\rm p}}}{4}
 -(i \omega_T) \Sigma_{2,{\rm p}}  \right) e^{-2 i \omega_T \tau} \right) \, ,
\eeq
where the functions $A_{2,{\rm c}},A_{2,{\rm l}},A_{2,{\rm p}}, \Sigma_{2,{\rm c}}, \Sigma_{2,{\rm p}}$ are evaluated at
$\rho=1$. 
The area of an element of the apparent horizon $\mathcal{A}_a$ at order $\lambda^2$
is then: 
\beq
\frac{\mathcal{A}_a}{\mu^3}=1+3 \lambda^2 \left( \tilde{\Sigma}_2 -\delta \rho_a \right)=
\eeq
\beq
=1+3 \lambda^2 \left(
\Sigma_{2,{\rm c}} -\frac{A_{2,{\rm c}}}{4} -\frac{A_{2,{\rm l}}}{4} \tau + {\rm Re} \left( \left( -\frac{A_{2,{\rm p}}}{4} +(1+i \omega_T) \Sigma_{2,{\rm p}} \right) e^{-2 i \omega_T \tau} \right)
\right) \, .
\eeq
Using the expression~(\ref{Aperiodica}), we find:
\beq
\frac{\mathcal{A}_a}{\mu^3}={\rm constant} +\lambda^2 \left( \frac{\omega_T^2 |\phi_1(1)|^2}{8} \, \tau + {\rm Re} \left( -\frac{i \omega_T \phi_1(1)^2}{16} e^{-2 i \omega_T \tau} \right) \right) \, .
\eeq
The area of the apparent horizon always increases as a function of
time, independently of the value of $ \phi_1(1)$ and $\omega_T$; 
the relative coefficient of the periodic and of the linear part in
time is such that in every cycle there is an instant at which the time
derivative of $\mathcal{A}_a$ vanishes. 

\begin{figure}[ht]
\begin{center}
$\begin{array}{c@{\hspace{.2in}}c} \epsfxsize=6.5cm
\epsffile{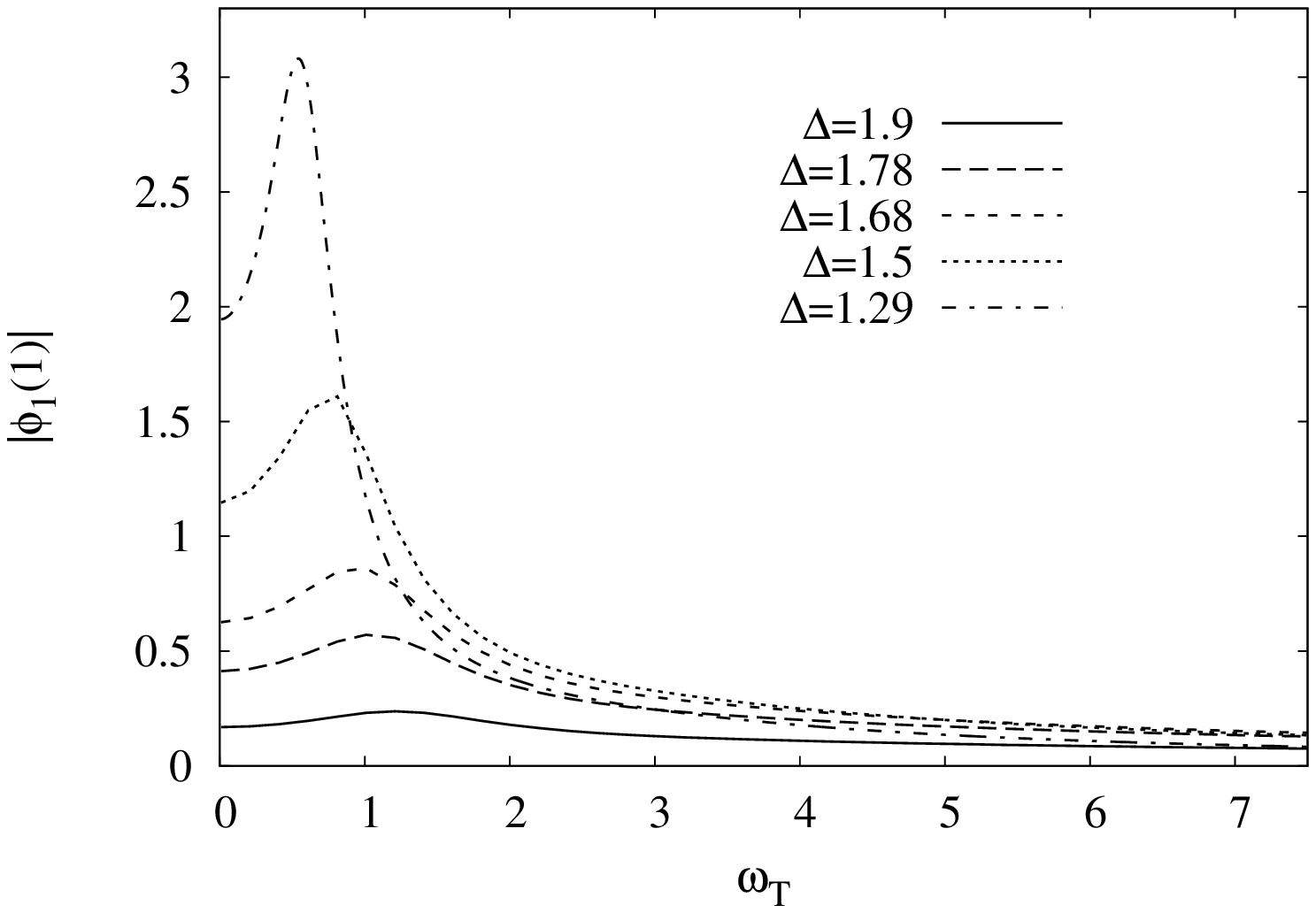} &
     \epsfxsize=6.5cm
    \epsffile{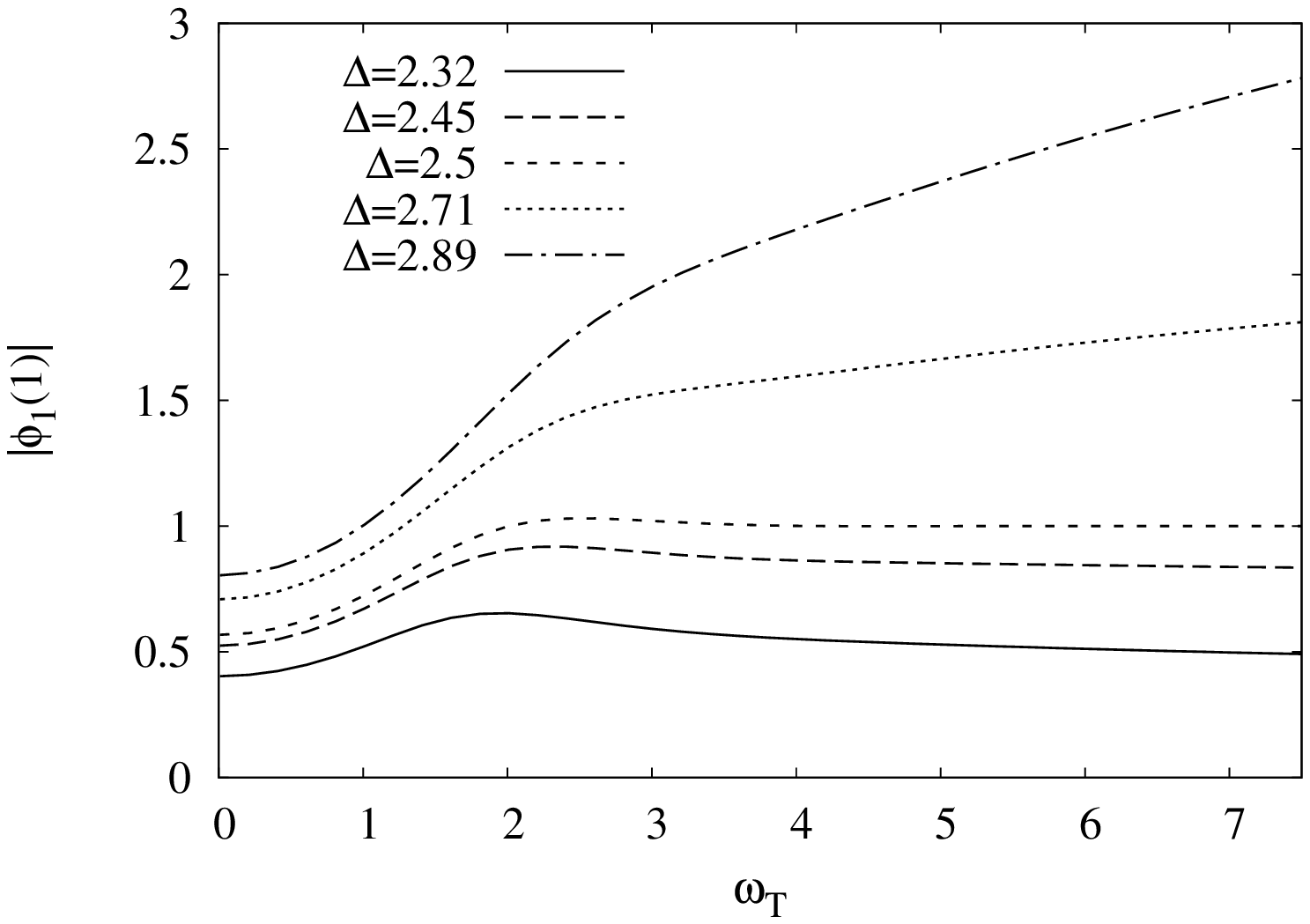}
\end{array}$
\end{center}
\caption{\footnotesize
The value of $|\phi_1(1)|$ as a function of frequency for various
values of $\Delta$. 
The entropy produced per unit time is proportional to $\omega_T^2
|\phi_1(1)|^2$. 
}  \label{modulo-entropy-figs}
\end{figure}

\begin{figure}[ht]
 \centerline{\includegraphics[width=6.5cm]{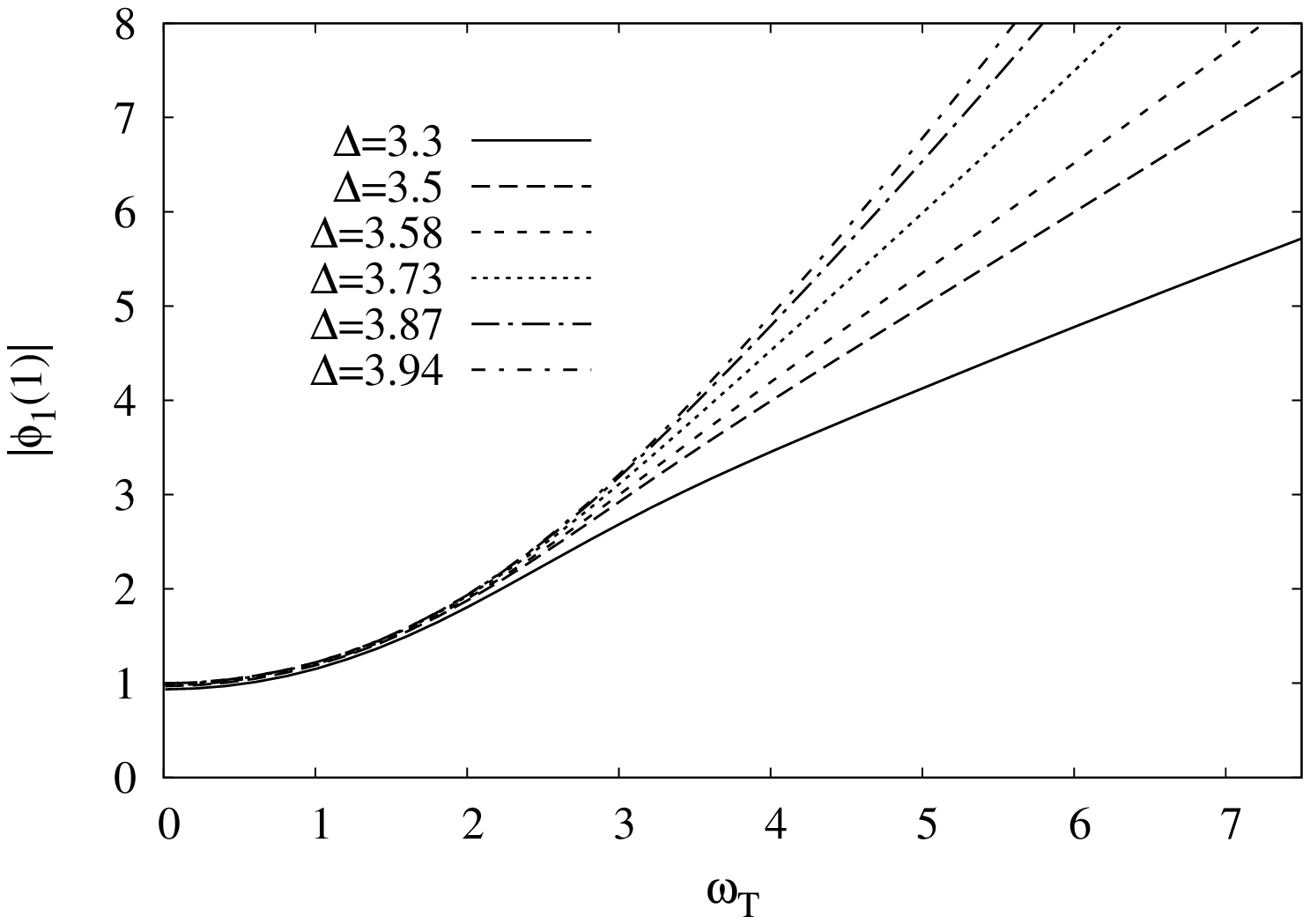}}
 \caption{\footnotesize The value of $|\phi_1(1)|$ as a function of
   frequency for various values of $\Delta$. 
The entropy produced per unit time is proportional to $\omega_T^2
|\phi_1(1)|^2$.} 
\label{modulo-entropy-figs-2}
\end{figure}

The increase of area in a cycle is proportional to $\omega_T |\phi_1(1)|^2$.
 Some plots of $|\phi(1)|$ as a function of $\omega_T$ for various values of $\Delta$ are shown in 
figs.~\ref{modulo-entropy-figs},\ref{modulo-entropy-figs-2}.
By comparing eq.~(\ref{cardellino}) and eq.~(\ref{upupa}),
we find that the value of $|\phi_1(1)|^2$ is directly related to ${\rm Im} \, \chi$: 
\beq
|\phi_1(1)|^2 = \frac{2 |\Delta-2| \,  {\rm Im } \, \chi_\Delta(\omega_T)}{\omega_T} \, ,
\eeq
(in our conventions, this expression is valid for $\Delta \neq 2$; for
$\Delta=2$, it gets modified to $|\phi_1(1)|^2 = {\rm Im } \, \chi_2
(\omega_T)/\omega_T $). 
The increase in area and in energy in a cycle are given by:
\beq
\frac{\delta \mathcal{A}_a}{\mathcal{A}_a}=\frac{\lambda^2}{2} \pi |\Delta-2| {\rm Im } \, \chi_\Delta \, , \qquad
\frac{\delta \mathcal{E}}{\mathcal{E}}=\frac{2 \lambda^2}{3} \pi |\Delta-2| {\rm Im } \, \chi_\Delta \, . 
\eeq
Taking the area of apparent horizon \cite{Chesler:2008hg,Figueras:2009iu}  as a representative of entropy density
 ($\mathcal{S}_a=\mathcal{A}_a/(4 G_N)$),
we see that the variation of entropy in a cycle is:
$\delta \mathcal{S}_a / \mathcal{S}_a
=\frac{3}{4} \delta \mathcal{E}/\mathcal{E}$.
This is the same variation in entropy as 
 one would expect from the equation of state $\mathcal{S} \propto \mathcal{E}^{3/4}$ of the undeformed CFT at equilibrium
for a variation in internal energy equal to the work done on the system.
This is consistent with thermalization.
It should be stressed that  this result can be made valid for arbitrary values of  $\omega_T$ 
by considering the limit of a sufficiently small amplitude $\xi_0$.
This is done in a way that keeps the total work per unit of volume performed
 on the system  much smaller than the initial energy density.
For $\omega_T \gg 1$, this does not overlap with
 the regime of validity of the hydrodynamic approximation,
 which instead is an expansion around $\omega_T=0$ and is valid for arbitrary amplitude $\xi_0$.
The hydrodynamic description of a driven system in the case of $\Delta=4$ was studied,
at  second order in the boundary derivative expansion, in \cite{Bhattacharyya:2008ji};
this corresponds to keeping just the linear and the quadratic order in $\chi_4(\omega_T)$,
see eq.~(3.19) in \cite{Bhattacharyya:2008ji}. 

\begin{figure}[ht]
\begin{center}
$\begin{array}{c@{\hspace{.2in}}c} \epsfxsize=6.5cm
\epsffile{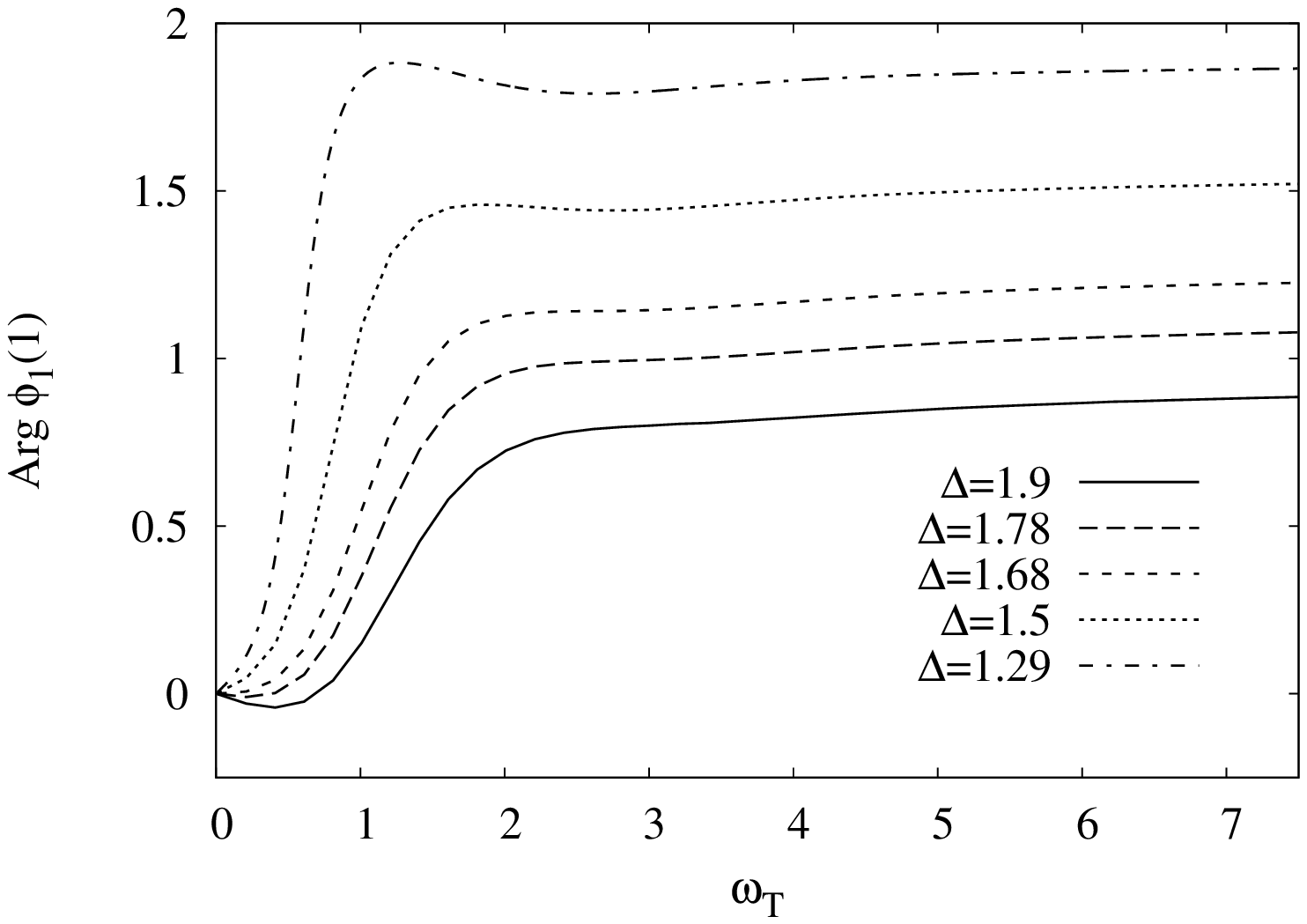} &
     \epsfxsize=6.5cm
    \epsffile{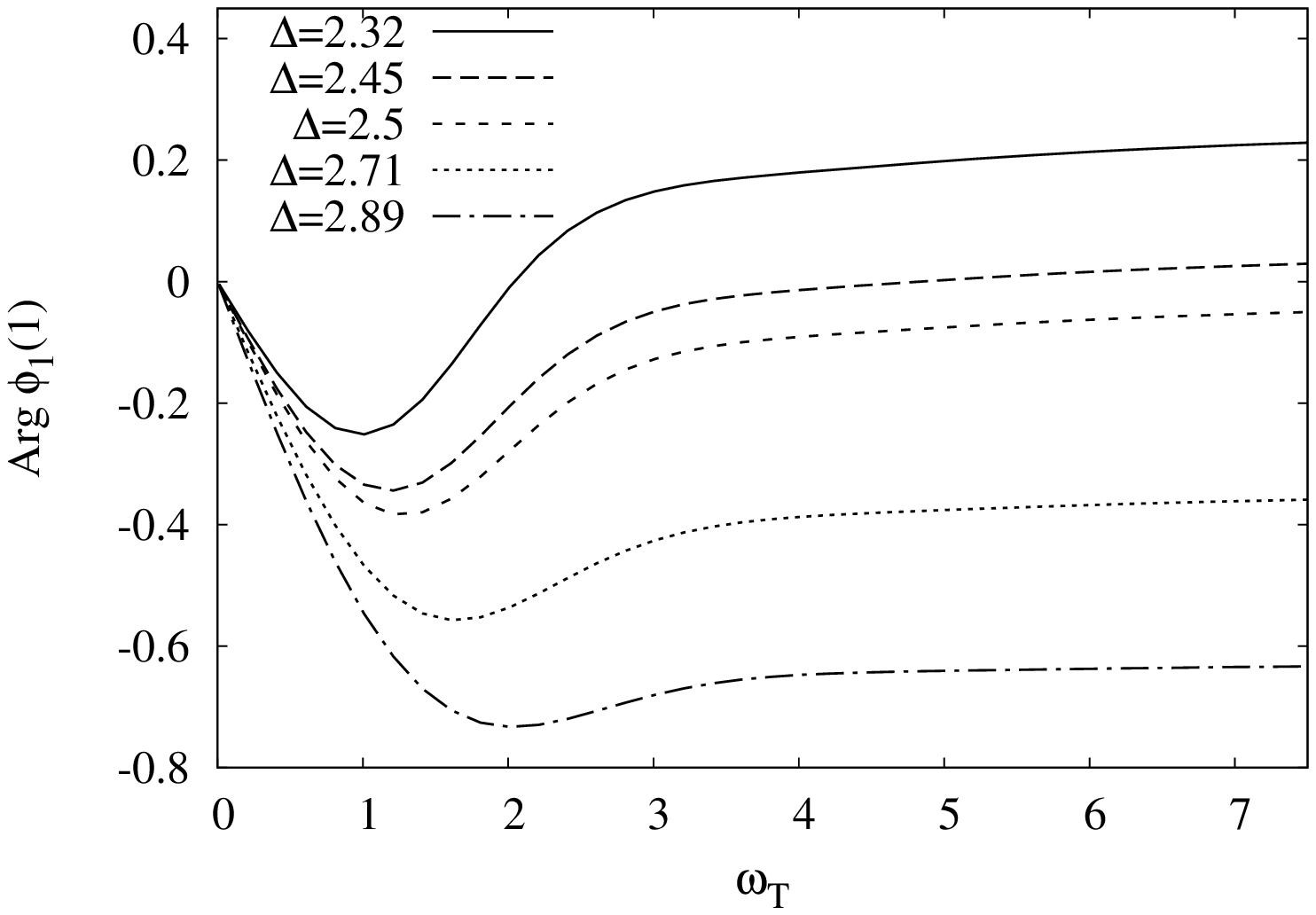}
\end{array}$
\end{center}
\caption{\footnotesize
  The value of ${\rm Arg} \, \phi_1(1)$ as a function of frequency for
  various values of $\Delta$. 
}  \label{phase-entropy-figs}
\end{figure}

\begin{figure}[ht]
 \centerline{\includegraphics[width=6.5cm]{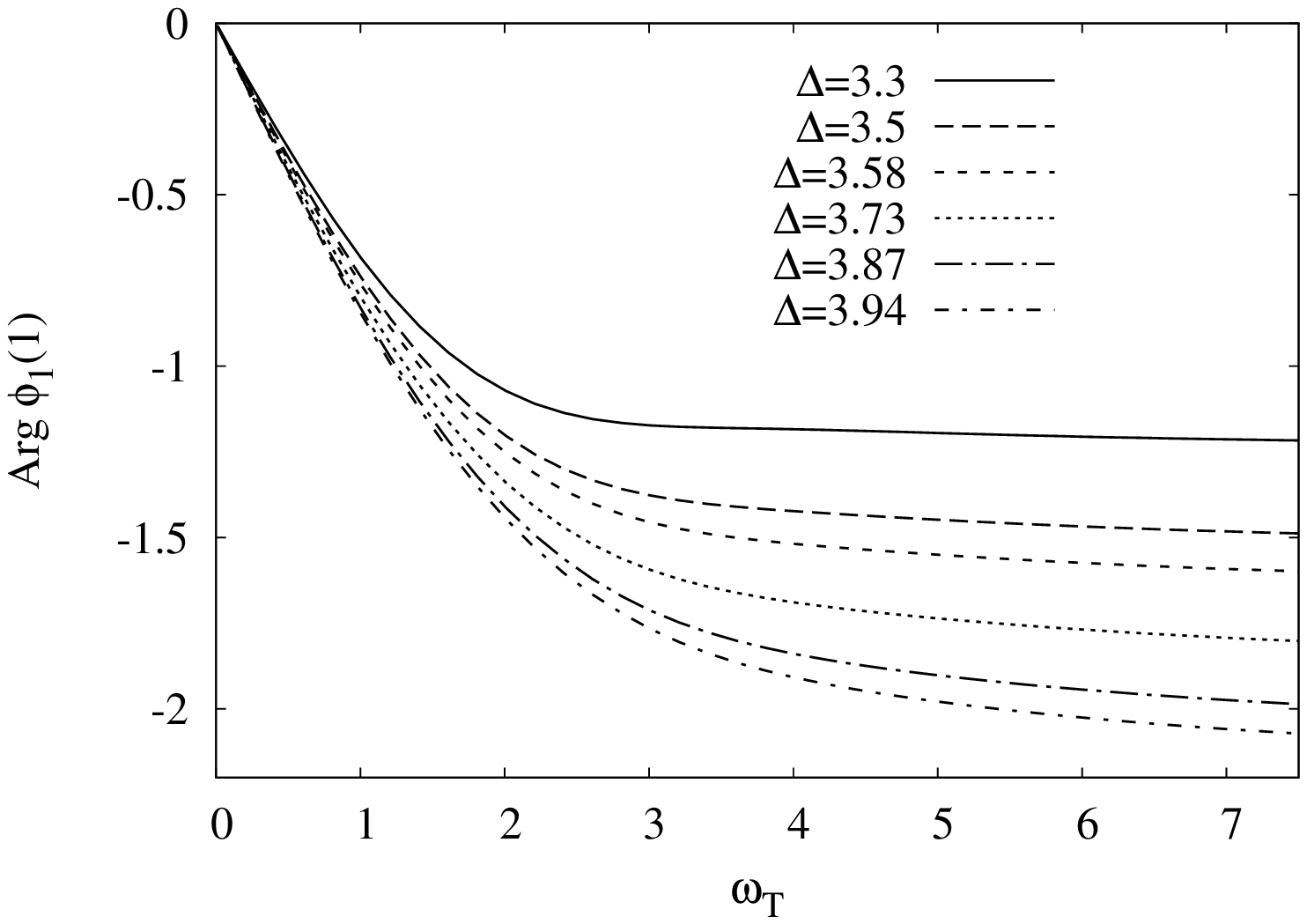}}
 \caption{\footnotesize  The value of ${\rm Arg} \, \phi_1(1)$ as a
   function of frequency for various values of $\Delta$.} 
\label{phase-entropy-figs-2}
\end{figure}

The quantity ${\rm Arg} (\phi_1(1))$ gives the difference
between the phase associated with the source and that given by the moment  at which the time derivative of $\mathcal{A}_a$ vanishes. 
In the limit $\omega_T\to 0$, the evolution of $\mathcal{A}_a$ is in
phase with the source and then ${\rm Arg} (\phi_1(1))\to 0$;
 this is the limit in which the hydrodynamic approximation is also valid. 
Some plots as functions of frequency, for several values of $\Delta$,
are shown in
figs.~\ref{phase-entropy-figs},\ref{phase-entropy-figs-2}; from the
numerical calculations, we find that this phase difference approaches
a constant value (which can have either sign depending on $\Delta$) at
large $\omega_T$.

\subsection{Event horizon}

The event horizon corresponds to a null geodesic which separates
trapped null geodesics from the ones that can escape to the
boundary. In the unperturbed geometry, it sits at $\rho(\tau)=1$. 
At order $\lambda^2$, the geodesic equation receives the correction:
\beq
\frac{d \rho}{d \tau} = - \frac{1-\rho^4+\lambda^2 \rho^2 \tilde{A}_2(\tau,\rho)}{2} \, .
\eeq
Let us expand the position of the event horizon as $\rho_e=1+\lambda^2
\delta \rho_e(\tau)+ \mathcal{O}(\lambda^4)$; 
the leading correction $\rho_e$ satisfies:
\beq
2 \frac{d \, \delta \rho_e}{d \tau} = 4 \delta \rho_e -\tilde{A}_2 (\tau,\rho=1)\, ,
\label{orizzonte-ventoso}
\eeq
whose solution is:
\beq
\delta \rho_e=\left( \frac{A_{2,{ \rm c}}}{4} +\frac{A_{2,{\rm l}}}{8} \right)  +\left( \frac{A_{2,{\rm l} }}{4}  \right)  \tau +{\rm Re} \left( \frac{A_{2,{\rm p}} }{4(1+i \omega_T)} e^{-i \omega_T \tau} \right) \, .
\label{posizione-orizzonte-eventi}
\eeq
Note that  one could add a function of the form $C e^{2 \tau}$, with $C$ an arbitrary constant, to the solution
  $\delta \rho_e$ in eq.~(\ref{posizione-orizzonte-eventi}), and still get 
a solution of eq.~(\ref{orizzonte-ventoso}). The curves with $C>0$ correspond to trapped  null  geodesics,
while the ones with $C<0$ correspond to  null geodesics which can escape to the boundary.

\begin{figure}[ht]
 \centerline{\includegraphics[width=6.5cm]{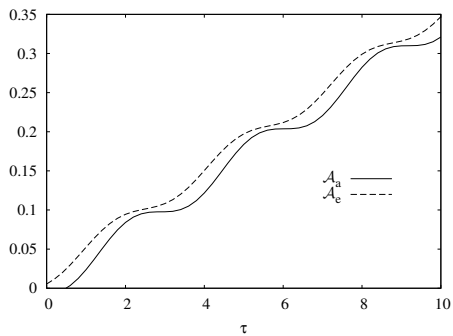}}
 \caption{\footnotesize
Comparison of the time evolution of the area $\mathcal{A}_a$ of the
apparent horizon (solid) and the event horizon $\mathcal{A}_e$
(dashed). 
As expected, the event horizon is always larger than the apparent one;
we used the illustrative values $\Delta=2.32$ and $\omega_T=1$. 
In every cycle there is an instant in which the time derivative of
$\mathcal{A}_a$ vanishes, independently of $\omega_T$ and
$\Delta$. The time derivative of $\mathcal{A}_e$ is always strictly
positive definite. 
}
\label{apparent-vs-event}
\end{figure}

This is different than the position of the apparent horizon;
the only term which is the same is the one linear in time. 
The area of the event horizon $\mathcal{A}_e$ is
\beq
\frac{\mathcal{A}_e}{\mu^3}=
1+3 \lambda^2 \left(
\Sigma_{2,{\rm c}} -\frac{A_{2,{\rm c}}}{4} -\frac{A_{2,{\rm l}}}{8} 
-\frac{A_{2,{\rm l}}}{4} \tau + {\rm Re} \left( \left( -\frac{A_{2,{\rm p}}}{4(1+i \omega_T)} + \Sigma_{2,{\rm p}} \right) e^{-2 i \omega_T \tau} \right)
\right) \, .
\eeq
which, using (\ref{Aperiodica}), can be written as:
\beq
\frac{\mathcal{A}_e}{\mu^3}={\rm constant} +\lambda^2 \left( \frac{\omega_T^2 |\phi_1(1)|^2}{8} \tau + {\rm Re} \left( -\frac{i \omega_T \phi_1(1)^2}{16(1+i \omega_T)} e^{-2 i \omega_T \tau} \right) \right) \, ,
\eeq
and it increases monotonically in time, as expected from the area
theorem of general relativity. 
The time derivative $\dot{\mathcal{A}}_e$ is always strictly positive;
this is different from the area of the apparent horizon
$\mathcal{A}_a$, for which $\dot{\mathcal{A}}_a=0$ once for every
cycle. 
The difference between the two areas is:
\beq
\frac{\mathcal{A}_e- \mathcal{A}_a}{\mu^3} =\lambda^2 \left(
 \frac{\omega_T^2 |\phi_1(1)|^2}{16} + {\rm Re} \left( -i \frac{\omega_T \phi_1(1)^2}{16} \frac{i \omega_T}{1+i \omega_T}  e^{-2 i \omega_T \tau }\right) \right) \, ,
\eeq
which is bigger than zero. This is consistent with the expectation that the area of
event horizon is bigger than that of the apparent one at a given moment in
time. A comparison in an illustrative example is shown in
fig.~\ref{apparent-vs-event}. 
Since the two definitions of horizon area  show the same linear growth in time, the corresponding increase
of entropy density $\delta \mathcal{S}=\delta \mathcal{A}/(4 G_N)$ in a cycle is in both cases the same as the one given by the equation of state of the undeformed CFT:
$\delta \mathcal{S}/\mathcal{S}=\frac{3}{4} \delta \mathcal{E}/ \mathcal{E}$.

 It should be emphasized that the calculation of the position of the event horizon in eq.~(\ref{posizione-orizzonte-eventi})
is valid only in the regime where the total work  per unit of volume done on the system
remains much smaller than the initial energy density. It is also important that a large number of cycles
is performed by the source in such a way that the initial transient period can be neglected.
Both these conditions can be achieved by keeping the amplitude  $\xi_0$ of the source  sufficiently small.

\subsection{Geodesics and equal-time two-point functions}

We will probe the back-reacted geometry 
by computing the renormalized equal-time Wightman function of 
an irrelevant scalar operator with large dimension $\Delta_p$.
This quantity is directly related to the space-like geodesics of the
geometry which join points of the same time at the boundary. 
For calculations in the case of the Vaidya metric see
\cite{Balasubramanian:2011ur,Aparicio:2011zy,Balasubramanian:2012tu}.

\begin{figure}[ht]
\begin{center}
$\begin{array}{c@{\hspace{.2in}}c} \epsfxsize=6cm
\epsffile{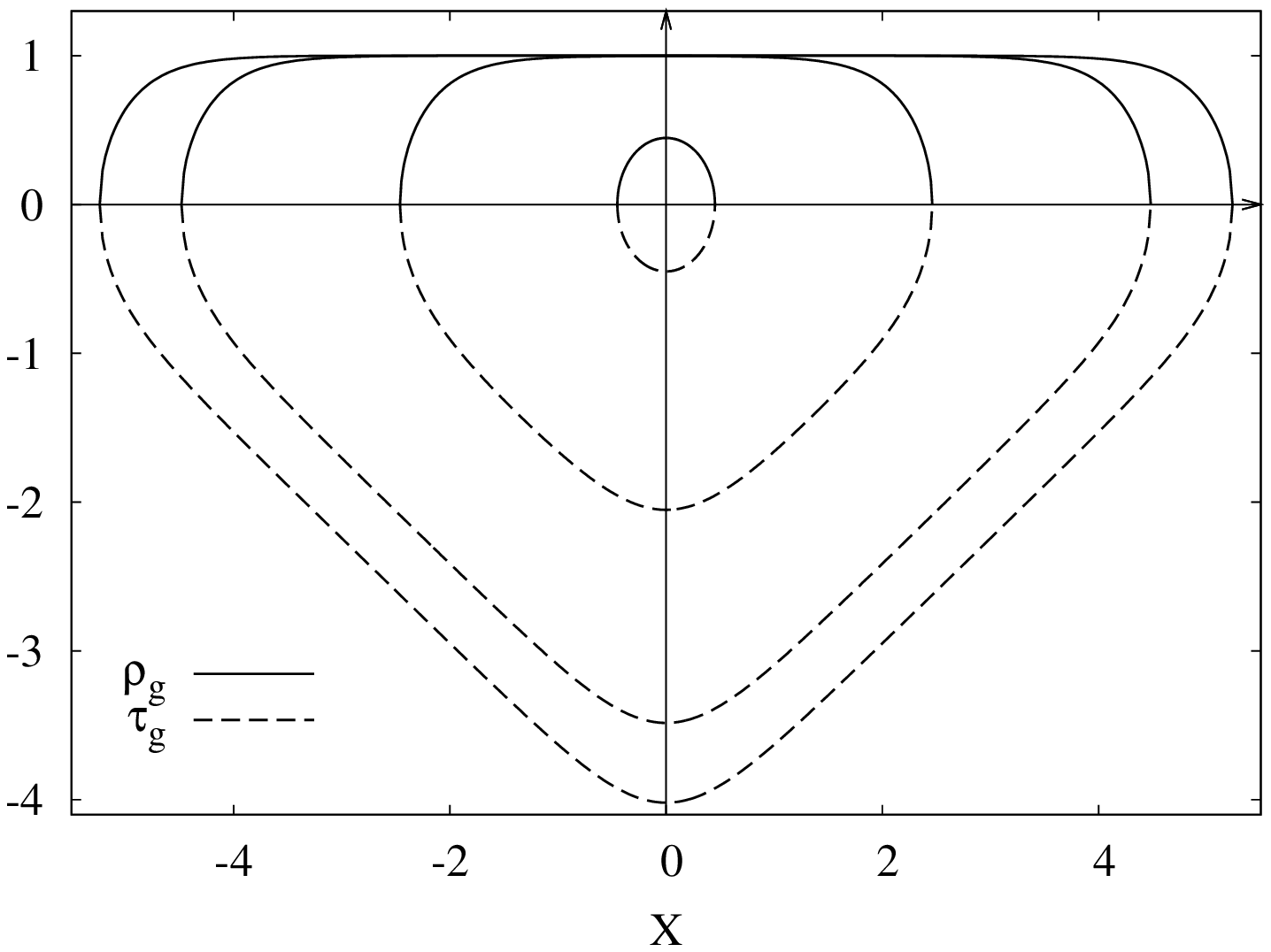} &
     \epsfxsize=6cm
    \epsffile{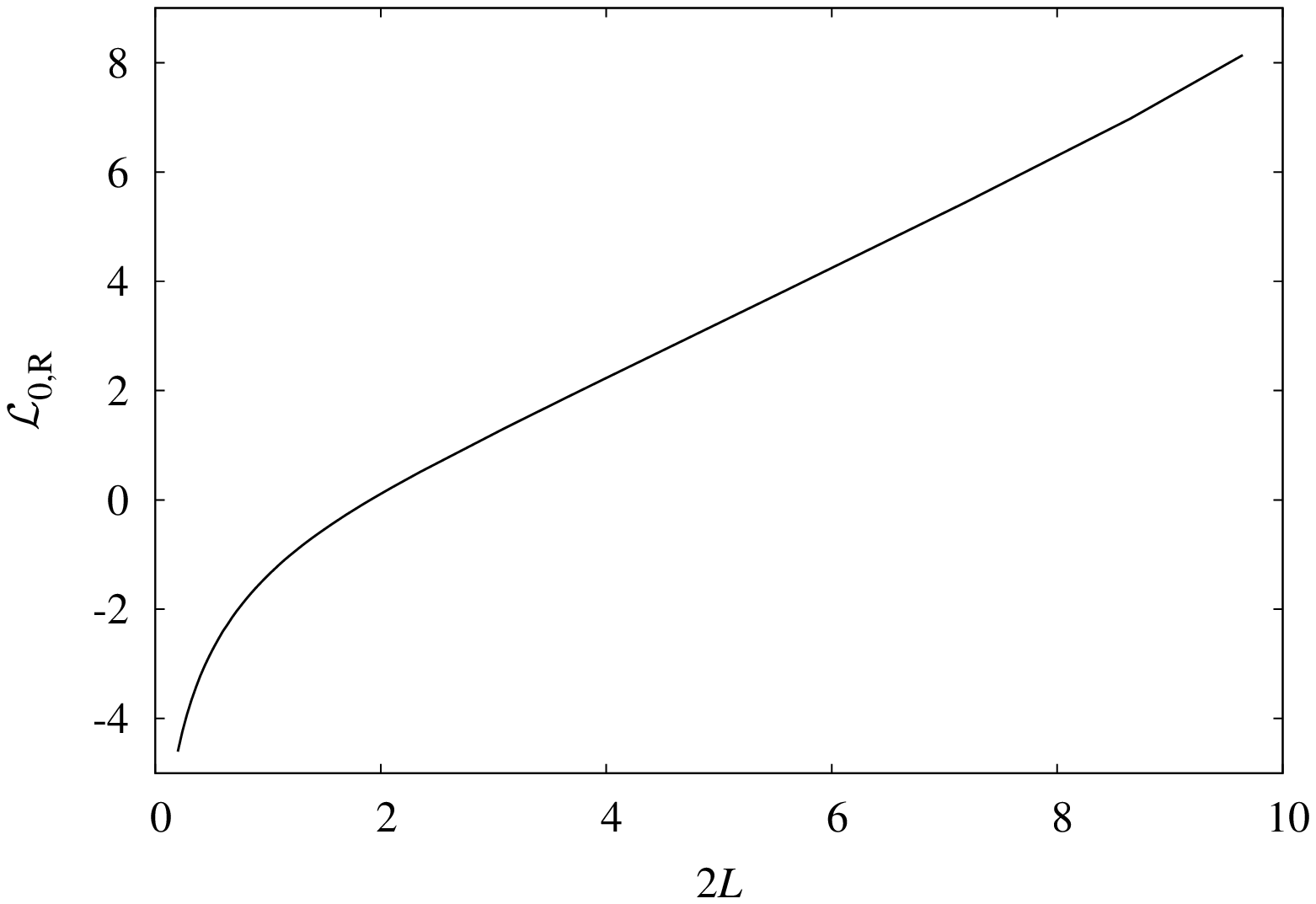}
\end{array}$
\end{center}
\caption{\footnotesize
Left: Examples of geodesics of the unperturbed metric, with
$\rho_g(X)$ (solid line) and $\tau_g(X)$ (dashed line). 
Right: $\mathcal{L}_{0,R}$ as a function of $L$ for the AdS$_5$
black hole. 
}  \label{geodesiche-unperturbed}
\end{figure}

Let us first discuss the $\lambda=0$ case, and
parametrize the geodesics with the coordinate $X$ and the functions $\tau_g (X)$, $\rho_g(X)$. 
We recall that all the distance and time scales are measured in units of the inverse temperature $T^{-1}$.
The geodesic length at zeroth order reads:
\beq
\mathcal{L}_0=\int \frac{\sqrt{\Xi_g}}{\rho_g} \, d X \, ,
\qquad
\Xi_g \equiv (\rho_g^4-1)(\tau_g')^2 -2 \tau_g' \rho_g' +1 \, .
\label{geolength}
\eeq
Translation invariance of $\tau$ and $X$ gives the conserved quantities: 
\beq
  \Xi_g  \rho_g^2  = (\rho_g^*)^2 \, , \qquad (\rho_g^4-1) \tau_g' - \rho_g' = K \, .
\eeq
Geodesics which join points of the same time at the boundary satisfy
the relation $\tau_g'(\rho_g^*)=0$, where $\rho_g^*$ is the value of
$\rho_g$ at the midpoint; for this class of geodesics $K=0$ and 
\beq
\Xi_g = 1+\frac{(\rho_g')^2}{1-\rho_g^4} \, , \qquad
\tau_g'=\frac{\rho_g'}{\rho_g^4-1} \, .
\eeq
This family of geodesics  can be
parametrized by the boundary length in the $X$ direction $2L$ or,
alternatively, by the maximum of $\rho_g$, which is $\rho_g^*$; this class of
geodesics does not penetrate the horizon ($\rho_g<1$, and $\rho_g^*
\to 1$ for $L \rightarrow \infty$). We consider geodesics stretching
from $X=-L$ and $X=L$ at the boundary; 
the integration constant for $\tau_g$ is chosen in such a way that
$\tau_g(\pm L)=0$. Some examples of these geodesics are shown in
fig.~\ref{geodesiche-unperturbed}. 

Near the boundary,  $\rho\to 0$, the asymptotics of the geodesics are:
\beq
\rho_g \approx \sqrt{2 \rho_g^* (L-X)} \, , \qquad \rho_g' \approx -\tau_g' \approx -\sqrt{\frac{\rho_g^*}{2(L-X)}} \, .
\label{geo-asy}
\eeq
The geodesic length is then a divergent quantity; it is convenient to
define a renormalized length 
\beq \mathcal{L}_{0,R}=\mathcal{L}_0+2 \log (\rho_{\rm cut} /2) \, ,
\eeq
where $\rho_{\rm cut}$ is a cut-off.
The renormalized equal-time Wightman function of an operator
$\mathcal{O}_{\Delta_p}$ with large dimension, $\Delta_p\gg 1$, 
is proportional to:
\beq
 \langle \mathcal{O}_{\Delta_p} (t,\vec{x} )  \mathcal{O}_{\Delta_p} (t,\vec{x}' ) \rangle_{\rm ren} \propto e^{-\Delta_p  \mathcal{L}_{0,R}} \, ,
 \label{tucano}
\eeq
where $\vec{x}=(L,0,0)$ and $\vec{x}'=(-L,0,0)$.
The numerical result as a function of $X$ for the unperturbed AdS$_5$
black hole is shown in fig.~\ref{geodesiche-unperturbed}.

In the limit $L\gg 1$, the quantity $ \mathcal{L}_{0,R}$ is
dominated by a configuration which is situated mostly nearby the horizon, that is at
$\rho_g\approx 1$; expanding $\rho_g(X)=1+\tilde{\rho}_g(X)$,  
one obtains the following geodesic equation to leading order in
$\tilde{\rho}_g$:
\beq
\tilde{\rho}_g^2+2 (\tilde{\rho}_g')^2 -4 \tilde{\rho}_g \tilde{\rho_g''} = 0 \, ,
\eeq
which can be solved exactly:
\beq
\tilde{\rho}_g=-\epsilon \cosh^2(\sqrt{2} X) \, , \qquad \epsilon \approx \frac{1}{\cosh^2(\sqrt{2} L)} \, .
\label{appr1}
\eeq
Then one can extract an approximation for $\tau_g$:
\beq
\tau_g \approx \frac{1}{2} \log \frac{\cosh \sqrt{2} X}{\cosh \sqrt{2} L} \approx \frac{-L + |X|}{\sqrt{2} } \, .
\label{appr2}
\eeq
Substituting back in eq.~(\ref{geolength}), one obtains that 
$\mathcal{L}_{0,R}$ scales like $2 L$. 

In order to compute the leading order correction in $\lambda^2$ to the
geodesic length $\mathcal{L}_R$, it is not necessary to compute
also the leading order correction $\mathcal{O}(\lambda^2)$ to the
geodesic. It suffices to evaluate the  change
in length of the unperturbed geodesic; the contribution of the correction
to the geodesic will be of higher order, due
to the fact that the leading order variation of the action vanishes on
solutions of the Euler-Lagrange equation. 

We can then expand the renormalized geodesic length as follows: 
\beq
 \mathcal{L}_R (\tau^*)= \mathcal{L}_{0,R} +
\lambda^2  \int\frac{\tilde{\Sigma}_2(\rho_g,\tau_g+\tau^*)  - \rho_g \tilde{A}_2(\rho_g,\tau_g+\tau^*)   (\tau_g')^2  /2 }
{\sqrt{ \Xi_g } } dX + \mathcal{O}(\lambda^4)  \, ,
\label{phish}
\eeq
where $\tau^*$ is the time at the boundary.
From eq.~(\ref{geo-asy}) one can check that there are no further
divergences in (\ref{phish}) at order $\lambda^2$.
It is important to remember that the functions
$\tilde{A}_2(\tau,\rho)$ and $\tilde{\Sigma}_2(\tau,\rho)$ are given
by eq.~(\ref{ansaziano}) in the time window $\tau_0^*<\tau<\tau_f^*$;
for time $\tau<\tau_0^*$ the system is in equilibrium and
$\tilde{A}_2=\tilde{\Sigma}_2=0$ (near $\tau \approx \tau_0^*$ there
is a short transient period which we ignore).

The geodesic probes a time in the past which is proportional to the
length $L$, see eq.~(\ref{appr2}) and
fig.~\ref{geodesiche-unperturbed}. 
For boundary times $\tau^*$ such that
\beq
\tau^* - \tau_0^*>-\tau_g(0) \approx \frac{L}{\sqrt{2}} \, ,
\eeq
so that we can use the expressions (\ref{ansaziano}) along
all the geodesic extension, we can decompose the quantity
$\mathcal{L}_R (\tau^*)$ into a constant, a linear and a periodic
part in $\tau^*-\tau_0^*$: 
\beq
 \mathcal{L}_R= \mathcal{L}_{0,R} + \lambda^2 \left(  \mathcal{L}_C+(\tau^*-\tau_0^*)  \mathcal{L}_L 
+ {\rm Re} ( \mathcal{L}_P e^{-2 i \omega_T (\tau^*-\tau_0^*)})\right) \, ,
\label{ella}
\eeq
where
\beq
 \mathcal{L}_C=\int \frac{\Sigma_{2,c}-(A_{2,{\rm c}} + \tau_g A_{2,{\rm l}}) \rho_g \tau_g'^2/2 }
 {\sqrt{ \Xi_g }} \, dX \, , 
\qquad
 \mathcal{L}_L=-\int \frac{ A_{2,{\rm l}} \rho_g \tau_g'^2 }
 {2 \sqrt{ \Xi_g }} \, dX \, ,
\eeq
\beq
 \mathcal{L}_P=\int  \frac{\Sigma_{2,{\rm p}}-A_{2,{\rm p}}  \rho_g \tau_g'^2/2 }
 {\sqrt{ \Xi_g }}  e^{-2 i \omega_T \tau_g } dX \, ,
\eeq
where  $A_{2,{\rm c}},A_{2,{\rm l}},A_{2,{\rm p}}, \Sigma_{2,{\rm c}}, \Sigma_{2,{\rm p}}$ 
 are functions of $\rho_g$.

It is convenient to split:
\beq
 \mathcal{L}_R=  \mathcal{L}_{R, eq} + \delta \mathcal{L}_{R} \, , \qquad
 \mathcal{L}_{R, eq}= \mathcal{L}_{0,R} + \lambda^2  (\tau^*-\tau_0^*)  \mathcal{L}_L  \, ,
\eeq
\[
\delta \mathcal{L}_{R}=
 \lambda^2 \left( \mathcal{L}_C + {\rm Re} ( \mathcal{L}_P e^{-2 i \omega_T (\tau^*-\tau_0^*)})\right) \, .
\]
The quantity $\mathcal{L}_{R, eq}$ corresponds to the geodesic length
in the  conformal field theory at equilibrium, 
with an energy density given by the initial energy plus the total work done on the system:
\beq
 \mathcal{E}(\tau^*)=\mathcal{E}(\tau_0^*)+\frac{ \omega_T  \mathcal{W}_c}{2 \pi} (\tau^*-\tau_0^*) \, .
 \label{enel}
 \eeq

The deviations from the equilibrium are parametrized by $\delta \mathcal{L}_{R}$;
this quantity has a constant part $\mathcal{L}_C$ and a periodic part $\mathcal{L}_P$.
The ratio $\tau_{d,g}=\mathcal{L}_C/\mathcal{L}_L$ can be interpreted
as a time delay in the thermalization of the two-point function.
 In the limit 
$L\gg 1$, using the approximations (\ref{appr1},\ref{appr2}), one
finds 
\beq
\mathcal{L}_L=\frac{\omega_T |\Delta-2| {\rm Im } \chi_\Delta}{6} L + \mathcal{O}(L^0) \, , \qquad 
\mathcal{L}_C=-\frac{\omega_T |\Delta-2| {\rm Im} \chi_\Delta}{12 \sqrt{2}}  L^2 + \mathcal{O} (L)\, .
  \label{ritardazzo}
\eeq
In the regime $L\gg 1$, the delay $\tau_{d,g}$  is linear in $L$: 
\beq
\tau_{d,g}\approx - L/(2 \sqrt{2}) \, .
\label{pappagallo}
\eeq
From this result one can deduce that thermalization time becomes longer with the scale $L$.
This kind of behavior in strongly coupled systems has been observed before in
 \cite{AbajoArrastia:2010yt,Albash:2010mv,Balasubramanian:2010ce,Balasubramanian:2011ur}.
An analogous situation takes place in
quenches \cite{Calabrese:2005in,Calabrese:2006rx,Calabrese:2007rg}  
where, due to causality, $ \mathcal{L}_R$ is linear in time
after the quench up to times of order $L/v$, where $v$ is the maximal 
speed of propagating signals. This is true also for other quantities, 
like the entanglement entropy (see the next section). 
For realization of the same situation in Vaydia geometry, see 
e.g.~\cite{AbajoArrastia:2010yt,Balasubramanian:2011ur,Aparicio:2011zy}.

\begin{figure}[ht]
\begin{center}
$\begin{array}{c@{\hspace{.2in}}c} \epsfxsize=6cm
\epsffile{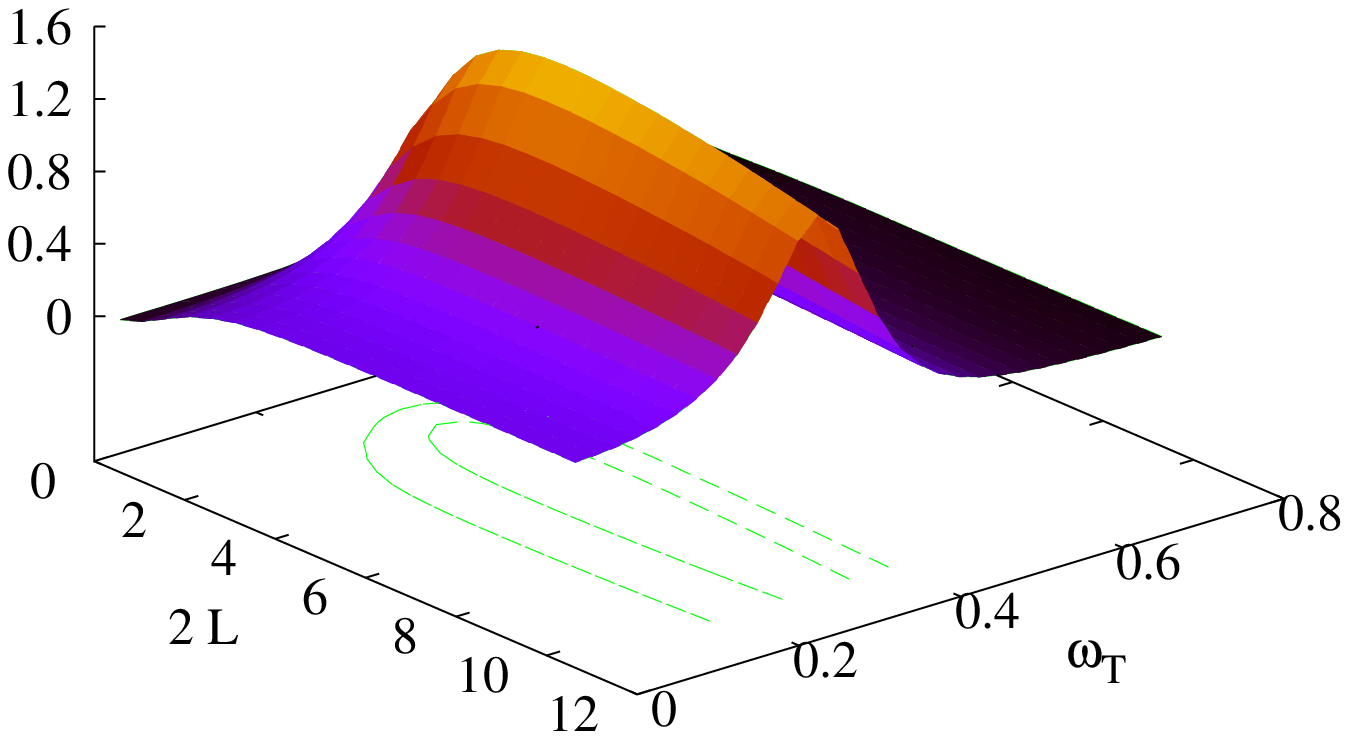} &
     \epsfxsize=6cm
    \epsffile{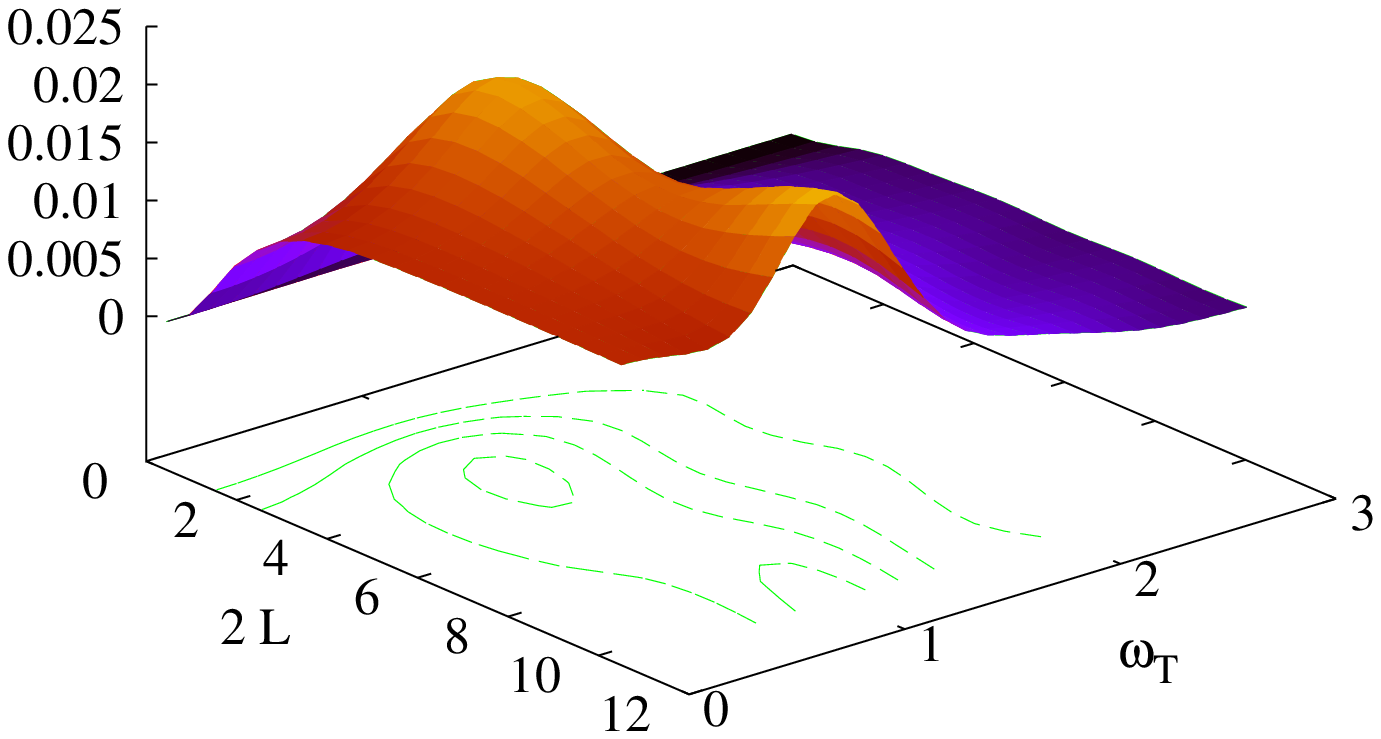}
\end{array}$
\end{center}
\caption{\footnotesize
$|\mathcal{L}_P|$ as a function of $L,\omega_T$ for $m^2=-3.2$,
  $\Delta=1.10$, (left)   $m^2=-3.9$,  $\Delta=1.68$ (right).  
}  \label{ge1}
\end{figure}

\begin{figure}[ht]
\begin{center}
$\begin{array}{c@{\hspace{.2in}}c} \epsfxsize=6cm
\epsffile{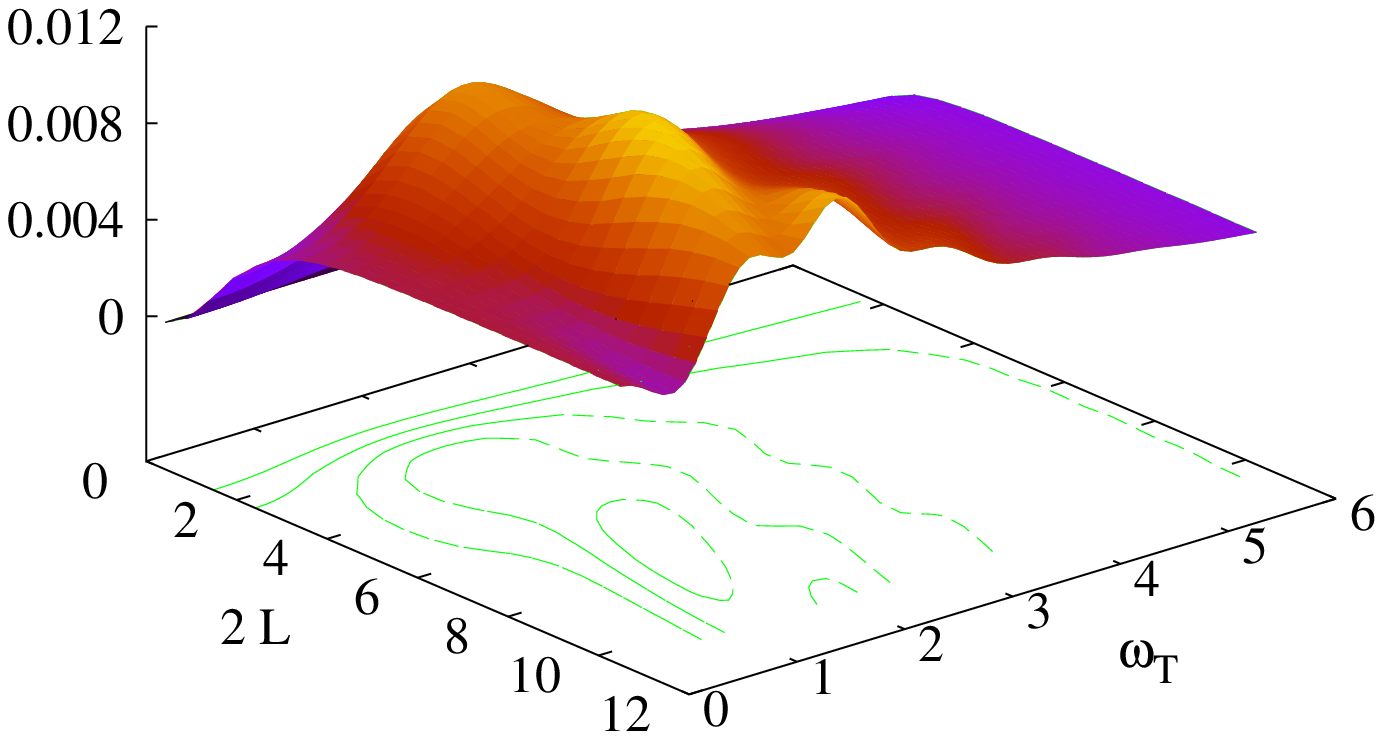} &
     \epsfxsize=6cm
    \epsffile{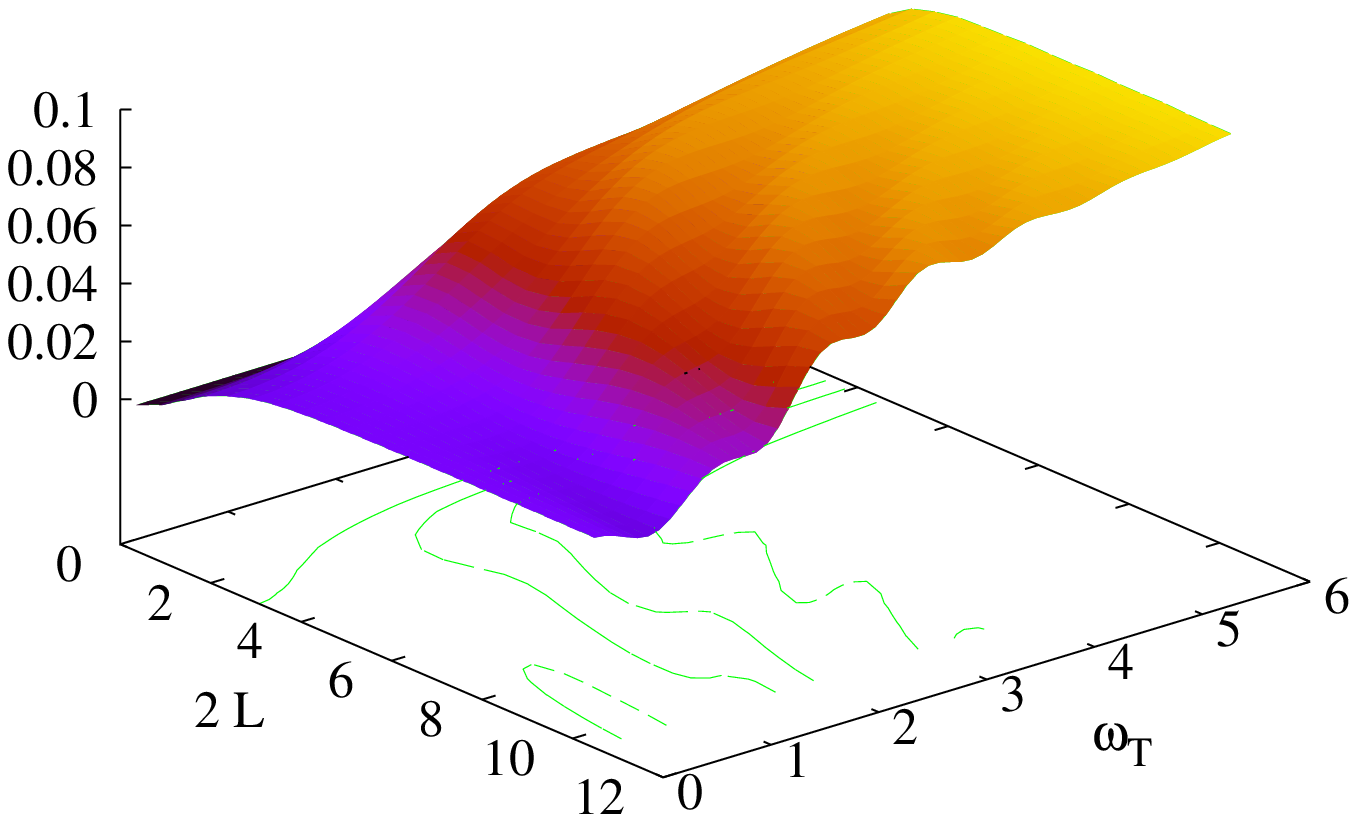}
\end{array}$
\end{center}
\caption{\footnotesize
$|\mathcal{L}_P|$ as a function of $L,\omega_T$ for $m^2=-3.9$,
  $\Delta=2.31$  (left), $m^2=-3.2$ $\Delta=2.89$ (right).  
}  \label{ge2}
\end{figure}

\begin{figure}[ht]
\begin{center}
$\begin{array}{c@{\hspace{.2in}}c} \epsfxsize=6cm
\epsffile{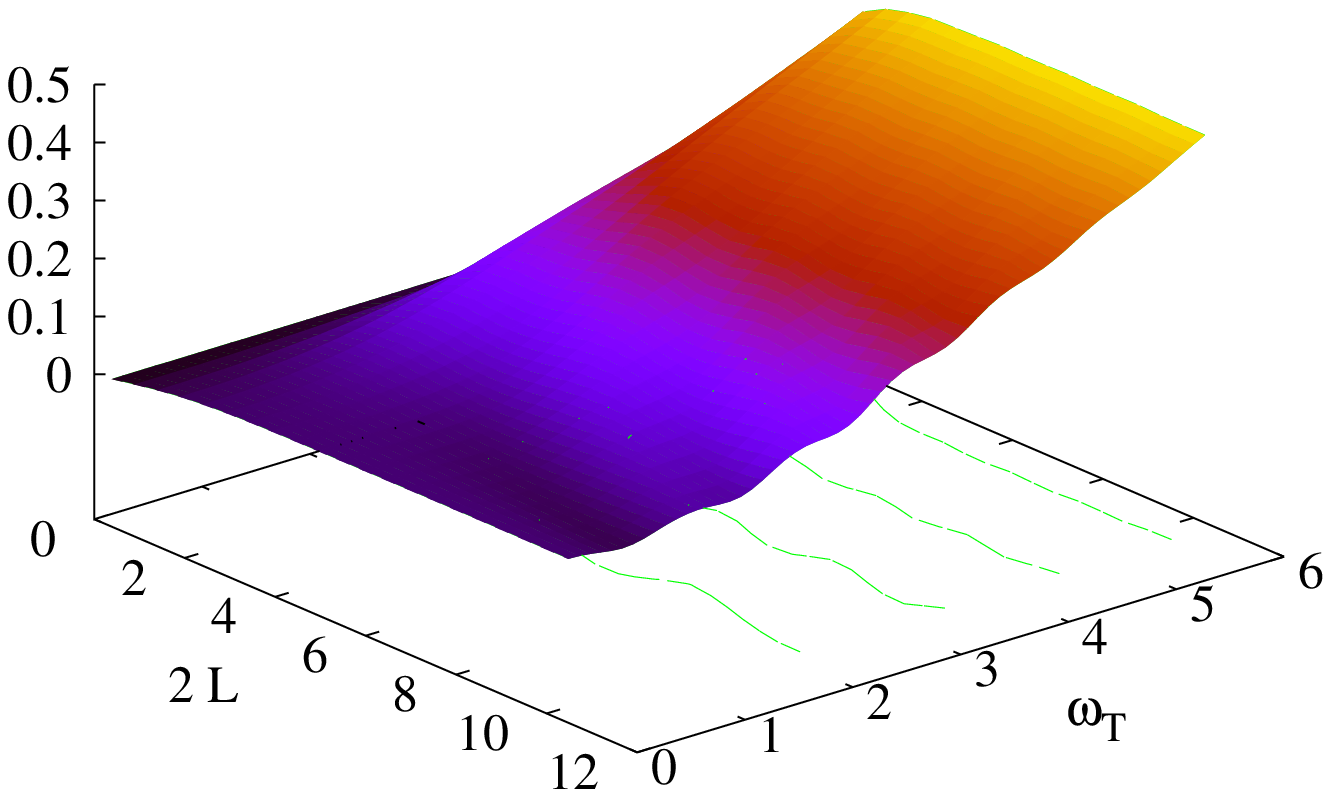} &
     \epsfxsize=6cm
    \epsffile{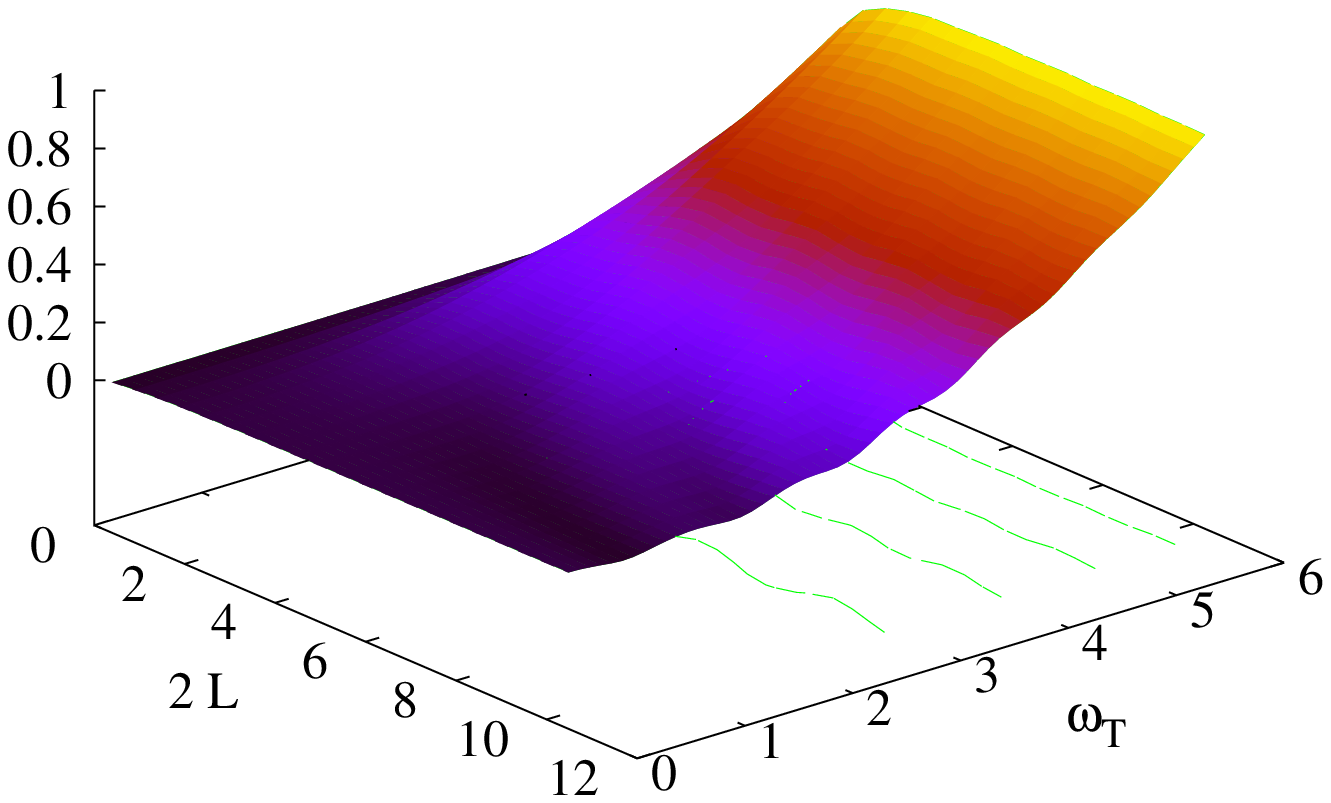}
\end{array}$
\end{center}
\caption{\footnotesize
$| \mathcal{L}_P|$ as a function of $L,\omega_T$ for $m^2=-2.0$,
  $\Delta=3.41$  (left), $m^2=-0.8$  $\Delta=3.79$ (right).  
}  \label{ge3}
\end{figure}

The periodic part of the geodesic length $\mathcal{L}_P$ can be interpreted as a periodic oscillation around the equilibrium value;
in the limit $L\gg 1$, $\mathcal{L}_P$ is negligible compared to the equilibrium value:
 $\mathcal{L}_P/\mathcal{L}_{R,eq} \rightarrow 0$.
 This can be checked by inserting the approximations
(\ref{appr1}, \ref{appr2}) into eq.~(\ref{phish}). This gives 
\beq
 \mathcal{L}_P \approx  2  
 \left( \Sigma_{2,{\rm p}} -\frac{A_{2,{\rm p} }}{4}  \right)   \int_0^L e^{-i(X-L)\sqrt{2} \omega_T } \, dX =
-\sqrt{2} i \frac{e^{i \sqrt{2} L \omega_T }-1}{\omega_T}  \left( \Sigma_{2,{\rm p}} -\frac{A_{2,{\rm p}}}{4}  \right)     \, ,
\label{cincilla}
\eeq
where $A_{2,{\rm p}}$, $\Sigma_{2,{\rm p}}$ are evaluated at $\rho=1$.
This contribution does not scale linearly with $L$ for $L\gg 1$; for
this reason the present contribution is of the same order as the boundary
effects. This quantity  $\mathcal{L}_P$ was computed numerically
as a function of $\omega_T$ and $L$, see figs.~\ref{ge1}, \ref{ge2},
\ref{ge3}. It approaches to a constant at
large $L$, with some wiggles obeying $\delta \omega_T  \times  \delta
L\approx 1$. The analytic expression eq.~(\ref{cincilla}) 
  gives a qualitative explanation for the wiggles in the
figs.~\ref{ge1}, \ref{ge2}, \ref{ge3}.

In this section we studied the time evolution of the equal-time two-point
function of a probe operator with large dimension, see eq.~(\ref{tucano}).
The delay in thermalization 
is proportional to the size $L$ of the region which is probed, see eq.~(\ref{pappagallo});
this is true in the limit $L \gg 1$ in units of $T^{-1}$. 
This result does not depend on the dimension of the driving operator $\Delta$
and on the frequency $\omega_T$. The two-point function has also an oscillatory term with frequency $\omega_T$;
in the regime $L\gg 1$ this term is negligible compared to
the equilibrium value. The dependence of the amplitude of this oscillatory term
on $\omega_T$, $\Delta$ and $L$ is shown in figs.~\ref{ge1}, \ref{ge2}, \ref{ge3}.

\subsection{Entanglement entropy}

After discussing geodesics, we skip the two-dimensional probes, 
i.e.~Wilson loops, and we study the time evolution of the entanglement 
entropy of a spherical region on the boundary.
The results will turn out to be similar to those found using geodesics
as a probe.
A proposal to compute the entanglement entropy in strongly coupled
theories with an AdS dual was introduced in \cite{Ryu:2006bv} and
generalized to the time-dependent case in \cite{Hubeny:2007xt}. For
the derivation of the time-independent case, see
\cite{Casini:2011kv,Lewkowycz:2013nqa}. 
For a review of entanglement entropy in AdS/CFT see
\cite{Nishioka:2009un}. 

In the gravity dual, the entanglement entropy is computed as follows. 
Consider first the boundary $\partial \mathcal{B}$ of the region of
space $\mathcal{B}$ whose entanglement entropy we wish to compute and
then construct the extremal surface (for AdS$_5$ it is a 3-dimensional
surface) in AdS space which ends on $\partial\mathcal{B}$; the
entanglement entropy of $\mathcal{B}$ is then given 
by the volume of this surface. In this section we will choose a region
$\mathcal{B}$ which is a sphere with radius $L$. 
The evolution of the entanglement entropy in the case of a Vaidya metric 
was studied in \cite{Hubeny:2007xt,AbajoArrastia:2010yt,Albash:2010mv,Balasubramanian:2010ce, Balasubramanian:2011ur}.

\begin{figure}[ht]
\begin{center}
$\begin{array}{c@{\hspace{.2in}}c} \epsfxsize=6cm
\epsffile{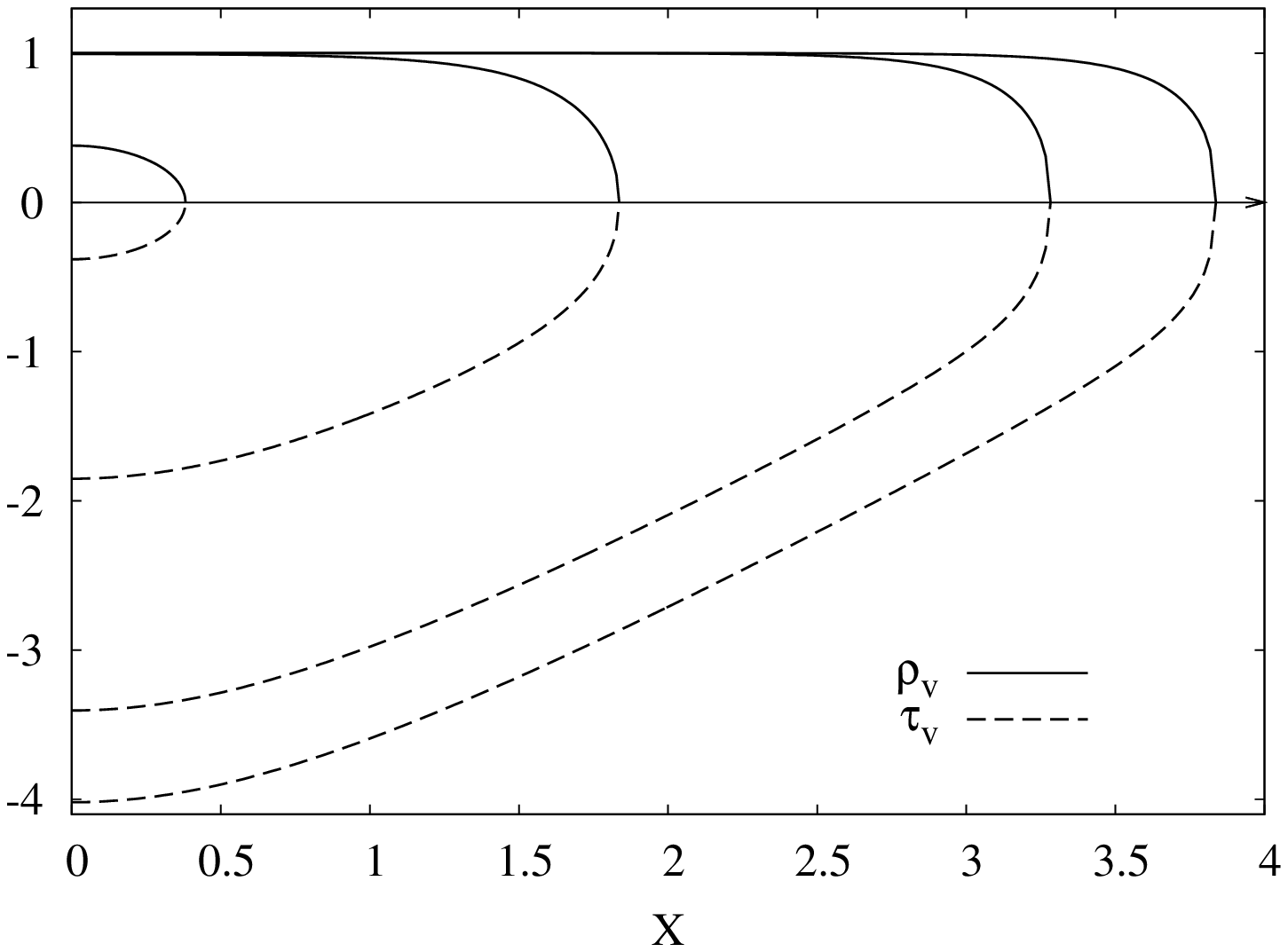} &
     \epsfxsize=6cm
    \epsffile{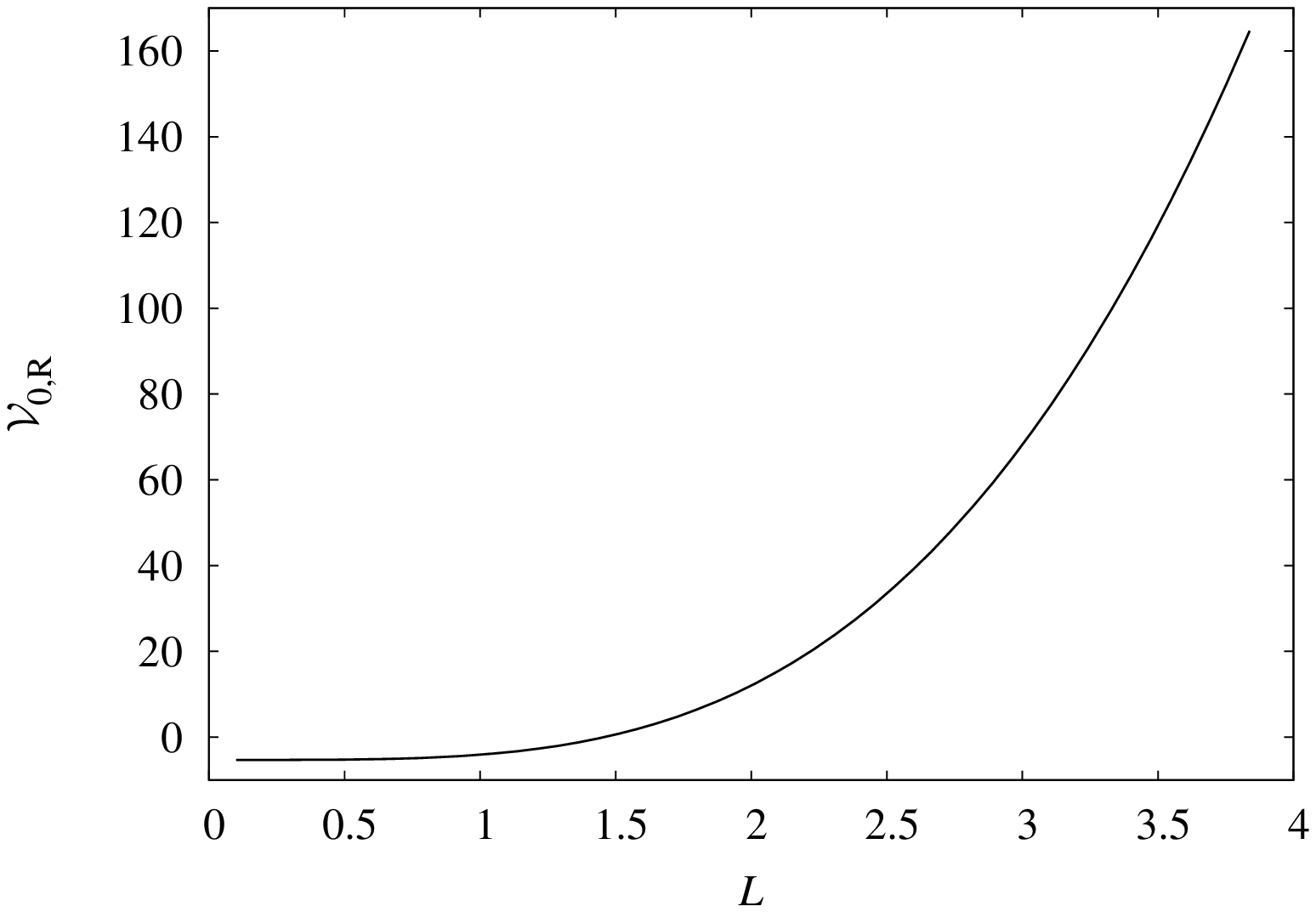}
\end{array}$
\end{center}
\caption{\footnotesize
Left: Examples of minimal surfaces in the unperturbed metric, with
$\rho_v(X)$ in solid lines and $\tau_v(X)$ in dashed lines. 
Right: $\mathcal{V}_{0,R}$ as a function of $L$ for an AdS$_5$ black 
hole. 
}  \label{eentropy-unperturbed}
\end{figure}

Let us first consider the extremal 3-surface parametrized in terms of
$\rho_v(X)$, $\tau_v(X)$ and ${\vec{X}}=X ( \sin \theta \cos \varphi,
\sin \theta \sin \varphi, \cos \theta)$.  
The minimal volume for $\lambda^2=0$ is:
\beq  
\mathcal{V}_0=4 \pi \int \frac{X^2 \sqrt{\Xi_v}}{\rho_v^3} \, dX \,  ,
\qquad
\Xi_v \equiv (\rho_v^4-1)(\tau_v')^2 -2 \tau_v' \rho_v' +1 \, .
\label{evolume}
\eeq
Translation invariance of $\tau$ gives a conserved quantity:
\beq
\frac{X^2 (\tau_v' (\rho_v^4-1)-\rho_v')}{\rho_v^3  \sqrt{\Xi_v}} =K \, .
\eeq
For the class of minimal surfaces that we are considering, $K=0$; this
gives $\tau_v'=\rho_v'/(\rho_v^4-1)$ as for the geodesics considered
in the previous section. We choose the integration constant for  
$\tau_v$ in a such a way that $\tau_v(L)=0$.
Examples of minimal surfaces are shown in
fig.~\ref{eentropy-unperturbed}. 
Near the boundary one can use an approximate solution:
\beq
\rho_v \approx \sqrt{2 L (L-X)} \, , \qquad \rho_v'\approx-\tau_v' \approx -\sqrt{ \frac{L}{2(L-X)}  } \, .
\label{pezzata}
\eeq
By insertion of this expansion, one can check that $\mathcal{V}_0$ has
divergences, which we can regularize by subtracting the divergent
part: 
\beq
 \mathcal{V}_{0,R} = \mathcal{V}_0 -2 \pi \left( \frac{L^2}{\rho_{\rm cut}^2} +\log \frac{\rho_{\rm cut}}{\sqrt{2} L} \right) \, ,
\eeq
where $\rho_{\rm cut}$ is some UV cutoff. This quantity is shown in
fig.~\ref{eentropy-unperturbed}. 

In the limit $L\to \infty$ (where lengths are measured in units of
inverse temperature), $ \mathcal{V}_{0,R}$ should reduce to the
thermal entropy of the region, which is an extensive quantity. 
Let us check this limit, which is dominated by a minimal surface
sitting at $\rho_v \approx 1$ for most of its extension. 
We denote by $\rho_v^*$ the maximum of $\rho_v$; in the limit that we
are considering $1-\rho_v^*=\epsilon$ is a very small positive value. 
Let us introduce $\tilde{\rho}_v=\rho_v-1$, then the leading order
equation for small $\tilde{\rho}$ is: 
\beq
24 X \tilde{\rho}_v^2 - 4 \tilde{\rho}_v \tilde{\rho}_v' + X (\tilde{\rho}_v')^2 -2 X \tilde{\rho}_v \tilde{\rho}_v'' = 0 \, .
\eeq
This differential equation can be solved analytically; the solution gives:
 \beq
 \tilde{\rho}_v=- \frac{\epsilon}{6} \frac{\sinh^2 (\sqrt{6} X)}{X^2} \, , \qquad
 \rho_v \approx 1- \frac{\epsilon}{6} \frac{\sinh^2 (\sqrt{6} X)}{X^2} \, ,
 \label{appro}
 \eeq
where the value of $\epsilon$ can be extracted from the requirement
that $\rho_0$ (computed using the approximation (\ref{appro})) vanishes:
 \beq
 \epsilon = \frac{6 L^2}{\sinh^2 (\sqrt{6} L)} \, .
 \eeq
One can find $\tau_0(X)$, by solving the equation
$\tau_v'=\rho_v'/(\rho_v^4-1)$; it turns out that in the limit $L\gg
1$ in units of inverse temperature, it is a good approximation to use: 
 \beq
 \rho_v=1 \, , \qquad \tau_v=\sqrt{\frac{3}{2}} (X-L)  \, .
 \label{appr-ee}
 \eeq
By insertion into eq.~(\ref{evolume}), one can indeed check that the
extensive part of $ \mathcal{V}_{0,R}$ reduces to the thermal
entropy in the $L\to\infty$ limit.

As before, it is not necessary to compute the change in the minimal
volume as a function of $\lambda^2$; it suffices to evaluate the
$\mathcal{O}(\lambda^2)$ action on the unperturbed minimal surface: 
\beq
\mathcal{V}=\mathcal{V}_0+   \lambda^2   4 \pi \int \frac{X^2 \sqrt{\Xi_v}}{\rho_v^3}
 \left( \frac{-\rho_v^2 \tilde{A}_2 \tau_v'^2 +2 \rho_v \tilde{\Sigma}_2}{2 \Xi_v}
+2 \rho_v \tilde{\Sigma}_2  \right)  \, dX  +\mathcal{O}(\lambda^4)  \, ,
\label{eeeq}
\eeq
where $\tilde{A}_2,\tilde{\Sigma}_2$ are taken as functions of
$\rho_0$ and $\tau_0+\tau^*$, where $\tau^*$ is the time on the
boundary. 

By insertion of eq.~(\ref{pezzata}), one can check that there is an
extra divergence for $\Delta>3$, proportional to $\rho_{\rm
  cut}^{2(3-\Delta)}$, (which is a $\log \rho_{\rm cut}$ divergence
for $\Delta=3$): 
 \beq
  \mathcal{V}_{{\rm ct},1}= \lambda^2 \pi L^2   \frac{  \alpha_{\Delta_-}  -4  \sigma_{\Delta_-} }{\Delta-3}
      \rho_{\rm cut}^{2(3-\Delta)}=
 - \lambda^2 \pi   L^2  \frac{a_{\Delta_-}^2 (4-\Delta)}{6 (2 \Delta-7)(\Delta-3)}     \rho_{\rm cut}^{2(3-\Delta)}  \, ,
 \label{cct1}
 \eeq
where we have used $\sigma_{\Delta_-}=\frac{a_{\Delta_-}^2
  (4-\Delta)}{12 (2 \Delta-7)}$ and $\alpha_{\Delta_-}=2
\sigma_{\Delta_-}$.  
The coefficients $(a_\Delta,\alpha_\Delta,\sigma_\Delta)$ are defined in eqs.~(\ref{scalare-espanso},\ref{sese1},\ref{sese2}).
For $\Delta=7/2$, the function $\tilde{\Sigma}_2$ has also a term
proportional to $\log \rho$, so an extra log divergence is expected;
for $\Delta>7/2$ there is another power-like divergence in addition to
the one canceled by $ \mathcal{V}_{\rm ct,1}$; the following
extra counter-term is thus needed: 
 \beq
   \mathcal{V}_{{\rm ct},2}=\lambda^2 \pi L^2  \frac{2   \alpha_{\Delta_-+1}  -8  \sigma_{\Delta_-+1} +
    {\rm Re} \left[ ( \alpha_{\Delta_-,p}  -8  \sigma_{\Delta_-,p} ) (2 i \omega_T ) e^{-2 i \omega \tau^*}   \right] }{2  \Delta- 7}
    \rho_{\rm cut}^{7-2 \Delta}
     \label{cct2}
  \eeq
  \[  
=\lambda^2 \frac{\pi L^2  \rho_{\rm cut}^{7-2 \Delta}}{3 (2 \Delta-9)(2 \Delta-7)}  {\rm Re} \left[ i \omega_T e^{-2 i \omega_T \tau^*}  \right] \, .
 \]
where we have used $\alpha_{\Delta_-+1} =\frac{\Delta-3}{3 (7-2 \Delta)}
a_{\Delta_-} \dot{a}_{\Delta_-} $, and $\sigma_{\Delta_-+1}
=\frac{5-\Delta}{6 (2 \Delta-9)} a_{\Delta_-} \dot{a}_{\Delta_-} $.
We have to subtract these time-dependent divergent pieces in order to
obtain a finite result. 
In the static case, this kind of divergent contributions to
the entanglement entropy were studied in \cite{Hung:2011ta}; 
here we find new divergences which are flagged by their dependence on $\omega_T$.
These divergent contributions do not depend on the state of theory, but only on
the deformation in the Lagrangian.

In a similar way to $\mathcal{L}_R (\tau^*)$ in eq.~(\ref{ella}),
for boundary times $\tau^*$ such that
\beq
\tau^* - \tau_0^*>-\tau_v(0) \approx \sqrt{\frac{3}{2}} L \, ,
\eeq
where $-\tau_v(0)$ is the maximal time in the past which is reached by the surface,
we can decompose the quantity $ \mathcal{V}_R(\tau^*)$ such that
it has a constant part, a linear part in $\tau^*$ and a periodic part: 
\beq
\mathcal{V}_R= \mathcal{V}_{0,R} + \lambda^2 \left(  \mathcal{V}_C+ (\tau^*-\tau_0^*) \mathcal{V}_L 
+ {\rm Re} ( \mathcal{V}_P e^{-2 i \omega_T (\tau^*-\tau_0^*)})\right) \, ,
\eeq
where
\beq
 \mathcal{V}_C=4 \pi \int
 \frac{X^2 \sqrt{\Xi}}{\rho_v^3}
 \left( \frac{-\rho_v^2 (A_{2,{\rm c}}+\tau_v A_{2,{\rm l}}) \tau_v'^2 +2 \rho_v \Sigma_{2,{\rm c}}}{2 \Xi_v}
+2 \rho_v \Sigma_{2,{\rm c}}  \right)  \, dX 
+  \mathcal{V}_{{\rm ct}}
\, ,
 \eeq
\beq
 \mathcal{V}_L=-4 \pi \int
 \frac{X^2 }{\rho_v}
 \frac{ A_{2,{\rm l}} \tau_v'^2}{2\sqrt{\Xi_v}}  \, dX \, ,
\eeq
\beq
 \mathcal{V}_P= 4 \pi \int
 \frac{X^2 \sqrt{\Xi}}{\rho_v^3}
 \left( \frac{-\rho_v^2 A_{2,{\rm p}} \tau_v'^2 +2 \rho_v \Sigma_{2,{\rm p}}}{2 \Xi_v}
+2 \rho_v \Sigma_{2,{\rm p}}  \right) e^{-2 i \omega_T \tau_v} \, dX + \mathcal{V}_{\rm ct}
\, . \label{giraffa}
 \eeq
 In these expressions
 $A_{2,{\rm c}},A_{2,{\rm l}},A_{2,{\rm p}}, \Sigma_{2,{\rm c}}, \Sigma_{2,{\rm p}}$ 
 are functions of $\rho_v$ and the
counterterms in eqs.~(\ref{cct1},\ref{cct2})  are included.

In analogy to the case of the two-point function, it is convenient to split:
\beq
 \mathcal{V}_R=  \mathcal{V}_{R, eq} + \delta \mathcal{V}_{R} \, , \qquad
 \mathcal{V}_{R, eq}= \mathcal{V}_{0,R} + \lambda^2  (\tau^*-\tau_0^*)    \mathcal{V}_L  \, ,
\eeq
\[
\delta \mathcal{V}_{R}=
 \lambda^2 \left( \mathcal{V}_C + {\rm Re} ( \mathcal{V}_P e^{-2 i \omega_T (\tau^*-\tau_0^*)})\right) \, .
\]
The quantity $\mathcal{V}_{R, eq}$ corresponds to the
equilibrium value of the entanglement entropy
in the undeformed conformal field theory at equilibrium, 
with an energy density given by eq.~(\ref{enel}).

The deviations from the equilibrium are parametrized by $\delta \mathcal{V}_{R}$.
The ratio $\tau_{d,e}=\mathcal{V}_C/\mathcal{V}_L$ can be interpreted
as a time delay in the thermalization of the entanglement entropy.
For large $L$, using the approximations (\ref{appr-ee}): 
\beq
 \mathcal{V}_L=\frac{\pi \omega_T |\Delta-2| {\rm Im} \chi_\Delta}{3} L^3 + \mathcal{O}(L^2) \, ,
\qquad
 \mathcal{V}_C=-\frac{\pi \omega_T |\Delta-2| {\rm Im} \chi_\Delta}{12} \sqrt{\frac{3}{2}} L^4+ \mathcal{O} (L^3)\, . 
\eeq
The delay $\tau_{d,e}$  is linear in $L$, in the regime $L\gg 1$: 
\beq
\tau_{d,e}\approx -\sqrt{\frac{3}{2}} \frac{L}{4} \, .
\label{pappagallino}
\eeq

\begin{figure}[ht]
\begin{center}
$\begin{array}{c@{\hspace{.2in}}c} \epsfxsize=6cm
\epsffile{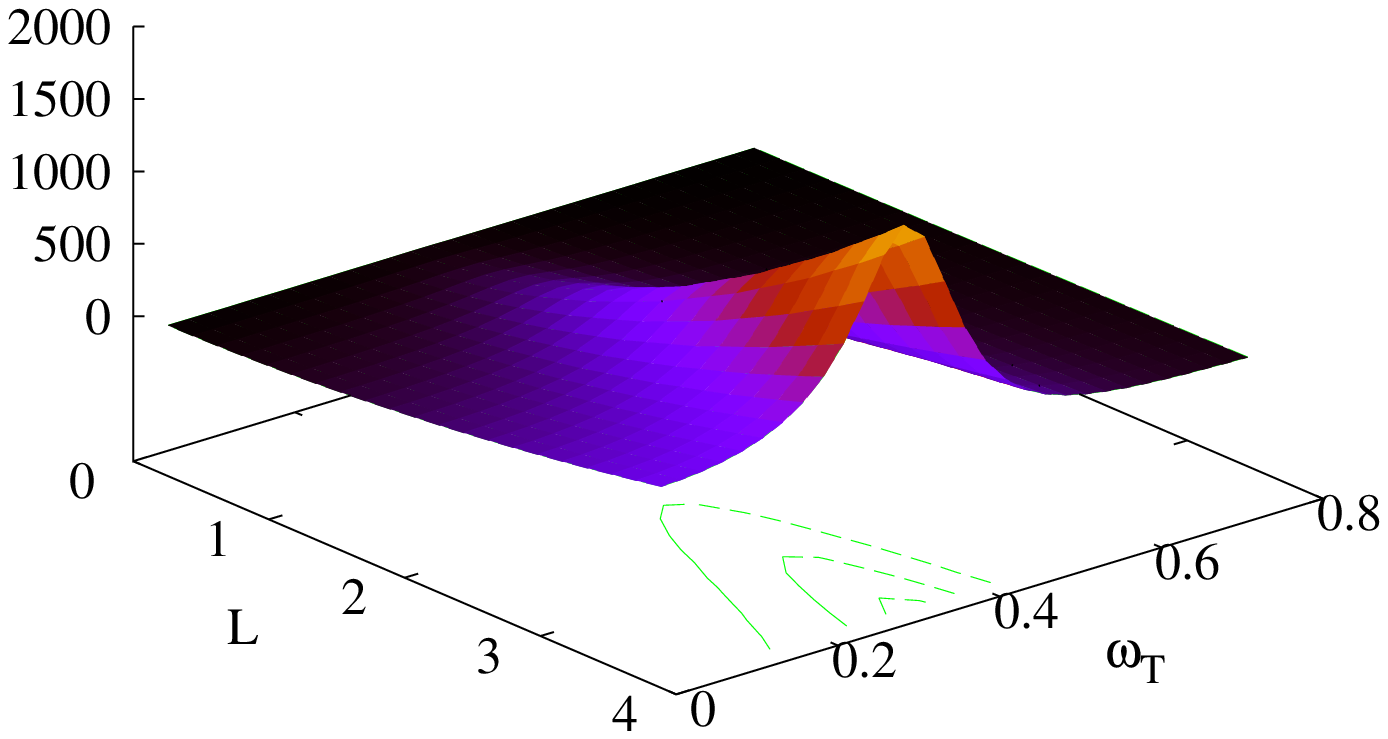} &
     \epsfxsize=6cm
    \epsffile{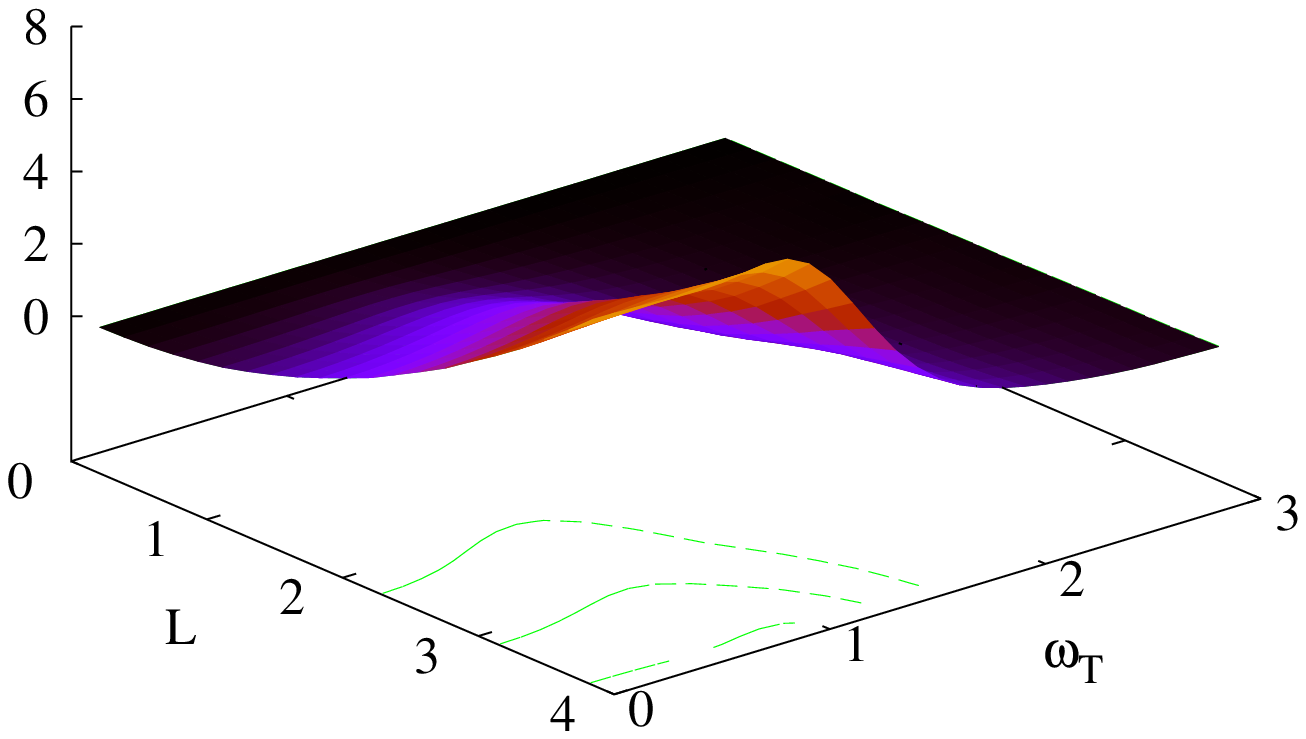}
\end{array}$
\end{center}
\caption{\footnotesize
$|\mathcal{V}_P|$ as a function of $L,\omega_T$ for $m^2=-3.2$,
  $\Delta=1.10$, (left)   $m^2=-3.9$,  $\Delta=1.68$ (right).  
}  \label{ee1}
\end{figure}

\begin{figure}[ht]
\begin{center}
$\begin{array}{c@{\hspace{.2in}}c} \epsfxsize=6cm
\epsffile{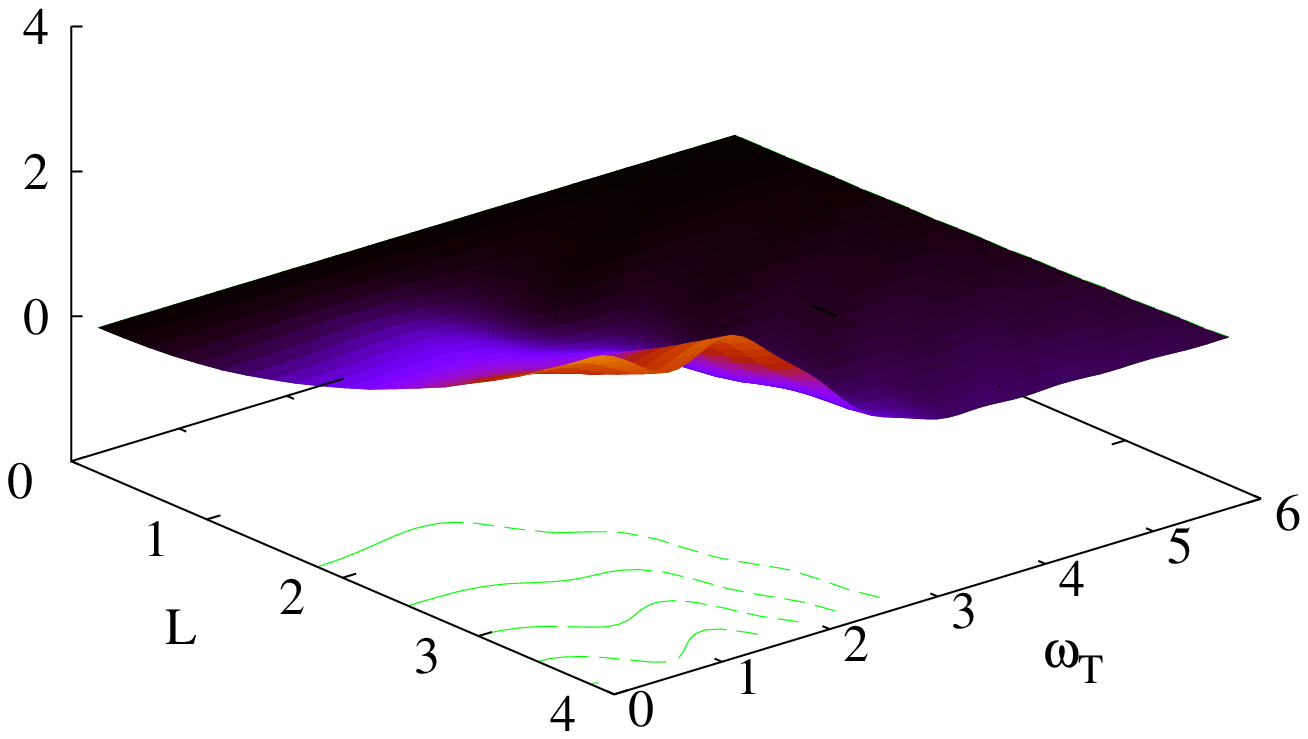} &
     \epsfxsize=6cm
    \epsffile{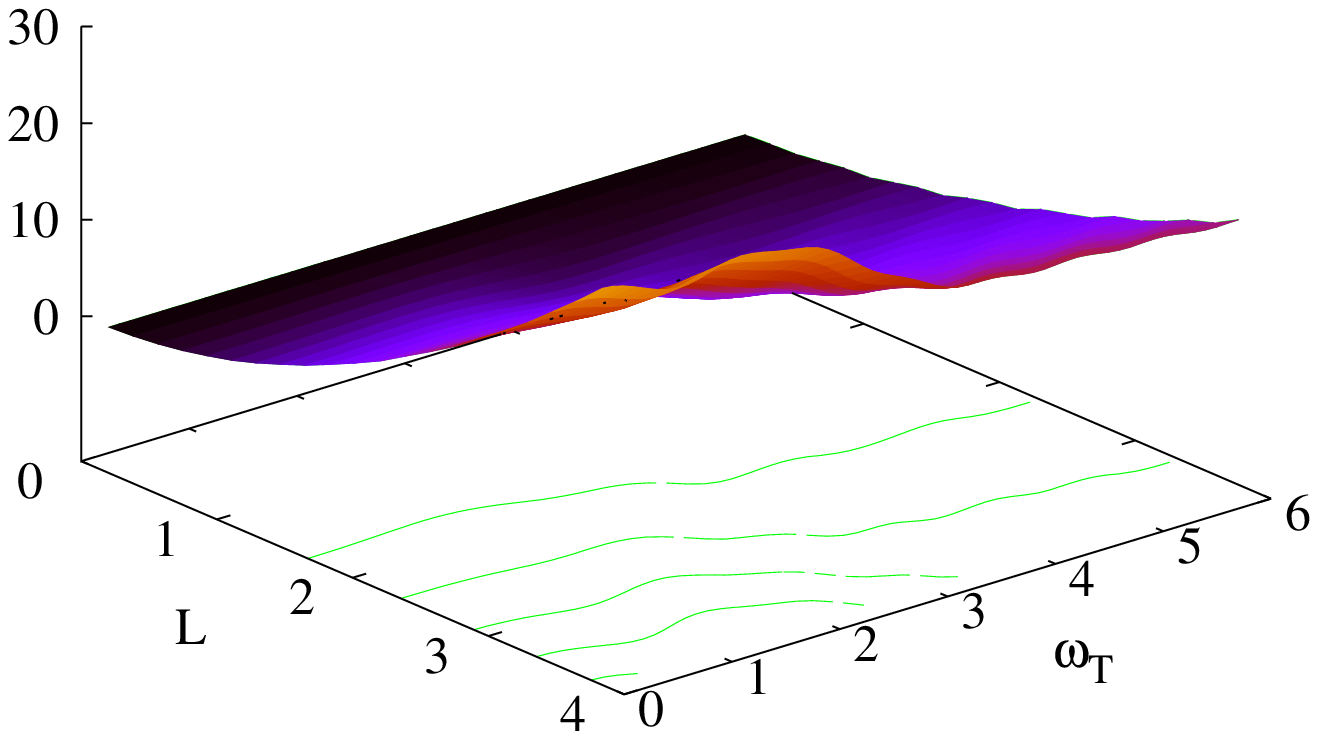}
\end{array}$
\end{center}
\caption{\footnotesize
$| \mathcal{V}_P|$ as a function of $L,\omega_T$ for $m^2=-3.9$,
  $\Delta=2.31$  (left), $m^2=-3.2$ $\Delta=2.89$ (right).  
}  \label{ee2}
\end{figure}

\begin{figure}[ht]
\begin{center}
$\begin{array}{c@{\hspace{.2in}}c} \epsfxsize=6cm
\epsffile{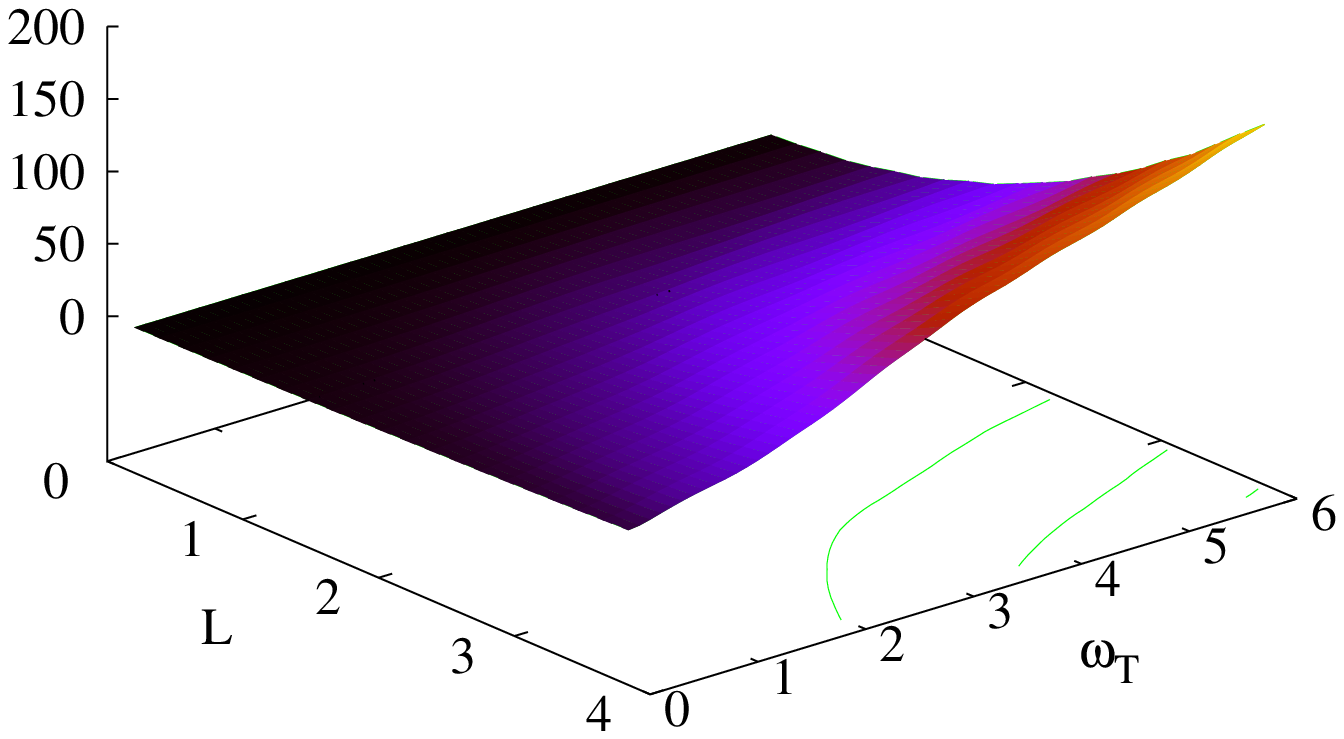} &
     \epsfxsize=6cm
    \epsffile{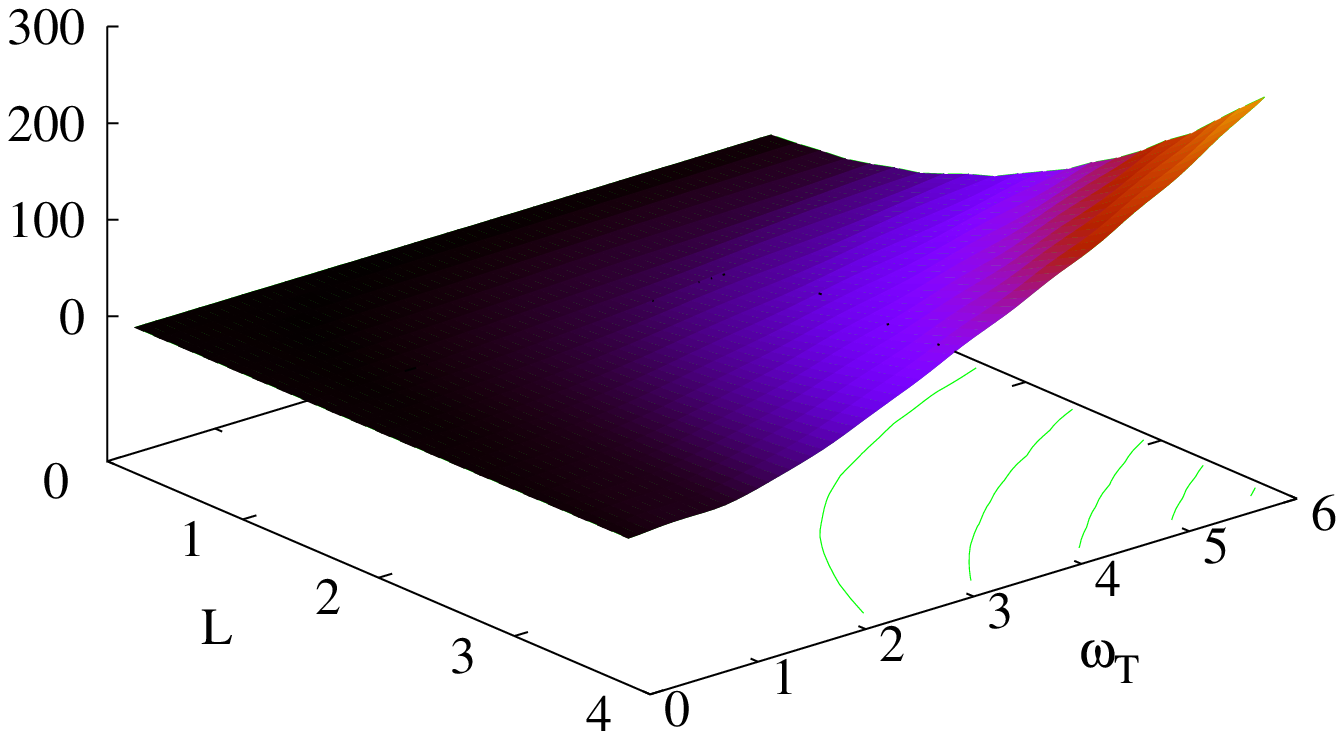}
\end{array}$
\end{center}
\caption{\footnotesize
$| \mathcal{V}_P|$ as a function of $L,\omega_T$ for $m^2=-2.0$,
  $\Delta=3.41$  (left), $m^2=-0.8$ $\Delta=3.79$ (right).  
}  \label{ee3}
\end{figure}

As in the case of the two-point functions,
the periodic part  $\mathcal{V}_P$ is negligible compared to the equilibrium value in the limit $L\gg 1$.
This can be checked substituting the approximations for $\rho\approx 1$ of
eq.~(\ref{appr-ee}) into eq.~(\ref{giraffa}): 
 \beq
 \mathcal{V}_P \approx
12 \pi   \left( \Sigma_{2,{\rm p}} -\frac{A_{2,{\rm p}}}{4}\right)   \int_0 ^L  e^{-2 i \omega_T ( \sqrt{\frac{3}{2}}(X-L)  )}  X^2  \, d X = 
\eeq
\[
= \frac{2 \pi}{3} \left( \Sigma_{2,{\rm p}} -\frac{A_{2,{\rm p}}}{4}\right) \frac{i \sqrt{6} e^{i \sqrt{6} L \omega_T} +3 i \sqrt{6} L^2 \omega_T^2 + 6 L \omega_T -i \sqrt{6}}{ \omega_T^3} \, ,
 \]
where $A_{2,{\rm p}}, \tilde{\Sigma}_{2,{\rm p}}$ are evaluated at $\rho=1$.
This expression scales as $L^2$  for $L\gg 1$, and so it is of the same order as the boundary effects.  
The quantity $\mathcal{V}_P$ was computed numerically,
 see figs.~\ref{ee1}, \ref{ee2}, \ref{ee3}. 

In this section the time evolution of the entanglement entropy was studied.
As for the two-point functions, for $L \gg 1$,
the delay in thermalization is proportional to the size $L$ of the region which is probed,
see eq.~(\ref{pappagallino}). This result does not depend on $\Delta$ and $\omega_T$.
The entanglement entropy has also an oscillatory term, which is negligible
compared to the equilibrium value  in the regime
$L \gg 1$. The dependence of this quantity
as a function of  $\omega_T$, $\Delta$ and $L$ is shown in
figs.~\ref{ee1}, \ref{ee2}, \ref{ee3}. 

\section{Energy fluctuations \label{sec:fluctuations}}

The external periodic perturbation can introduce new
 dynamical features in  the behavior of the system; in particular
 the energy fluctuations may exhibit a phase transition in their behavior as a function of dimension
 of the driving operator $\Delta$. In  \cite{nature} the case of a thermally isolated driven system was analyzed.
 It turns out that the resulting energy fluctuations have a universal distribution
 different from the one given by the Gibbs measure
 \footnote{Energy fluctuations for a system in a heat bath are
determined by the Gibbs measure: $\sigma_{E,eq}^2 = T^2 C_v$, where
$T$ is the temperature and $C_v$ the specific heat. This relation is
independent of the details of the interactions between the system and
the thermal environment, and is valid both for classical and quantum
systems. Let us denote by $\mathcal{E}$ the energy per unit volume of the
system, and by $E=\mathcal{E} V$, where $V$ is the volume, the total energy. 
For example, in the case  of the strongly coupled $\mathcal{N}=4$ SYM in four
dimensions, in thermal equilibrium: 
\[
E=\frac{3 \pi^2 N_c^2}{8} T^4 V \, , \qquad C_V=\frac{3}{2} \pi^2 N_c^2 T^3 V \, ,
\qquad
\frac{\sigma_{E,{\rm eq}}^2}{E^2} = \frac{32}{3 \pi^2}  \frac{1}{N_c^2 V T^3} \, .
\]}.
 Two qualitatively different regimes are predicted, depending on the
details of the protocol by which the time-dependent Hamiltonian is employed.

This behavior depends on two universal parameters: 
\begin{enumerate}
\item  the power $\gamma$ 
which appears in the entropy $S=E^\gamma$, several systems have this behavior over a significant range of energies.
\item the power $s$ which appears in the energy dependence of the work made per cycle
by the driving force, $W=E^s$.
\end{enumerate}
As shown in section \ref{sec:periodic_cft},  the protocol resulting from a periodic perturbation
of a CFT in flat space is such that the work done in each cycle $W_c$ is proportional to $E^s$
in the $\omega \ll T$ limit, where:
\beq
s=\frac{2 \Delta -d-1}{d} \, .
\label{essatto}
\eeq
The two regimes then depend on $\eta=2(s-1) +\gamma$:
for $\eta<0$ the   variance  of the energy $\sigma_E$ is such that
$\sigma_E^2/\sigma_{E,eq}^2$ approaches a constant after a large number
of cycles, while, for $\eta>0$,  $\sigma_E^2/\sigma_{E,eq}^2 \propto E^\eta$. 
The transition between these two regimes occurs when $\eta=0$; for
this value $\sigma_E^2/\sigma_{E,eq}^2 \propto \log E$. 
Close to this transition, a divergent time scale is required in order
to reach the asymptotic regime. This is somehow similar to a second-order phase transition, with the
diverging time scale being analogous to the divergent correlation
length. It would be interesting to study this transition 
in a strongly-coupled CFT and to explore how 
the deviation from the Gibbs measure manifests itself in the bulk gravity dual.

In the case of a conformal field theory in flat $d$-dimensional space,
$\gamma=(d-1)/d$; our choice of protocol is determined by the
dimension of the operator $\Delta$ and by the frequency $\omega$. 
In the limit $\omega \ll T$, $s$ is given by the following expression: 
\beq
 \eta=\frac{4 \Delta- 3 d -3}{d} \, .
\eeq
The parameter $\eta$ becomes positive for $\Delta>3(d+1)/4$;
i.e.~$\Delta>15/4=3.75$ in four dimensions (corresponding to a scalar
with mass $m^2=-15/16$ in AdS$_5$)  
and $\Delta>3$ in three dimensions (corresponding to a massless scalar
in AdS$_4$).

$\mathcal{N}=4$ SYM on a sphere is a typical example of a system in which $S(E)\propto E^\gamma$
with different exponents $\gamma$ for different energy ranges.
In the micro canonical ensemble there are four different regimes \cite{Banks:1998dd}:
free gravitons ($\gamma=9/10$), free strings ($\gamma=1$), small black holes (BH)
($\gamma=8/7$) and  large BH ($\gamma=3/4$).
In each of these phases, depending on the detailed features of the
protocol, it should be possible to achieve both $\eta>0$ and $\eta<0$,
and this should correspond to a different kind of quantum
gravitational behavior in the bulk.
The result eq.~(\ref{essatto}) has been derived in the Poincar\'e patch, so
it is actually applicable only to the case of a large BH.
To get a sense of what is possible, we can apply the result of eq.~(\ref{essatto})
which is valid for $\gamma=3/4$, also to the other three regimes.
This yields the critical $\Delta$ for which the transition occurs:
\beq
\Delta_c=\frac{9}{2}-\gamma \, .
\eeq
 The  spectrum of $\mathcal{N}=4$ SYM 
 (which contains operators of dimension $\Delta=2,3,4$)
  then allows both for the $\eta>0$ and $\eta<0$ regimes.

\section{Conclusions \label{sec:conclusion}}

In this paper we have used the AdS/CFT correspondence to explore how a
strongly-coupled $d=4$ CFT in a 
thermal state reacts to a deformation by a relevant operator with
dimension $\Delta$ which is periodic in time; we studied the problem
to leading order in the deformation parameter $\xi_0$. 
We computed the amount of energy which is dissipated in the system as a function of
$\Delta$ and $\omega$ in the linear response regime.
The leading order backreaction on the metric is studied;
it turns out that, as a function of time, it can be decomposed into a constant, a linear and a periodic
part, see eqs.~(\ref{ansaziano},\ref{ansazianooo}).

The entropy can be monitored in terms of the area of the horizon. The entropy increase in a cycle 
is the same as the one that is found from the equation of state of the undeformed CFT,
given the amount of energy dissipated in a cycle. This is true both for the entropy
defined in terms of the event and of the apparent
horizons, which turn out to be different within short-time behavior.

We have studied the time evolution of the two-point functions of a probe operator with large dimension
and of the entanglement entropy of a spherical region.
Both the quantities increase in time proportionally to the energy dissipated in the
system. There is a delay in achieving the equilibration, which
is linear in the size $L$ of the region which is probed.
In the limit $L >> 1$ in units of the inverse temperature,
the delay in the thermalization of the two-point function and
of the entanglement entropy is respectively: 
\beq
\tau_{d,g}\approx - \frac{L}{2 \sqrt{2}} \, , \qquad
\tau_{d,e}\approx -\sqrt{\frac{3}{2}} \frac{L}{4} \, .
\eeq
This result does not depend on the dimension $\Delta$ of the driving operator and on the frequency $\omega_T$.
Both the quantities have also a term which is periodic in time, 
which is subleading as a function of the size of region $L$ compared to the equilibrium value.

The thermalization time becomes longer the further one ventures into the IR domain.
This feature was observed in other studies of different out of equilibrium systems,
see e.g. \cite{AbajoArrastia:2010yt,Albash:2010mv,Balasubramanian:2010ce,Balasubramanian:2011ur,Aparicio:2011zy,Balasubramanian:2011at,Allais:2011ys,Balasubramanian:2012tu}.
 All this body of work lends extra credence to the thermodynamical features of black holes.

From the results of \cite{nature}, we expect a transition in the time evolution of energy
fluctuation to occur as a function of the protocol parameters $\omega_T$ and $\Delta$.
In the regime of small $\omega_T$, we expect that for $\Delta>15/4$
the variance of the energy has a run-away behavior of the type 
$\sigma_E^2/\sigma_{E,eq}^2 \propto E^\eta$, where $\sigma_{E,eq}$ is the Gibbs-like
variance and $\eta>0$. It would be interesting to understand the bulk dual of such
a run-away behavior;  it could be that this is related to
 Hawking radiation leaking out of the boundary due to coupling to the external source,
 as in \cite{Rocha:2008fe,Rocha:2009xy}. We leave this as a topic for future investigation.

\acknowledgments

The authors thank Guy Bunin, Pasquale Calabrese, Jan de Boer,  Pau Figueras,  Carlos Hoyos,
Zohar Komargodski, Giuseppe Policastro, Erik Tonni, David Vegh and Helvi Witek for useful discussions.
The work of S.~Elitzur is partially supported by 
the Israel Science Foundation Center of Excellence.
The work of  S.~B.~Gudnason and E.~Rabinovici is
partially supported by 
the American-Israeli Bi-National Science Foundation,
by the Israel Science Foundation Center of Excellence and
 by the I Core Program of the Planning and Budgeting Committee and The Israel
Science Foundation "The Quantum Universe".

\appendix
\section{Appendix: some special values of $m^2$  \label{appendix:A}}

In this appendix we will consider separately the value of $m^2$
which gives integer or half-integer dimension $\Delta$.
Generically we can expand eq.~(\ref{wave-eq}) in powers of $k$,
replacing $\phi$ of the form 
\beq
\tilde{\phi}=\sum_k a_k(v) \rho^k + b_k(v) \rho^k \log \rho \, .
\eeq
Here $k$ is not necessarily integer, it is just a discrete label. 
The expansion is (note that it is exact, not just leading order): 
\beq
0=\sum_k  -\rho^{-2+k} \log \rho \left( k(k-4)-m^2\right)   b_k   
-\rho^{-2+k} \left( (k(k-4)-m^2) a_k +(2 k-4) b_k  \right) 
\eeq
\[
+ \rho^{-1+k} \log \rho (2k-3) \dot{b}_k +  \rho^{-1+k} \left( (2 k-3) \dot{a}_k 
+2 \dot{b}_k  \right)
 + \rho^{2+k} \log \rho  (k^2 b_k) +  \rho^{2+k} (k a_k +2 b_k)k \, .
\]
It turns out that the $b_k$ are all zero except for integer
dimension $\Delta$. 
For half-integer $\Delta$s there are still some differences with
respect to the generic case, concerning the metric backreaction. 
We will consider each of the five cases separately next.

 \subsection{$m^2=-4$}

For $m^2=-4$, all the coefficients are determined in terms
of $a_2,b_2$: 
\beq
b_3=\dot{b}_2 \, , \qquad b_4=\frac{3}{4} \dot{b}_3 \, , \qquad b_5= \frac{5}{9} \dot{b}_4 \, ,\qquad
b_k=\frac{(2 k-5) \dot{b}_{k-1} + 2 (k-4) b_{k-4} }{(k-2)^2} \, , \,  k\geq 6 \, ,
\eeq
\[
a_3=\dot{a}_2 +2 \dot{b}_2-2 b_3 \, , \qquad
a_4=\frac{3 \dot{a}_3 +2 \dot{b}_3 - 4 b_4}{4} \, ,  \qquad
a_5=\frac{5 \dot{a}_4 +2 \dot{b}_k -6 b_5}{9} \, ,
\]
\[
a_k=\frac{(2 k-5) \dot{a}_{k-1} + (k-4)^2 a_{k-4}   +(4- 2 k ) b_k + 2 \dot{b}_{k-1} }{(k-2)^2} \, , \,  k\geq 6 \, .
\]
The functions $\tilde{A}_2$, $\tilde{\Sigma}_2$ can be expanded as follows:
\beq
\tilde{A}_2=\sum_{j=0}^{\infty} \alpha_j \rho^{2+j}+\bar{\alpha}_j \rho^{2 +j} \log \rho +\bar{\bar{\alpha}}_j \rho^{2+j} \log^2 \rho \, ,
\eeq
\beq
\tilde{\Sigma}_2=
\sum_{j=0}^{\infty} \sigma_j \rho^{3+j}+\bar{\sigma}_j \rho^{3 +j} \log \rho +\bar{\bar{\sigma}}_j \rho^{3+j} \log^2 \rho \, ,
\eeq
where the first of these coefficients are given by: 
\beq
\sigma_0=\frac{-72 a_2^2 +12 a_2 b_2 -13 b_2^2}{1296} \, ,
\qquad
\bar{\sigma}_0=\frac{b_2 (b_2-12 a_2)}{108} \, , \qquad
\bar{\bar{\sigma}}_0=-\frac{b_2^2}{18} \, ,
\eeq
\[
\bar{\alpha}_0=-\frac{2 a_2 b_2}{9} \, , \qquad
\bar{\bar{\alpha}}_0=-\frac{b_2^2}{9} \, .
\]
The coefficient $\alpha_0$ is determined by the differential equation:
\beq
\dot{\alpha}_0=-\frac{2}{9} a_2 \dot{a}_2 -\frac{10}{81} b_2 \dot{b}_2 +\frac{5}{27} \dot{a}_2 b_2 -\frac{4}{27} a_2 \dot{b}_2 \, .
\eeq
Holographic renormalization gives \cite{Buchel:2012gw}:
\beq
8 \pi G_N \mathcal{E}=
\frac{3}{2} \mu^4 -\lambda^2 \mu^4 \left( \frac{3}{2} \alpha_0
+\frac{5}{54} b_2^2 +\frac{1}{6} a_2^2-\frac{5}{18} a_2 b_2 \right) \, ,
\eeq
\beq
8 \pi G_N \mathcal{P}=\frac{1}{2} \mu^4 -\lambda^2 \mu^4 \left( \frac{1}{2} \alpha_0
+\frac{1}{18} a_2^2 - \frac{17}{324} b_2^2 +\frac{13}{54} a_2 b_2\right) \, ,
\eeq
\beq
8 \pi G_N \langle \mathcal{O}_\Delta \rangle=-\lambda \mu^2 \frac{a_2}{2}\, .
\eeq
There is an ambiguity in the counterterm in the renormalization  
procedure; here we use the one which corresponds to $\delta_1=0$
in the conventions of \cite{Buchel:2012gw}.
The source and the response function are then parametrized as follows:
\beq
b_{2}={\rm Re} \big( e^{- i \omega_T \tau}\big) \, , \qquad  
a_{2}=-{{\rm Re}} \big(\chi_2 (\omega_T) e^{-i \omega_T \tau}\big) \, ,
\eeq
which gives:
\beq
G_R(\omega_T)=\frac{\mu^2 \chi_2(\omega_T)}{16 \pi G_N} \, , \qquad
A_{2,l}=-\frac{ \omega_T \, {\rm Im} \, \chi_2}{6} \,  \rho^2 \, .
\label{green-delta-2}
\eeq

\subsection{$m^2=-15/4$}

For $m^2=-15/4$, both $a_{\Delta_-}=a_{3/2}$ and
$a_{\Delta_+}=a_{5/2}$ are free parameters; the remaining $a_{\Delta_+
  +k}$ are given by 
eq.~(\ref{all-the-as}). No logarithms appear in the expansion; some
aspects of the source and VEV are still peculiar, so we discuss it as
a separate case. 
The functions $\tilde{A}_2$, $\tilde{\Sigma}_2$ can be expanded as
follows: 
\beq
\tilde{A}_2=\alpha_{-1} \rho +\sum_{j=0}^{\infty} \alpha_j \rho^{2+j}\, ,
\qquad
\tilde{\Sigma}_2=
\sum_{j=0}^{\infty} \sigma_j \rho^{2+j}  \, ,
\eeq
The first terms are determined as:
\beq
\sigma_0=-\frac{a_{3/2}^2}{16} \, , \qquad \sigma_1=-\frac{5}{48} a_{3/2} a_{5/2} \, , 
\eeq
\beq
\sigma_2=\frac{-25 a_{5/2}^2 -42 a_{3/2} \dot{a}_{5/2}}{480} \, , \qquad
\alpha_{-1}=-\frac{a_{3/2}^2}{8} \, ,
\eeq
and the differential equation for $\alpha_0$ is:
\beq
\dot{\alpha}_0=\frac{-\dot{a}_{3/2}^2 -4 a_{3/2} \dot{a}_{5/2} +3 a_{3/2} \ddot{a}_{3/2}}{12} \, .
\eeq
If we choose the operator $\Delta_+$ as a source, we get:
\beq
8 \pi G_N \mathcal{E}=
\frac{3}{2} \mu^4 -\lambda^2 \mu^4 \left( \frac{3}{2} \alpha_0
+\frac{a_{3/2} a_{5/2}}{2}  -\frac{3}{8} a_{3/2} \dot{a}_{3/2} \right) \, ,
\eeq
\beq
8 \pi G_N \langle \mathcal{O}_\Delta \rangle= \frac{\lambda \mu^{5/2} (a_{5/2}-\dot{a}_{3/2})}{2} \, .
\eeq
\beq
a_{3/2}={\rm Re} \big( e^{- i \omega_T \tau}\big) \, , \qquad 
 a_{5/2}={\rm Re} \left[ (\chi_{5/2} (\omega_T)-i \omega_T) e^{-i \omega_T \tau} \right] \, .
\eeq
If instead we choose the operator $\Delta_-$ as a source:
\beq
8 \pi G_N \mathcal{E}=
\frac{3}{2} \mu^4 -\lambda^2 \mu^4 \left( \frac{3}{2} \alpha_0
+\frac{a_{3/2} \dot{a}_{3/2}}{8} \right) \, ,
\qquad
8 \pi G_N \langle \mathcal{O}_\Delta \rangle= -\frac{\lambda \mu^{3/2} a_{3/2}}{2} \, .
\eeq
\beq
a_{3/2}={{\rm Re}} (-\chi_{3/2} (\omega_T) e^{- i \omega_T \tau}) \, , \qquad 
 a_{5/2}={{\rm Re}} \left[  (1+i \omega_T \chi_{3/2}) e^{-i \omega_T \tau} \right] \, .
\eeq

\subsection{$m^2=-3$}

For $m^2=-3$, all the coefficients are determined in
terms of $a_1,a_3$: 
\beq
a_2=\dot{a}_1 \, , \qquad b_3=\frac{\dot{a}_2}{2} \, , \qquad b_k=\frac{(2 k -5)  \dot{b}_{k-1} +2 (k-4) b_{k-4} }{(k-1)(k-3)} \, ,
\eeq
\[
a_k=\frac{  (2 k -5)  \dot{a}_{k-1} + (k-4)^2 a_{k-4} +(4-2 k) b_k + 2 \dot{b}_k }{(k-1)(k-3)}  \, \,   {\rm for } \, k \geq 4 \, .
\]
The functions $\tilde{A}_2$, $\tilde{\Sigma}_2$ can be expanded as
follows: 
\beq
\tilde{A}_2=\alpha_{-1}+\sum_{j=0}^{\infty} \alpha_j \rho^{2+j}+\bar{\alpha}_j \rho^{2 +j} \log \rho +\bar{\bar{\alpha}}_j \rho^{2+j} \log^2 \rho \, ,
\eeq
\beq
\tilde{\Sigma}_2=
\sum_{j=0}^{\infty} \sigma_j \rho^{1+j}+\bar{\sigma}_j \rho^{1 +j} \log \rho +\bar{\bar{\sigma}}_j \rho^{1 +j} \log^2 \rho \, ,
\eeq
where the first coefficients are
\beq
\sigma_0=-\frac{a_1^2}{12} \, , \qquad \sigma_1=-\frac{a_1 \dot{a}_1}{9} \, , \qquad \sigma_2=\frac{-24 a_1 a_3 -16 \dot{a}_1^2 +2 a_1 \ddot{a}_1 }{288} \, ,
\eeq
\[
\qquad \bar{\sigma}_2=-\frac{a_1 \ddot{a}_1}{24} \, , \qquad
\alpha_{-1}=-\frac{a_1^2}{6} \, , \qquad \bar{\alpha}_0=\frac{\dot{a}_1^2 - a_1 \ddot{a}_1}{6} \, ,
\]
and $\bar{\bar{\sigma}}_{0,1,2}= \bar{\sigma}_{0,1} =
\bar{\bar{\alpha}}_0=0$. 
The coefficient $\alpha_0$ is determined by the equation:
\beq
\dot{\alpha}_0=\frac{6 a_3 \dot{a}_1 - 6 a_1 \dot{a}_3 -\dot{a}_1 \ddot{a}_1 + 4 a_1 a_1^{(3)} }{18} \, .
\eeq
Holographic renormalization gives \cite{Buchel:2012gw}:
\beq
8 \pi G_N \mathcal{E}=
\frac{3}{2} \mu^4 -\lambda^2 \mu^4 \left( \frac{3}{2} \alpha_0
+\frac{5}{24} \dot{a}_1^2 +\frac{a_1 a_3}{2} -\frac{a_1 \ddot{a}_1}{3}  \right) \, ,
\eeq
\beq
8 \pi G_N \mathcal{P}=\frac{1}{2} \mu^4 -\lambda^2 \mu^4 \left( \frac{1}{2} \alpha_0
+\frac{11}{72} \dot{a}_1^2 - \frac{a_1 a_3}{6} +\frac{a_1 \ddot{a}_1}{18 }\right) \, ,
\qquad
8 \pi G_N \langle \mathcal{O}_\Delta \rangle= \lambda \mu^3 a_3\, .
\eeq
There are also renormalization ambiguities; here we use the
conventions corresponding to $\delta_1=\delta_3=0$ and $\delta_2=-1/4$  
in the notation of \cite{Buchel:2012gw}.
The source and the response functions are:
\beq
a_{1}={\rm Re} \big( e^{- i \omega_T \tau}\big) \, , \qquad  
a_{3}={\rm Re} \big(\chi_3 (\omega_T) e^{-i \omega_T \tau}\big) \, .
\eeq

\subsection{$m^2=-7/4$}

For $m^2=-7/4$, both $a_{\Delta_-}=a_{1/2}$ and $a_{\Delta_+}=a_{7/2}$
are free parameters; $a_{3/2}=\dot{a}_{1/2}$, 
$a_{5/2}=0$ and the remaining $a_{\Delta_+ +k}$ are again given by
eq.~(\ref{all-the-as}); the first of these relations gives
$a_{3/2}=\dot{a}_1/2 $ and $a_{5/2}=0$. 
No logarithms appear in the leading order expansion of
$\tilde{\phi}_1$; however logs do appear in  
$\tilde{A}_2, \tilde{\Sigma}_2$:
\beq
\tilde{A}_2=\sum_{j=-3}^{\infty} \alpha_j \rho^{2+j}+\sum_{j=-3}^{\infty} \bar{\alpha}_j \rho^{2+j} \log \rho\, ,
\qquad
\tilde{\Sigma}_2= \sigma_l \log \rho + \sum_{j=0}^{\infty} \sigma_j \rho^{j+1}  \, .
\eeq
The leading terms are:
\beq
\sigma_l=-\frac{a_{1/2}^2}{24} \, , \qquad \sigma_0= -\frac{a_{1/2} \dot{a}_{1/2}}{8} \, , \qquad
\sigma_1= -\frac{\dot{a}_{1/2}^2}{16} \, , \qquad \sigma_2= -\frac{7 a_{1/2} a_{7/2}}{144}  \, ,
\eeq
\beq
\bar{\alpha}_{-3}=-\frac{a_{1/2}^2}{12} \, , \qquad \bar{\alpha}_{-2}=\frac{a_{1/2} \dot{a}_{1/2}}{6} \, , 
\eeq
\beq
\alpha_{-2}=-\frac{a_{1/2} \dot{a}_{1/2}}{3} \, , \qquad
\alpha_{-1}=\frac{-5 \dot{a}_{1/2}^2 +6 a_{1/2} \ddot{a}_{1/2}}{24} \, ,
\eeq
with $\alpha_{-3}=\bar{\alpha}_{-1}=\bar{\alpha}_0=0$.
The equation for $\alpha_0$ is:
\beq
\dot{\alpha}_0=
\frac{7}{9} \dot{a}_{1/2} a_{7/2} -\frac{2}{9} a_{1/2} \dot{a}_{7/2}
-\frac{1}{12} \ddot{a}_{1/2}^2 +\frac{1}{4} \dot{a}_{1/2} a_{1/2}^{(3)}  \, .
\eeq
The energy density and the VEV are:
\beq
8 \pi G_N \mathcal{E}=
\frac{3}{2} \mu^4 -\lambda^2 \mu^4 \left( \frac{3}{2} \alpha_0
+\frac{a_{1/2} a_{7/2}}{3} +\frac{\ddot{a}_{1/2}  a_{1/2}}{8} \right) \, ,
\eeq
\beq
8 \pi G_N \langle \mathcal{O}_\Delta \rangle= \frac{3 \lambda \mu^{7/2} (\alpha_{7/2} +\alpha_{1/2}^{(3)}/3) }{2} \, .
\eeq
The source and response functions are:
\beq
a_{1/2}={{\rm Re}} \big(e^{- i \omega_T \tau}\big) \, , \qquad 
a_{7/2}={{\rm Re}} \left[ ( \chi_{7/2}(\omega_T) -(-i \omega_T)^3/3 ) e^{- i \omega_T \tau}\right] \, .
\eeq

\subsection{$m^2=0$}

For $m^2=0$, all the coefficients are determined in terms
of $a_0,a_4$: 
\beq
a_1=\dot{a}_0 \, , \qquad a_2=\frac{\dot{a}_1}{4} \, , \qquad a_3= -\frac{\dot{a}_2}{3} \, , \qquad b_4= \frac{3}{4} \dot{a}_3 \, , 
\eeq
\[
b_k=\frac{(2 k-5) \dot{b}_{k-1} +2 (k-4) b_{k-4}}{k (k-4)} \, ,
\]
\[
a_k=\frac{(2 k-5) \dot{a}_{k-1} + (k-4)^2 a_{k-4} +(4-2 k) b_k +2 \dot{b}_{k-1}}{k (k-4)} \, \,   {\rm for } \, k \geq 5 \, .
\]
The functions $\tilde{A}_2$, $\tilde{\Sigma}_2$ can be expanded as
follows: 
\beq
\tilde{A}_2=\alpha_{-2}+\sum_{j=0}^{\infty} \alpha_j \rho^{2+j}+\bar{\alpha}_j \rho^{2 +j} \log \rho +\bar{\bar{\alpha}}_j \rho^{2+j} \log^2 \rho \, ,
\eeq
\beq
\tilde{\Sigma}_2=
\sum_{j=0}^{\infty} \sigma_j \rho^{1+j}+\bar{\sigma}_j \rho^{1 +j} \log \rho +\bar{\bar{\sigma}}_j \rho^{1 +j} \log^2 \rho \, .
\eeq
The first coefficients are:
\beq
\sigma_0=-\frac{\dot{a}_0^2}{12} \, , \qquad \sigma_2=1=-\frac{\dot{a}_0 \ddot{a}_0}{36} \, , \qquad
\sigma_2=\frac{2 \dot{a}_0 a_0^{(3)} -\ddot{a}_0^2}{288} \, , 
\eeq
\[
\sigma_3=\frac{-160 \dot{a}_0 a_4 +5 \ddot{a}_0 a_0^{(3)} -2 \dot{a}_0 a_0^{(4)}}{2400} \, ,
\]
\[
\bar{\sigma}_3=\frac{a_0^{(4)} \dot{a}_0}{240} \, , \qquad
\alpha_{-2}=-\frac{5 \dot{a}_0^2}{12} \, , \qquad \bar{\alpha}_0=\frac{\ddot{a}_0^2-2 \dot{a}_0 a_{0}^{(3)}}{24} \, ,
\]
and $\bar{\bar{\sigma}}_{0,1,2,3}=\bar{\bar{\sigma}}_{0,1,2}=\bar{\bar{\alpha}}_{0}=0$.
The equation which determines $\alpha_0$ is:
\beq
\dot{\alpha}_0=\frac{4}{3} a_4 \dot{a}_0 -\frac{1}{24} \ddot{a}_0 a_0^{(3)} +\frac{17}{144} \dot{a}_0 a_{0}^{(4)} \, .
\eeq
The energy density and the VEV are:
\beq
8 \pi G_N \mathcal{E}=
\frac{3}{2} \mu^4 -\lambda^2 \mu^4 \left( \frac{3}{2} \alpha_0
+\frac{21}{192} \ddot{a}_0^2  +\frac{17}{96}\dot{a}_0 a_0^{(3)}  \right) \, ,
\qquad
8 \pi G_N \langle \mathcal{O}_\Delta \rangle=  2 \lambda \mu^{4}  a_4  \, ,
\eeq
where renormalization ambiguities have been fixed in such 
a way that $\langle \mathcal{O}_\Delta \rangle$ is proportional to
$a_4$. 
The source and response functions are: 
\beq
a_{0}={{\rm Re}} \big(e^{- i \omega_T \tau}\big) \, , \qquad 
a_{4}={{\rm Re}} \left[  \chi_{4}(\omega_T) e^{- i \omega_T \tau} \right] \, .
\eeq

\section{Oscillating potential in a free field theory as an effective
  quench \label{appendix:B}}

In this appendix we consider first the toy example of a quantum
mechanical Hamiltonian corresponding to a free massive field theory in
$0+1$ dimensions. We split the mass-squared term into a constant
value, $\kappa^2$ for time $t<0$ and an oscillating function
$\kappa^2\cos(\omega t)$ for $t>0$. 

The Hamiltonian for time $t<0$ reads
\beq
H_- = 
\frac{\pi^2}{2} + \frac{\kappa^2}{2} \psi^2 \, ,
\eeq
while at time $t\geq 0$, we turn on the above mentioned time-dependent mass
\beq
H_+ = 
\frac{\pi^2}{2} + \frac{\kappa^2}{2} \cos(\omega t) \psi^2 \, ,
\eeq
where $\pi$ is the conjugate momentum of $\psi$. 
We will now scale the time coordinate $t\to \tfrac{2}{\omega}\tau$
and make a corresponding rescaling of the Hamiltonian which yields
\beq
H_- = \frac{\pi^2}{2} - q \psi^2 \, , \qquad
H_+ = \frac{\pi^2}{2} - q \cos(2\tau) \psi^2 \, , \qquad
q \equiv -\frac{2\kappa^2}{\omega^2} \, . 
\label{ququ}
\eeq
Let us first see what we expect to happen if $\omega\gg
\sqrt{2}\kappa$ or equivalently if $|q|\ll 1$. 
The effective Hamiltonian after time $t\geq 0$ can be calculated by
the method of separation of time scales \cite{Rahav:2003b}, \cite{Auzzi:2012ca} and is to order
$1/\omega^2$:
\beq
H_+^{\rm eff} = \frac{\pi^2}{2} + \frac{1}{8}\kappa^2 |q| \psi^2 \, .
\label{eq:Heff}
\eeq
Hence, we expect the system to quench at time $t=0$ with the harmonic
oscillator frequency jumping from $\kappa$ to
$\frac{1}{2}\kappa\sqrt{|q|}$, where $|q|\ll 1$ for the approximation 
to hold. 

Let us check our expectations. The equation of motion for the Hamiltonian $H_+$ reads
\beq
\ddot\psi + \kappa^2 \cos(\omega t) \psi = 0 \, ,
\eeq
which has the solution 
\beq
\psi(t) = \psi(0) C\left(\frac{\omega t}{2}\right)
+ \frac{2\pi(0)}{\omega} S\left(\frac{\omega t}{2}\right) \, .
\label{eq:phisol}
\eeq
The  Mathieu cosine $C$ and sine $S$  functions depend also on the standard parameters $q$ and $a$
 \cite{Abramowitz}. Here $a=0$ and $q$ is given in eq.~(\ref{ququ}).
The stability bands for the Mathieu functions are symmetric in $q\to
-q$. We have here adhered to the convention of taking $C(0)=1$ and
$S'(0)=1$ and we leave $q$ negative.

The large-$\omega$ limit of eq.~\eqref{eq:phisol} gives:
\beq
\lim_{\omega\gg\sqrt{2}\kappa} \psi(t) = 
\psi(0) \cos\left(\frac{\kappa^2}{\sqrt{2}\omega} t\right)
+\frac{\sqrt{2}\omega\pi(0)}{\kappa^2}
\sin\left(\frac{\kappa^2}{\sqrt{2}\omega} t\right)  
+ \mathcal{O}(q) \, , 
\eeq
where we used the fact that (for $a=0$ which yields the Mathieu exponent
$\nu=q/\sqrt{2}$, see \cite{Abramowitz}, \cite{Auzzi:2012ca})
\begin{align}
C(\tau) &= \frac{2-q\cos 2\tau}{2-q} \cos\nu\tau 
+ \mathcal{O}(q^2)  \qquad&
S(\tau) &= \frac{2-q\cos 2\tau}{\nu(2-q)} \sin\nu\tau 
+ \mathcal{O}(q^2) \\
&\simeq \left(1 + q \sin^2 \tau\right) \cos\nu\tau 
+ \mathcal{O}(q^2) \, , \qquad &
&\simeq \frac{1}{\nu}\left(1 + q \sin^2 \tau\right) \sin\nu\tau 
+ \mathcal{O}(q^2) \, .
\end{align}
This shows that the amplitude to leading order in the $1/\omega^2$
expansion is given by waves, completely analogously to the situation
of a quench \cite{Sotiriadis:2010si}. This is also what one would
expect by comparing the Hamiltonian $H_-$ with the effective
Hamiltonian $H_+^{\rm eff}$ to leading order in $1/\omega^2$. 
Subleading terms will however reveal the difference between a quench
and the suddenly turned-on oscillating mass term. 

Let us now calculate the propagator for the system at hand without
making any assumptions about $\omega$. We have
\begin{align}
\langle\Psi_0|\psi(t_1)\psi(t_2)|\Psi_0\rangle
&=\langle\Psi_0|\psi^2(0)|\Psi_0\rangle 
C\left(\frac{\omega t_1}{2}\right)
C\left(\frac{\omega t_2}{2}\right) 
+\langle\Psi_0|\pi^2(0)|\Psi_0\rangle 
\frac{4}{\omega^2}S\left(\frac{\omega t_1}{2}\right)
S\left(\frac{\omega t_2}{2}\right) \nonumber \\
& \phantom{=\ }
+\langle\Psi_0|\{\psi(0),\pi(0)\}|\Psi_0\rangle 
\frac{1}{\omega}\left[
C\left(\frac{\omega t_1}{2}\right)
S\left(\frac{\omega t_2}{2}\right) 
+S\left(\frac{\omega t_1}{2}\right)
C\left(\frac{\omega t_2}{2}\right) 
\right] \non
& \phantom{=\ }
+\frac{i}{\omega}\left[
C\left(\frac{\omega t_1}{2}\right)
S\left(\frac{\omega t_2}{2}\right) 
-S\left(\frac{\omega t_1}{2}\right)
C\left(\frac{\omega t_2}{2}\right) 
\right] \, , 
\end{align}
where we used the bracket $[\psi(0),\pi(0)]=i$.
Since at time $t=0$ we  have the harmonic oscillator, we
find
\beq
\langle\Psi_0|\psi^2(0)|\Psi_0\rangle = \frac{1}{2\kappa} \, , \qquad
\langle\Psi_0|\pi^2(0)|\Psi_0\rangle = \frac{\kappa}{2} \, , \qquad
\langle\Psi_0|\{\psi(0),\pi(0)\}|\Psi_0\rangle = 0 \, ,
\eeq
from which we obtain the propagator
\begin{align}
\Delta(t_1,t_2) &= 
\frac{1}{2\kappa}
C\left(\frac{\omega t_1}{2}\right)
C\left(\frac{\omega t_2}{2}\right)
+\frac{2\kappa}{\omega^2}
S\left(\frac{\omega t_1}{2}\right)
S\left(\frac{\omega t_2}{2}\right) \nonumber \\
& \phantom{=\ }
+\frac{i\epsilon}{\omega}
\left[
C\left(\frac{\omega t_1}{2}\right)
S\left(\frac{\omega t_2}{2}\right) 
-S\left(\frac{\omega t_1}{2}\right)
C\left(\frac{\omega t_2}{2}\right) 
\right] \, ,
\end{align}
where the sign $\epsilon=+1$ for $t_2>t_1$ while $\epsilon=-1$ for
$t_1>t_2$. This step implements the time-ordering of the operators in
the definition of the propagator. 
If we expand in $|q|\ll 1$, with the characteristic exponent $\nu = q/\sqrt{2}+\mathcal{O}(q^3)$, we obtain the
propagator of the form 
\begin{align}
\Delta(t_1,t_2) &=
A(t_1)A(t_2)\bigg[
\frac{1}{2\kappa}
\left(\frac{1}{\sqrt{2}}-\frac{\omega}{\kappa}\right)^2
\cos\frac{\kappa^2}{\sqrt{2}\omega}(t_1-t_2)
+\frac{1}{2\kappa}
\left(\frac{1}{2}-\frac{\omega^2}{\kappa^2}\right)
\cos\frac{\kappa^2}{\sqrt{2}\omega}(t_1+t_2) \non
& \phantom{=A(t_1)A(t_2)\bigg[\ }
+\frac{\omega}{\sqrt{2}\kappa^2}
\exp\left(-\frac{i\kappa^2}{\sqrt{2}\omega}|t_1-t_2|\right) \bigg]
+ \mathcal{O}(q^2) \, ,
\label{bytheway}
\end{align}
where
\beq
A(t) \equiv 1 + q \sin^2\left(\frac{\omega t}{2}\right) \, . 
\eeq
In the limit $|q|\ll 1$, the function $A(t)\rightarrow 1$;
this limit reproduces the propagator of a quench in a harmonic oscillator
(see e.g. eq.~(8) of \cite{Sotiriadis:2010si}),
in which the frequency is suddenly changed from $\kappa$ to $\kappa_f$.
The final frequency of the effective quench is:
\beq
\kappa_f = \frac{\kappa^2}{\sqrt{2} \omega} \, \, ,
\eeq
in agreement with the expectation from the effective Hamiltonian, see eq.~(\ref{eq:Heff}).
Notice that the $q$ corrections on top of the quench propagator are a)
small due to the smallness of $q$ and b) are very fast
$\omega^2\gg\sqrt{2}\kappa^2$ which is equivalent to $|q|\ll\sqrt{2}$.

Consider now the case of a free field theory. 
The Hamiltonian density is: 
\beq
H_- = \frac{\pi^2}{2} + \frac{1}{2}(\nabla\psi)^2 +
\frac{\kappa^2}{2}\psi^2 \, , \qquad
H_+ = \frac{\pi^2}{2} + \frac{1}{2}(\nabla\psi)^2 +
\frac{\kappa^2}{2}\cos(\omega t)\psi^2 \, .
\eeq
After a Fourier transform in the spatial directions,
the equation of motion  is:
\beq
\ddot{\psi} + \left[a - 2q\cos(2\tau)\right] \psi = 0 \, , \qquad
\psi = \psi(\tau,\vec{k}) \, ,
\label{eq:Mathieu}
\eeq
where we have rescaled time as $\tau\equiv\omega t/2$,
$\vec{k}$ is the spatial momentum 3-vector  and 
\beq
a=\nu^2 = \frac{4 \vec{k}^2}{\omega^2} \, , \qquad
q = -\frac{2 \kappa^2}{\omega^2} \, . 
\eeq 
Eq.~\eqref{eq:Mathieu} is the Mathieu equation with parameters
$(a,q)$. The large $\omega$ expansion of $C,S$ is:
\begin{align}
C(\tau) &= \left(1+\frac{\nu\sin 2\tau \tan\nu\tau -
  2\sin^2\tau}{2(\nu^2-1)} q \right) \cos\nu\tau 
+ \mathcal{O}(q^2) \, , \\
S(\tau) &= \left(1-\frac{\nu\sin 2\tau \cot\nu\tau -
  2\cos^2\tau}{2(\nu^2-1)} q \right) \frac{\sin\nu\tau}{\nu} 
+ \mathcal{O}(q^2) \, . 
\end{align}
The propagator for a given $\vec{k}$-vector thus reads
\begin{align}
\Delta_k(t_1,t_2) &= 
\frac{1}{2\widetilde{\kappa}} A(t_1,\nu) A(t_2,\nu) \cos |k|t_1 \cos |k|t_2 
+ \frac{\widetilde{\kappa}}{2k^2} B(t_1,\nu) B(t_2,\nu) \sin |k| t_1 \sin |k|t_2
\non &\phantom{=\ }
+ \frac{i\epsilon}{2|k|}\left[
  A(t_1,\nu) B(t_2,\nu) \cos |k|t_1 \sin |k|t_2 
  - A(t_2,\nu) B(t_1,\nu)\sin |k|t_1 \cos |k|t_2\right] \, ,
  \label{propagatorone}
\end{align}
where $\widetilde{\kappa}=\sqrt{\kappa^2+k^2}$ and
$\nu=2|k|/\omega$ while
\begin{align}
A(t) &= 1+\frac{\nu\sin \omega t \tan\frac{\nu\omega t}{2} -
  2\sin^2\frac{\omega t}{2}}{2(\nu^2-1)} q 
+ \mathcal{O}(q^2) \, , \\
B(t) &= 1-\frac{\nu\sin \omega t \cot\frac{\nu\omega t}{2} -
  2\cos^2\frac{\omega t}{2}}{2(\nu^2-1)} q 
+ \mathcal{O}(q^2) \, . 
\end{align}
In the limit $q \rightarrow 0$, the above propagator
(\ref{propagatorone}) again coincides with the one of a quenched free
field theory, see \cite{Sotiriadis:2010si}. 
For each momentum mode, the initial frequency of the effective quench
is $\widetilde{\kappa}$, while the final frequency is: 
\beq
\widetilde{\kappa}_f= |\vec{k}| \, .
\eeq
The order $q$ correction to the propagator has a more complex structure
compared to the quantum mechanical case, where it is contained in the
prefactor $A(t_1) A(t_2)$ of the propagator~(\ref{bytheway}). 

As discussed in \cite{Sotiriadis:2010si}, it is possible to define a
momentum-dependent effective temperature $T_k=\beta_k^{-1}$, where 
\beq
\beta_k=\frac{2}{\widetilde{\kappa}_f } \log \left|
\frac{\widetilde{\kappa}_f+\widetilde{\kappa}} 
{\widetilde{\kappa}_f-\widetilde{\kappa}} \right| \, . 
\eeq
At large times the system tends to a state with thermal-like
correlation functions, having an effective temperature which is a
function of the absolute value of the momentum $|\vec{k}|$.

\end{document}